\documentclass[useAMS,usenatbib]{mn2e}
\usepackage{natbib}
\usepackage{color}
\usepackage{amsmath, amssymb}
\usepackage{bm}\usepackage{pifont}
\usepackage{float}
\usepackage{graphicx}
\usepackage[usenames,dvipsnames]{xcolor}
 \usepackage{relsize}
\usepackage{subfigure}  %inclusion de varias imagenes
\usepackage[singlelinecheck=false % <-- important
]{caption}

\usepackage{soul} %strikeoutcommand \st{text}
\usepackage{epsfig}
\usepackage{enumerate}
\usepackage[shortlabels]{enumitem}
\usepackage[version=3]{mhchem}
\usepackage{caption}
\usepackage{varwidth}
\newsavebox\CBox
\usepackage{color}
\usepackage{lineno}
\usepackage{epsf,epsfig}
\usepackage{epstopdf} 
\usepackage{float}
\usepackage{rotating} 
\usepackage{epstopdf} % with I can solve the problem about the pictures.
\title[The mineral clouds on  HD\,209\,458b and  HD\,189\,733b]{The mineral clouds on  HD\,209\,458b and  HD\,189\,733b}
\author[Ch. Helling, G. Lee, I. Dobbs-Dixon, N. Mayne et al.]{{Ch. Helling$^{1}$\thanks{E-mail: ch80@st-andrews.ac.uk}, G. Lee$^{1}$, I. Dobbs-Dixon$^{2}$, N. Mayne$^{3}$, D.~S. Amundsen$^{3}$} 
\newauthor{J. Khaimova$^{4,1}$, A.A. Unger$^{1}$, J. Manners$^{3}$, D. Acreman$^{5}$, C. Smith$^{5}$}\\
$^{1}$SUPA, School of Physics and Astronomy, University of St Andrews, St Andrews KY16 9SS, UK\\
$^{2}$Department of Physics, NYU Abu Dhabi PO Box 129188 Abu Dhabi, UAE\\
$^{3}$Physics and Astronomy, College of Engineering, Mathematics and Physical Sciences, University of Exeter, EX4 4QL, UK\\
$^{4}$Brooklyn College, City University of New York, US\\
$^{5}$Met Office, Exeter, EX1 3PB, UK
}

\begin{document} 
\maketitle
%\linenumbers
\label{firstpage}

\begin{abstract}
   3D atmosphere model results are used to comparatively study the
   kinetic, non-equilibrium cloud formation in the atmospheres of 
     two example planets guided by the giant gas planets
   HD\,209\,458b and HD\,189\,733b. Rather independently of
     hydrodynamic model differences, our cloud modelling suggests that both
   planets are covered in mineral clouds throughout the entire
   modelling domain.  Both planets harbour chemically complex
   clouds that are made of mineral particles that have a
   height-dependent material composition and size. The remaining
   gas-phase element abundances strongly effects the molecular
   abundances of the atmosphere in the cloud forming
   regions. Hydrocarbon and cyanopolyyne molecules can be rather
   abundant in the inner, dense part of the atmospheres of
   HD\,189\,733b and HD\,209\,458b. No one value for metallicity and
   the C/O ratio can be used to describe an extrasolar planet.  Our
   results concerning the presence and location of water in relation
   to the clouds explain some of the observed differences between
   the two planets. In HD\,189\,733b, strong water features have been
   reported while such features are less strong %not as clear
    for HD\,209\,458b.
   By considering the location of the clouds in the two atmospheres,
   we see that obscuring clouds exist high in the atmosphere of
   HD\,209\,458b, but much deeper in HD\,189\,733b. We further
   conclude that the (self-imposed) degeneracy of cloud parameters in
   retrieval methods can only be lifted if the cloud formation
   processes are accurately modelled in contrast to prescribing them
   by  independent parameters.
\end{abstract}


%________________________________________________________________

\section{Introduction}\label{s:intro}

%   Cloud formation is a primary motivator to determine quantities of
%   chemical abundance, opacities, and albedo.  Results from Radiative
%   Hydrodynamic models that drive energy transport feed into the cloud
%   formation model giving local properties of dust grains around the
 %  exoplanet.

HD\,189\,733b and HD\,209\,458b are the two most observed extrasolar
planets. Both planets are highly irradiated due to their close
proximity to their host star. HD\,189\,733b forms a bow shock in the
wind of its host star (\citealt{llama2013}) from which the planetary
magnetic field strength can be derived. \cite{cauley2015} obtained
HIRES observations in H$\alpha$ which suggest a magnetic field
strength of 28G for the giant gas planet HD\,189\,733b.  {\it Chandra}
observations (0.1 - 5 keV) indicate a deeper X-ray transit than for optical
wavelengths (\citealt{popp2013}), but see \citealt{llama2015}
for the effect of stellar variability.
%, and the observation of Na D lines
%blue-shifted by $8\pm 2$ km s$^{-1}$ in the upper layers of the
%atmosphere above a cloud deck (\citealt{wytt2015}) were both
%interpreted as atmospheric mass loss on HD\,189\,733b.  HARPS
%observations enabled the spatial resolution of an eastward wind on
\cite{leca2010} presented the  first study of HD\,189\,733b mass loss based on HST transit observations, later confirmed in (\citealt{leca2012}). HARPS observations of blue-shifted Na D lines ($8\pm 2$ km s$^{-1}$) in the upper layers of the
atmosphere above a cloud deck (\citealt{wytt2015}) enabled the spatial resolution of an eastward wind on
HD\,189\,733b (\citealt{Louden2015}) indicating a globally circulating
atmosphere (\citealt{show2008,Dobbs-Dixon2013}). The presence of
H$_2$O is suggested by dayside emission spectrum
(\citealt{grill2008}), secondary eclipse emission spectrum
(\citealt{tod2014,cro2014}), transit absorption spectrum
(\citealt{Mcc2014}) and by ground-based high-resolution spectroscopy
(\citealt{birk2013}). \cite{swain2014} find strong evidence for CH$_4$
absorption in their NICMOS transmission spectra and interpret this
finding as a confirmation of earlier publications on the detection of
CH$_4$ in the atmosphere of HD\,189\,733b. \cite{rodler2013} present
the detection of CO at the HD\,189\,733b day-side from NIRSPEC
observations (see also \citealt{des2009,dek2013}). \cite{knu2012}
require a vertical mixing enhancement of CO to explain their Spitzer
observations. \cite{dek2013} could not confirm CO$_2$ NICMOS secondary
eclipse observation. The efforts of deriving observational constraints
on the atmosphere chemistry are challenged by the repeated detection
of a thick cloud layer on HD\,189\,733b through transit spectroscopy
(\citealt{gib2012,pont2013}). \cite{leca2008} were the first to
use MgSiO$_3$[s] as a possible abundant condensate with particle
size $\approx 10^{-2}\,\ldots\,0.1\mu$m based on transit observations
with ACS on Hubble.  Retrieval methods were applied to
reflectance spectra to provide further insight in chemical composition
of the atmosphere and cloud properties
(\citealt{bars2014,lee2014,benneke2015}).

%Observations of hydrogen and Mg atoms suggest a strong mass loss
%escaping from HD\,209\,458b's exosphere (\citealt{kis2014,bour2014}).
The first mass loss detection on HD\,209\,458b was made by
\cite{vida2003} and \cite{vida2004} suggesting the escape of HI, OI
and CII. \cite{vida2013} report absorption in MgI in a region where
atmospheric escape is suggested to take place.  4.5$\mu$m Spitzer/IRAC
data imply the presence of a hot spot shifted eastward by
$\approx40^o$ and a day-night-side temperature difference of
$\approx500$K (\citealt{zell2014}). The analysis of Spitzer/Infrared
Array Camera primary transit and secondary eclipse light curves
(\citealt{evans2015}) did not confirm the detection of water in
HD\,209\,458b, but suggest a depletion of atmospheric CO and no
thermal inversion layer to be detectable at Spitzer wavelengths (also
\citealt{dia2014}).  However, the Hubble observation at 1.4$\mu$m by
\cite{deming2013} provides evidence for H$_2$O absorption in
HD\,209\,458b which was confirmed by \cite{sing2016}.  \cite{schw2015}
observed the thermal dayside emission by high-resolution spectroscopy
of the carbon monoxide band at 2.3 $\mu$m. They do not find any
emission signature required to confirm a thermal inversion in the
atmosphere of HD\,209\,458b. \cite{hoe2015} suggest that their TiO
opacity data was insufficient to reach any conclusion regarding a
thermal inversion on HD\,209\,458b. These conclusions, however, are
based on the assumption that the featured molecules are associated
with a thermal inversion and not with another process that affects,
for example, their abundance. The presence of TiO and VO is derived
from STISS observation of HD\,209\,458b (\citealt{des2008}).  It is
suggested that TiO and VO exist in two distinct layers including one
at higher atmospheric layers above the $S=1$-level where potential
condensates would be thermally stable\footnote{The supersaturation
  ratio, S, equals 1 if the growth and the evaporation rates are
  equal, hence when thermal stability, or phase equilibrium, is
  achieved. The supersaturation ratio is defines as $S(T)= p_{\rm
    x}(T_{\rm gas}, p_{\rm gas} )/p_{\rm sat, s}(T_{\rm s})$ with
  $p_{\rm x}(T_{\rm gas}$ partial pressure of the gas species x, and
  $p_{\rm sat, s}(T_{\rm s})$ is the saturation vapour pressure of the
  (e.g. solid) material s.}.  This conclusion about the presence of
gaseous TiO or VO, however, can only be correct if the layer above
$S=1$ is warmer such that $S<1$, i.e. in a case of an outward
temperature increase. \cite{kos2013} conclude that the detection of
Si$^{2+}$ indicates that clouds involving Si do not form in the
atmosphere of HD\,209\,458b.

Both planets have been compared regarding the interplay between the
irradiation and the atmosphere dynamics and its effect on the
molecular gas composition (\citealt{agu2012,agu2014}), regarding their
mass loss (\citealt{bour2013}), and their cloud-free dynamic
atmosphere structure (\citealt{show2009}).  \cite{benneke2015}
suggests from his retrieval procedure that both planets, HD\,209\,458b
and HD\,189\,733b host a thick cloud deck. Our paper takes this
approach further to present a comparative study of the cloud
structures of HD\,209\,458b and HD\,189\,733b based on kinetic
simulation of the non-equilibrium cloud forming processes in
combination with results of 3D atmosphere structure simulations. 
  We, however, note that our present approach is limited by the need
  to run the cloud-formation and 3D RHD simulations independently.

%\noindent
%Modelling:\\*[-0.5cm]
%\begin{itemize}
%\item HD 189733b: Rauscher \& Menou (2013): ohmic dissipation
%\item GJ 1214b: \cite{char2015}
%\item HD 209b: Roger \& Showman (2014): Magnetohydrodynamic Simulations  
%\end{itemize}

\begin{table}
\caption{Exoplanet Data}
\centering
\begin{tabular}{l l l}
\hline\hline
Parameters &  HD\,189\,733b &  HD\,209\,458b  \\ [0.5ex] % inserts table
%heading
\hline % inserts single horizontal line
Mass & 1.13 $\pm$0.025 M$_J$ & 0.69 $\pm$0.017 M$_J$ \\ 
          &    \citep{Boisse2009}                & \citep{MNR:MNR19600}  \\
Semi-Major  & 0.03142    & 0.04747 \\
Axis                & $\pm$0.00052 AU & $\pm$0.00055 AU \\
                &  \multicolumn{2}{c}{\citep{2010MNRAS.408.1689S}}  \\
Orbital Period & 2.21857d & 3.52472 d \\
                       & \citep{refId0} & \citep{MNR:MNR19600} \\
Eccentricity & 0.041 & 0.0082  \\
                   & \citep{refId0} & \citep{MNR:MNR19600} \\
Radius & 1.138 $\pm$0.077 R$_J$ & 1.38 R$_J$  \\ 
            & \citep{0004-637X-677-2-1324}  & \citep{2010MNRAS.408.1689S} \\
Geometric & 0.4 $\pm$0.12 & 0.038 ($\pm$0.045)  \\
Albedo                              & \citep{2041-8205-772-2-L16} & \citep{2008ApJ...689.1345R} \\
Mean Density & 1800 kgm$^{-3}$  & 616 kgm$^{-3}$ \\[1ex]
T$_{\rm equ, planet}$ & $\sim$ 1000K &  $\sim$ 1500K\\
geometrical & & \\
 albedo & & \\
\hline
Parameters &  HD\,189\,733 &  HD\,209\,458  \\ [0.5ex] % inserts table
SpecTyp & K2V & G0V\\
               & \citep{Boisse2009} & \\
T$_{\rm eff, *}$ & $\sim$ 4800K & $\sim$ 6000K\\ 
               & \multicolumn{2}{c}{\citep{Boyajian2015}}\\
\hline 
\end{tabular}
\label{table:exoplanet} 
\end{table}

Every observation requires input from a model.   Interpretations of
all observations for HD\,189\,733b and HD\,209\,458b rely on various
modelling approaches which mostly only include a subset of physics
complexes like radiative transfer incl. gas and cloud opacities,
gas-phase chemistry (kinetic or LTE), hydrodynamics (HD),
magnetohydrodynamics (MHD), and cloud formation. The radiative
transfer problem and the gas-phase chemical abundances have achieved
rather high standards as core problems of atmosphere modelling and
data analysis (see review by \citealt{marley2015ARAA}, also
\citealt{benneke2015}). HD and MHD have a long standing tradition in
meteorology and in solar physics, respectively.  Cloud formation, in
contrast, is almost always parameterised in terms of number and sizes
of particles, settling and mixing efficiencies in complex simulations
like as part of radiative transfer simulations or circulation models
(see e.g. \citealt{char2015}). Parameterisation approaches have been
guided by terrestrial studies, which however, treat an environment
that does not compare overly well with giant gas planets. Main
differences are the chemical composition of the atmospheric gas and
the need to treat the formation of condensation seeds in situ (see
review by \citealt{helling2013RSPTA}, also Sect. 3 in
\citealt{helling2014AARv}). We therefore aim to provide some
  guidance for future cloud parameterisations based on the examples
  HD\,189\,733b and HD\,209\,458b for other planets as needed in
  preparation of JWST.

This paper applies a kinetic cloud formation model to study and
compare the cloud structures of the two example extrasolar giant gas planets
HD\,189\,733b \& HD\,209\,458b based on global circulation models of
their dynamic atmospheres (\citealt{Dobbs-Dixon2013, Mayne2014}) which were run for
600 and 1800 Earth days, respectively.
All local cloud properties (number, size and material composition of cloud
particles, depleted element abundances, cloud opacity) are determined
by the local thermo- and hydrodynamic properties. We aim to
demonstrate and to compare the consistently calculated cloud
structures for both planets and the chemical feedback on the
background atmosphere gas. Section~\ref{s:approach} outlines our
approach. Section~\ref{s:localTD} presents the local thermodynamic
properties of the atmospheres of HD\,189\,733b and HD\,209\,458b in
comparison. Section~\ref{s:cloudstruc} discusses the resulting local
cloud properties for the two giant gas planets with focus on the
changing cloud material and grain sizes across the globe as major
opacity input. Section~\ref{s:gaschem} compares the impact of cloud
formation on the local chemistry for HD\,189\,733b and HD\,209\,458
and how it changes with longitude as result of the underlying
atmosphere structure. The changing element abundances and the
molecular composition are discussed with the resulting inhomogenous
metallicity and C/O ratio throughout an atmosphere with clouds. We also
discuss the abundance of larger hydrocarbon and cyanopolyyne molecules
in the inner atmosphere.  Section~\ref{s:opacity} compares the
opacities for the day and the night side of both
planets. Section~\ref{s:disc} presents our discussion and Sect.~\ref{s:concl} concludes this paper.

\section{Approach}\label{s:approach}

We apply the two-model approach outlined in \cite{Lee2015}. We extract
1D (T$_{\rm gas}$(z), p$_{\rm gas}$(z), v$_{\rm z}(z)$)-trajectories
from 3D HD atmosphere simulations across the globe and use these
structures as input for our kinetic, non-equilibrium cloud-formation
code {\sc Drift}
(\citealt{Woitke2003,Woitke2004,Helling2006,Helling2008}). The 3D
results for HD\,189\,733b are from the 3D radiation-hydrodynamic
simulation by \cite{Dobbs-Dixon2013}, the 3D results for HD\,209\,458b
are from the 3D GCM simulations by \cite{Mayne2014} and
%the further development in 
\cite{Amundsen2014}. We refer to the individual papers regarding more
details on the 3D simulations and also the cloud formation
modelling. A summary of the different modelling parts and of our
assumptions is given below.   The approach that this paper follows
  has the limitation of not taking into account the potential effect
  of horizontal winds on the cloud formation. We note, however, that
  the cloud formation processes are determined by the local
  thermodynamic properties which are the result of the 3D dynamic
  atmosphere simulations in our approach. Horizontal winds would
  affect the cloud formation profoundly if the horizontal wind
  time-scale would be of the order of the time-scales of the
  microscopic cloud formation processes. We have demonstrated that
  this is not the case in \cite{Lee2015}. \cite{Woitke2003} have shown
  that latent heat release in negligible for the condensation of the
  materials considered here. However, a complete and consistent
  coupling of our non-equilibrium cloud formation model with a full 3D
  radiative hydrodynamics is most desirable to study the cloud
  formation affects in considerably more details than our present
  approach allows. We further note that some of the results from
  \cite{Lee2015} are reproduced in the present paper for the benefit
  of the reader.

\begin{figure*}
\centering
\includegraphics[scale=0.45]{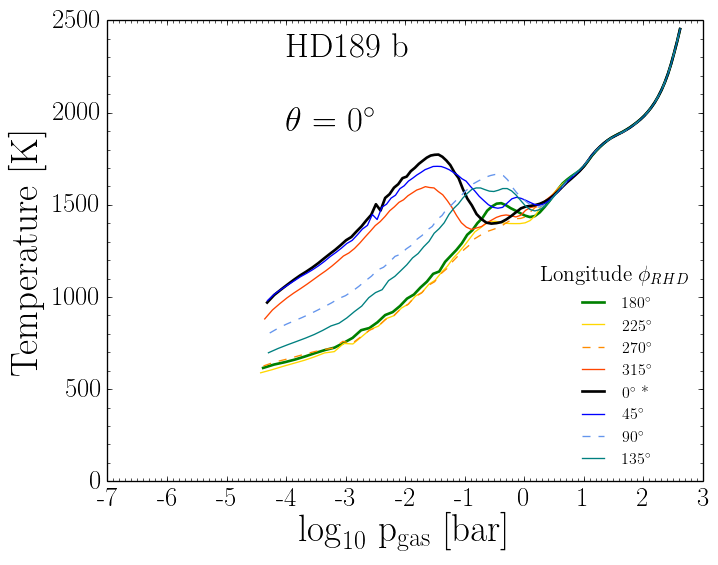}
\includegraphics[scale=0.45]{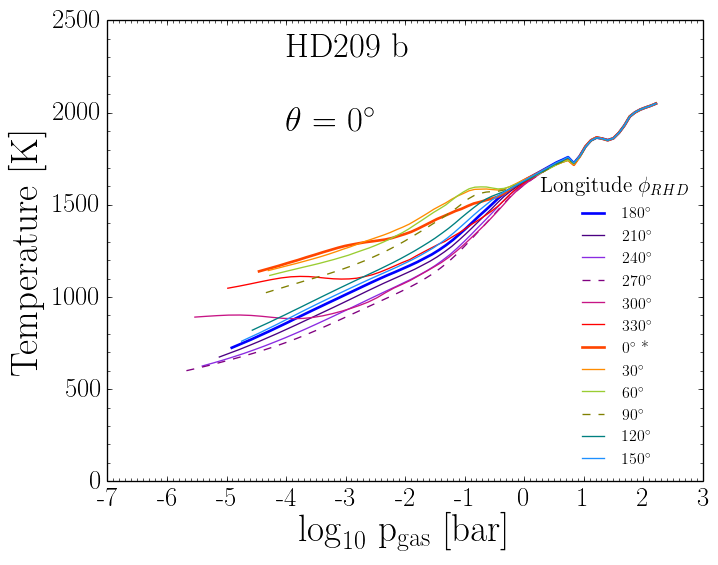}
\caption{Longitudinal and vertical changes of the local gas
  temperature, T$_{\rm gas}$ [K], and gas pressure, p$_{\rm gas}$ [bar]
  along the equator ($\theta=0^{\circ}$) for HD\,189\,733b (left) and
  HD\,209\,458b (right). The (T$_{\rm gas}$, p$_{\rm gas}$)-profiles
  are results of 3D atmosphere simulations by \citep{Dobbs-Dixon2013}
  for HD\,189\,733b (left) and by \citet{Mayne2014a} for HD\,209\,458b
  (right). The parameter for the planets and their host stars are summarized in 
  Table~\ref{table:exoplanet}.}
\label{fig:Tp}
\end{figure*}

\begin{figure*}
\centering
\includegraphics[scale=0.47]{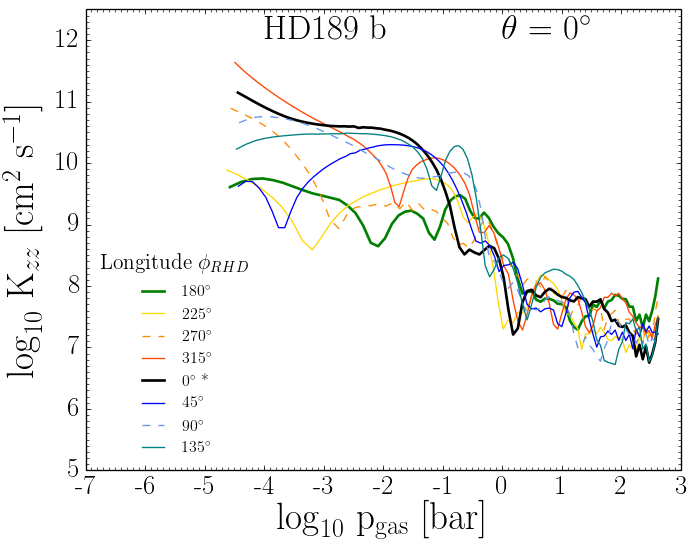}
\includegraphics[scale=0.47]{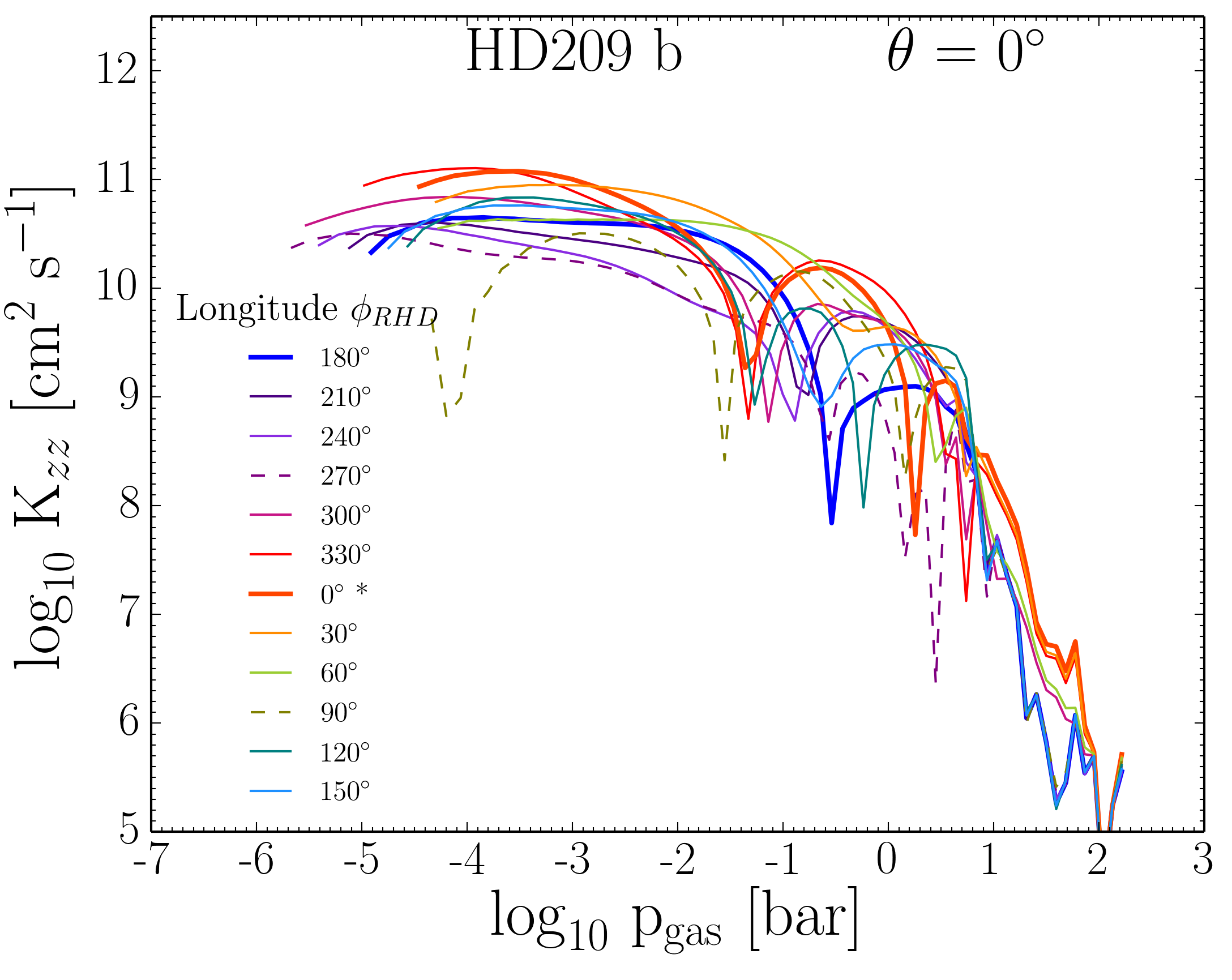}
\caption{The local mixing parameter, $K_{\rm zz}=H_{\rm P}\cdot v_{\rm
    z}$ [cm$^2$ s$^{-1}$], for HD\,189\,733b (left) and HD\,209\,458b
  (right) along the equator ($\phi=0^{\circ}$) for different
  latitudes. $v_{\rm z}$ and $H_{\rm P}$ are results from
  the same models as in Fig.~\ref{fig:Tp}.}
\label{fig:Kzz}
\end{figure*}

\subsection{Kinetic formation of cloud particles from oxygen-rich gases}

Cloud formation in extrasolar atmospheres requires the formation of
seed particles because  giant gas planets have no crust from which
sand or ash particles are diffused upwards or injected into the
atmosphere by volcanic eruptions.  These condensation seeds provide a
surface onto which other materials can condense more easily as surface
reactions are considerably more efficient than the sum of chemical gas
phase reactions leading to the formation of the seed. The formation of
the first surface out of the gas phase proceeds by a number of
subsequent chemical reactions that eventually result in small seed
particles. Such a chain of chemical reactions can proceed by adding a
molecular unit during each reaction step (e.g.,
\citealt{jeong2000,plane2013}). \cite{gou2012}, for example, show that
condensation occurs from small gas-phase constituent like MgO and SiO,
which will lead to the formation of bigger units like Mg$_2$SiO$_4$
during the condensation process. It is important to realise that big
molecules like Mg$_2$SiO$_4$ or Ca$_4$Ti$_3$O$_{10}$ do not exist in
the gas phase (compare also Sect.~\ref{s:gaschem}). Here, the concept
of homogeneous nucleation is applied to model the formation of TiO$_2$
seed particles (\citealt{helling2013RSPTA,lee2015a}) which allows us
to calculate the nucleation rate, $J_*$ [cm$^{-3}$ s$^{-1}$], which
determines the number of cloud particles, n$_{\rm d}$ [cm$^{-3}$], and
therefore influences the total cloud surface.
% A peak of the cloud
%particle number density (n$_{\rm d}$ [cm$^{-3}$]) coincides with the peak of the
%nucleation rate but the number density remains high towards higher
%temperature. This is a clear sign that cloud particles fall into the
%atmosphere and therefore do exist below the seed formation region.
The cloud particle size, $\langle a\rangle$ [cm], is determined by the
efficiency of surface growth and evaporation reactions. These
reactions are determined by the composition of the gas phase that
provides the number density for surface reactions (Table 1 in
\citealt{Helling2008}) and leads to the formation of a substantial
material mantle on top of the seed. This mantle determines the mass,
volume ($V_{\rm tot}$) and main chemical composition, $V_{\rm
  s}/V_{\rm tot}$ ($V_{\rm s}$ - volume fraction of material s), of
the cloud particles. Many materials can be simultaneously thermally
stable in a gas but these materials change depending on the
carbon-to-oxygen ratio and the abundance ratios of other elements. In
principle, all thermally stable materials grow simultaneously on a
seed particle as long as the key species (e.g. TiO for TiO$_2$[s];
SiO, Mg, SiS, Mg(OH)$_2$ for Mg$_2$SiO$_4$[s]; CaO, CaS, TiO for
CaTiO$_3$[s]) are present in the gas phase. Our cloud model includes
12 solid species that form from 60 gas-solid surface reactions (Table
1 in \citealt{Helling2008}). These are the solids that are expected to
provide the most efficient growth materials and include silicates
(MgSiO$_3$[s], Mg$_2$SiO$_4$[s], SiO[s], SiO$_2$[s]), metals and metal
oxides (MgO[s], Fe[s], FeO[s], Fe$_2$O$_3$[s], Al$_2$O$_3$[s],
TiO$_2$[s], CaTiO$_3$[s]) and others (FeS[s]). Internal rearrangement
may lead to the formation of Ca$_2$Al$_2$SiO$_7$[s] or
MgAl$_2$O$_4$[s] which are long living minerals that, however, can not
form directly from the gas phase. The life time of any of the cloud
particles might, however, be too short for such rearrangement
processes to have any effect.

Once cloud particles have formed, they are affected by gravity and
friction with the surrounding gas. The equilibrium between friction
and gravity determine how strongly the cloud particles are coupled to
the gas and, hence, how fast the cloud particles fall through the
atmosphere (\citealt{Woitke2003}).  This gravitational settling
determines the vertical extension of a cloud. The cloud particles will
continue to grow and to change their material composition during their
descent into denser and warmer atmospheric regions. Mineral cloud
particles can reach a size for which they fall faster than they can
grow, hence, the cloud particle will not change any more in size and
composition (\citealt{Woitke2004}).  The cloud formation has a strong
impact on the local chemistry by depleting those elements that
participate in the formation of the cloud particles. This will result
in an inhomogeneous metallicity and C/O ratio of the atmosphere (see
Sect.~\ref{ss:elmab}).

If nucleation and growth become inefficient, large-scale processes
like gravitational settling determine the global cloud formation
(Sect. 6 in \citealt{Woitke2003}). The loss of elements over
large-scales is counter balanced by the (vertical) mixing processes (see
Sect.~2.4.2. in \citealt{Lee2015}). Different approaches are applied
in the literature to account for vertical mixing. For example,
\cite{acker2001} use an equation where gravitational settling is
balanced by cloud material diffusion. The local amount of cloud
material is then derived for a given settling parameter ($f_{\rm
  sed}$) and diffusion coefficient ($K_{\rm zz}$, see pg 5 in
  \citealt{Lee2015} for a discussion). It remains to prescribe the
material composition and a grain size distribution (like Eq. 2 in
\citealt{mor2015}). The approach presented in \cite{Woitke2003}, and
utilized in this paper, applies to force balance between friction and
gravity to derive a size-dependent drift velocity which is required to
determine a drift dependent growth term in the moment equations. The
vertical mixing efficiency enters by a mixing time scale that is
derived from the local gas velocity. Beside the local element
abundances, $T_{\rm gas}(z)$ and $\rho_{\rm gas}(z)$ (local gas
density, [g cm$^{-3}$]) determine if atmospheric clouds can form, to
which sizes the cloud particles grow and of which materials they will
be composed. The gravitational settling through an atmosphere is
determined by the local gas density, $\rho_{\rm gas}(z)$, and the
cloud particle size.

The mathematical model is formulated in terms of moment equations
which describes the above physical processes
(\citealt{Woitke2003,Helling2006,Helling2008,helling2013RSPTA}).

We assume that both planets, HD\,189\,733b and HD\,209\,458b, have an
oxygen-rich atmosphere of approximately solar element composition. We
use the solar element abundances as boundary and initial values for
the cloud formation simulation.

\subsection{Chemical gas composition}
We apply a chemical equilibrium (thermochemical equilibrium) routine
to investigate the chemical composition of the atmospheres. We use the
1D (T$_{\rm gas}(z)$, p$_{\rm gas}(z)$) trajectories and element
abundances $\epsilon_{\rm i}(z)$ (i=O, Ca, S, Al, Fe, Si, Mg) depleted
by the cloud formation processes.  All other elements are assumed to
be of solar abundance. Given that collisional gas-phase processes
dominate in the atmospheric part of interest, we apply a chemical
equilibrium routine that allows us to provide first estimates of the
abundances of carbon-bearing macro-molecules and small PAHs. This
approach allows us to study the gas-phase abundances at the inner
boundary of future kinetic considerations, and to look at species not
presently included in most of the current networks (e.g. C$_6$H$_6$).
A combination of $199$ gas-phase molecules (including $33$ complex
carbon-bearing molecules), $16$ atoms, and various ionic species were
used under the assumption of LTE. \cite{Bilger2013} is an extension of
the gas-phase chemistry routine used so far in our dust cloud
formation according to \citet{Helling1996}. The \citet{Grevesse2007}
solar composition is used for calculating the gas-phase chemistry
outside the metal depleted cloud layers and before cloud formation.
No solid particles were included in the chemical equilibrium
calculations in contrast to equilibrium condensation models. The
influence of cloud formation on the gas phase composition results from
the reduced element abundances due to cloud formation and the cloud
opacity impact on the radiation field, and hence, on the local gas
temperature and gas pressure.

We demonstrate that the gas phase composition is affected by cloud
formation to a degree that it should be taken into account also as
inner boundary and initial values for kinetic gas phase calculations
for irradiated, giant gas planets.

\subsection{3D atmosphere simulation}\label{ss:GCM209}

\subsubsection{HD\,209\,458b  (\citet{Mayne2014a})}
The simulations of HD\,209\,458b we use as input were run using the
adapted UK Met Office General Circulation Model (GCM), called the
Unified Model (UM). The testing and adaptation of this model to hot
Jupiter conditions is detailed in \citet{Mayne2014a,Mayne2014b} and
\citet{Amundsen2014}. Briefly, the model solves the fully compressible
Euler equations, including a height varying gravity. The equations are
solved on a longitude-latitude-height grid using a semi-Lagrangian,
semi-implicit scheme \citep[see][for more details]{Mayne2014a}. The
lower and upper boundaries are both free-slip and impermeable.  To
represent the convective flux from the interior, an additional net
upward flux is added to the black body emission from the bottom
boundary (i.e. additional to the emission required to balance the
downward radiation). This convective flux corresponds to an intrinsic
temperature of 100K as determined from internal evolution models (see
\citealt{Amundsen2014} for details).  
%The lower boundary includes a
%blackbody emission (with $T_{\rm eff}=100\,$K) accounting for heat
%escaping from the interior, and t
At the upper (outer) boundary a so-called {\it ghost layer} is
included where the atmosphere is extrapolated to account for
absorption, scattering and emission above the dynamically modeled
domain.  The radiative transfer is solved using the two-stream
approximation and opacities treated using the correlated-k method
\citep[see][for more details]{Amundsen2014}. The opacities are derived
using analytic abundances for chemical equilibrium, and no additional
opacity sources are added.
% to, for example, produce a temperature inversion. 
The planetary parameters are selected to match
observational constraints of HD\,209\,458b as given in
Table~\ref{table:exoplanet}. The planet radius, however, is set to
1.259\,R$_{\rm jup}$ to take into account the atmosphere contribution.
Full details of the model setup can be found in
\citet{Amundsen2015thesis}.

\subsubsection{HD\,189\,773b (\citet{Dobbs-Dixon2013})} 
The simulations of HD\,189\,733b were calculated using the model
described in \cite{Dobbs-Dixon2013}. This model solves the fully
compressible Navier-Stokes equations coupled to a two-stream radiative
transfer scheme averaging over 30 wavelength bins. The equations are
solved on a latitude, longitude, radius grid. The temperature and
pressure dependent molecular opacities (calculated assuming solar
composition) are supplemented by both a grey component and a Rayleigh
scattering tail to approximate the effects of clouds. Detailed
observational diagnostics have been carried out and the best-fit model
(utilized here) has a kinematic viscosity of $\nu=10^7$
cm$^2$s$^{-1}$. Similar to the HD\,209\,458b model, the lower boundary
is slip-free and impermeable. A heat flux from the interior
corresponding to a blackbody with a temperature of $170$ K is assumed
to be flowing through the interior boundary. The exterior boundary is
imposed at a density of $\rho=10^{-9}$g\,cm$^{-3}$, is assumed to be
isothermal, and allows outflow. Heating is incorporated through a
spatially varying direct heating term in each of the wavelength
bins. Please refer to \cite{Dobbs-Dixon2013} for more details.

\smallskip
The 3D atmosphere simulations utilised here solve the full 3D
  Navier-Stockes equations (table 2 in \citet{Mayne2014a}) coupled
  with a solution of the radiative transfer problem. The largest
  difference between the codes arises from the different methods that
  are employed to solve the radiative transfer as input to the energy
  equation. The largest uncertainty arrises from the need to prescribe
  opacity values as in both cases the solution of the full gas-phase
  chemistry for a potentially large number of frequencies is
  computational unfeasible.
% It might be interesting in this context, that \cite{kitz2016} show that a reduced number of %sampling points does allow an good-quality solution of the radiative transfer in a planetary %atmosphere with a complex chemical structure. 
The differences in the radiative transfer treatment will affect the details of the local values, while the global properties maybe less affected.

\subsection{Atmosphere boundaries}
The hydrodynamic and radiative transfer boundary conditions determine
the computational domain of every atmosphere simulation. In the
present cases, they determine the vertical extension towards lower
pressures and temperature differently for the two cases considered here:

-- All GCMs place a top boundary at either a set height or a pressure
based usually on some value that assures numeric stability. The height
is increased until the contribution of the {\it ghost layer} to
the derived thermodynamic structure (T$_{\rm gas}$, p$_{\rm gas}$) of
the atmosphere is not significant, providing the pressure and
temperature at the upper/outer boundary. This means that any
significant heating will be resolved within the atmospheric domain
simulated.  It will provide the pressure minimum and the
temperature at the top of the modelled atmosphere.

\medskip
We present our comparison results for three specific
latitude-longitude ($\theta$, $\phi$) coordinates\footnote{We remark that the
  coordinate system used in meteorology refers to $\lambda$ as the
  longitude and $\phi$ as latitude.} but provide the full
comparison in Sect.~\ref{ss:chemequ}. The chosen ($\theta$,
$\phi$)-pairs represent the substellar point ($\theta=0^{\circ}$,
$\phi=0^{\circ}$), and the two terminators (($\theta=0^{\circ}$,
$\phi=90^{\circ}$) and ($\theta=0^{\circ}$, $\phi=270^{\circ}$)). Our
($\theta$, $\phi$)-quotations follow \cite{Dobbs-Dixon2013} and have
been used in \cite{Lee2015}. For visualization see Fig. 1 in
\cite{Lee2015}. The day/night side coordinates used here are as follows:
\begin{itemize}
\item Days side:\\
HD\,189\,733b: $\phi=315^o, 0^o, 45^o$\\
HD\,209\,458b: $\phi=300^o, 330^o, 0^o, 30^o, 60^o$

\item Day/night terminator:  $\phi=270^o, 90^o$

\item Night side\\
HD\,189\,733b: $\phi=135^o, 180^o, 225^o$\\
HD\,209\,458b: $\phi=120^o, 150^o, 180^o, 210^o, 240^o$
\end{itemize}

\section{The general dynamical regime of the hot Jupiters}

The dominant dynamical feature in hot Jupiter atmospheres is the
super-rotating equatorial jet seen in the simulations of both
HD\,209\,458b and HD\,189\,733b. The stellar radiative forcing coupled
to the slow planetary rotation, drives equatorial Rossby and Kelvin
waves that act to pump angular momentum to the equator creating and
maintaining super-rotation at the equator and retrograde jets at
mid-latitude (\citealt{show2011,tsai2014}). These jets, which can
reach speeds of Mach $1.5$, advect significant energy throughout the
atmosphere. Though the horizontal velocities are not directly considered
in this paper, the pressure-temperature profiles that we use are
significantly influenced by the jet structure.

For both our target planets the general flow regime includes a
prograde equatorial jet flanked by a retrograde high latitude flow,
for the low pressure strongly radiatively forced atmosphere. The outer
or low pressure atmosphere also exhibits strong day night contrasts
(and in HD\,189\,733b's case a thermal inversion discussed later), which
decreases as we move deeper into the higher pressure regions.

\section{The local thermodynamic structures of  HD\,189\,733b and  HD\,209\,458b}\label{s:localTD}

The local thermodynamic structure (T$_{\rm gas}(z)$, p$_{\rm gas}(z)$)
determines the cloud formation because the cloud formation processes
(nucleation, growth/evaporation) depend on local properties only. 

The 3D atmosphere simulations utilised here suggest for both planets,
HD\,189\,733b (left) and HD\,209\,458b (right) in Fig.~\ref{fig:Tp}, a
distinct day-night difference of $\approx 500$K in the upper
atmosphere.  The local gas temperature, T$_{\rm gas}$, is
independent on longitude, $\phi$, in the inner (deeper), high-pressure
part of the atmosphere for p$_{\rm gas}>5$bar in both giant gas
planets. The velocity, however, changes dependent on longitude as the
vertical mixing parameter, $K_{\rm zz}$, demonstrate in
Fig.~\ref{fig:Kzz}.  The HD\,209\,458b
(T$_{\rm gas}(z)$, p$_{\rm gas}(z)$) structures are generally smoother
than in HD\,189\,733b across all latitudes ($\phi$).

The local, height-dependent thermodynamic structures of HD\,189\,733b
and HD\,209\,458b (Fig.~\ref{fig:Tp}) differ substantially for
different longitudes as both planets differ in mass ($M_{\rm HD189b}>
M_{\rm HD209b}$) and radii ($R_{\rm HD189b}< R_{\rm HD209b}$), hence
both planets have different bulk densities, $\rho_{\rm bulk,
  HD189b}\gg\rho_{\rm bulk, HD209b}$, and for the atmospheric scale
height, $H$, follows that $H_{\rm bulk, HD189b}< H_{\rm bulk,
  HD209b}$.  HD\,209\,458b orbits its host star at a somewhat larger
distance than HD\,189\,733b. The result is that the
local (T$_{\rm gas}(z)$, p$_{\rm gas}(z)$)-profiles along the equator
differ considerably between the two planets, as does the local
vertical mixing. These differences will affect the
cloud structure of the planets.
% We therefore point out some of the most important features in the
%(T$_{\rm gas}(z)$, p$_{\rm gas}(z)$)-profiles that distinguish the
%two planets.
The (T$_{\rm gas}(z)$, p$_{\rm gas}(z)$) structures span different
ranges in gas pressure with HD\,189\,733b reaching deeper into the
high-pressure regime than HD\,209\,458b.  The extension of
HD\,209\,458b into lower pressures compared to HD\,189\,733b is a
result of the choice of the outer boundary conditions for both
simulations.  

Temperature inversion inside atmospheres are generally interesting
features as they can develop a feedback with the local velocity
structure. Temperature inversions (or pressure inversion) can occur as
a result of the hydrodynamic evolution like for the present simulation
of HD\,189\,733b or as result of an increase in local opacity
(e.g. due to C$_2$ and C$_3$ \citealt{hell2000} or TiO/VO
\citealt{fort2008}). Unlike for C$_2$ and C$_3$ in the case of AGB
stars, TiO and VO had to be artificially injected at appropriate
atmospheric heights as both molecules would not be present in the gas
phase in the upper atmosphere (compare Fig.~\ref{fig:molabun18920900}). In the present 3D atmosphere simulation for HD\,189\,733b, a
temperature inversion develops without the need of an an additional
opacity source.  The HD\,209\,458b atmosphere simulation does not
suggest such a pronounced temperature inversion inside the
atmosphere. If an opacity source is artificially added in form of
TiO/VO, also the HD\,209\,458b models develops a temperature inversion
(\citealt{Showman2009}) at $\sim 10^{-3}$ bar on the day
side. However, also the add-on-free-TiO/VO HD\,209\,458b model suggest
a small temperature inversion at longitude $\phi=30^o$ and latitude
$\theta=0^o$ at $p_{\rm gas}\approx 0.1$bar. A tendency of an outward
temperature increase occurs for $\phi=300^o$ and $\phi=330^o$ at
$\theta=0^o$.  Comparing system parameters for the two planets
(Table~\ref{table:exoplanet}) shows that HD\,209\,458b orbits its host
star at a larger distance than HD\,189\,733b, hence HD\,209\,458b
receives less irradiation.

There are two effects working together to create the temperature
inversions seen in some of the profiles for HD\,189\,733b: stellar
heating at low pressures and advective cooling at higher
pressures. The peak in the temperature profiles at p$_{\rm gas}\approx
5\cdot 10^{-2}$ bars is associated with the peak in the net
stellar-energy deposition. In the absence of dynamics, the radiation
would diffuse inward until the region at higher pressures became
roughly isothermal (e.g. \citealt{guill2010}). However, including
dynamics results in a return flow of gas from the night side that is
cooler then the overlying gas. The radiative timescale increases with
depth, becoming comparable to the advective timescale at approximately
0.5 bars. The radiative heating of the newly arriving cooler gas is
insufficient to create an isothermal region that extends throughout
the region.  Balancing the radiative heating and advective cooling
yields the decrease in temperature in the region around 0.5 bars.

\subsection{Day/night temperature difference}

The day/night side temperature difference, $\Delta T_{\rm day/night}$,
are largest in the upper, low pressure part for both giant gas planets
due to the effect of stellar irradiation. $\Delta T_{\rm day/night}
\rightarrow 0$ in the high-pressure atmospheres for $p_{\rm gas}>1$bar
for both giant gas planets.

The day/night side temperature difference for HD\,209\,458b is nearly
similar for all longitudes in the atmosphere model applied
here. Comparing $\phi=0^o$ vs $\phi=180^o$ at the equator ($\theta=0$)
results in $\Delta T_{\rm day/night} \approx 500$K for HD\,209\,458b
in the upper, low pressure part of the atmosphere model. This is,
within error bars, the value that was suggested by \cite{zell2014}
base on 4.5$\mu$m Spitzer/IRAC data for HD\,209\,458b. The day/night
side temperature difference is larger for HD\,189\,733b in the
middle-atmosphere resulting in $\Delta T_{\rm day/night} \approx 700$K
at the maximum of the temperature inversion at $p_{\rm
  gas}=10^{-2}$bar. The day/night temperature difference at the
equator has decreased to $\Delta T_{\rm day/night} \approx 500$K for
$\phi=0^o$ vs $\phi=180^o$ similar to HD\,209\,458b.

\subsection{Vertical mixing}
The local vertical velocity is used to determine the vertical mixing,
$K_{\rm zz}$, required for the cloud formation module. This approach
is similar to the diffusion approach advocated by \cite{acker2001} who
specify mixing by a constant diffusion coefficient, $K_{\rm
  zz}=H^2/\tau_{\rm mix}$. \cite{char2015} show that the 1D diffusive
ansatz is likely to overestimate the vertical mixing resulting in an
overestimation of particle abundance and particle size at high
altitudes. \cite{char2015} provide a gas pressure dependent
parameterisation for $K_{\rm zz}$ comparable to \cite{par2013}.
\cite{Lee2015} (their Fig. 3)  suggest that such parameterisations
are very approximate. Here we use the vertical velocity component from
the 3D simulations for both planets directly.  A test has shown that the use of the  \cite{par2013} values mainly expands the cloud height towards lower pressures.  We, however, note that the need for a vertical mixing representation results for every 
  1D approach, hence also for ours that we presently apply to provide a first
  insight into kinetically calculated cloud properties of giant gas
  planets like HD\,189\,733b and HD\,209\,458b. This mixing approximation will become obsolete in fully coupled 3D
  radiative-hydrodynamic models with cloud formation feedback.

Figure~\ref{fig:Kzz} shows that the vertical mixing is stronger in the
inner (deeper) atmospheric regions in HD\,189\,733b compared to
HD\,209\,458b. This is were element enhancement
through cloud particle evaporation will occur. The terminator regions
and the substellar points show different mixing efficiencies between
the two planets despite relatively comparable (T$_{\rm gas}(z)$,
p$_{\rm gas}(z)$) structures. The vertical mixing at the substellar
point is stronger on HD\,189\,733b than on HD\,209\,458b in the upper
atmospheric parts.

Figure~\ref{fig:Kzz} further suggests that the global vertical mixing
for different longitudes across the planet is homogeneously more
vigorous in the atmosphere of HD\,209\,458b (right) which is stronger
irradiated by its hotter host star than HD\,189\,733b (left). This is
different to the day/night side differences in mixing of
HD\,189\,733b. Consequently, the optically thin atmospheric layers
will always be more vigorously mixed in HD\,209\,458b compared to
HD\,189\,733b indicating a stronger variability in potential spectral
features. The less irradiated HD\,189\,733b show a stronger
day/night side variation in mixing activity with the night side
showing considerably less efficient mixing than HD\,209\,458b.

\begin{figure*}
\centering
\includegraphics[scale=0.65]{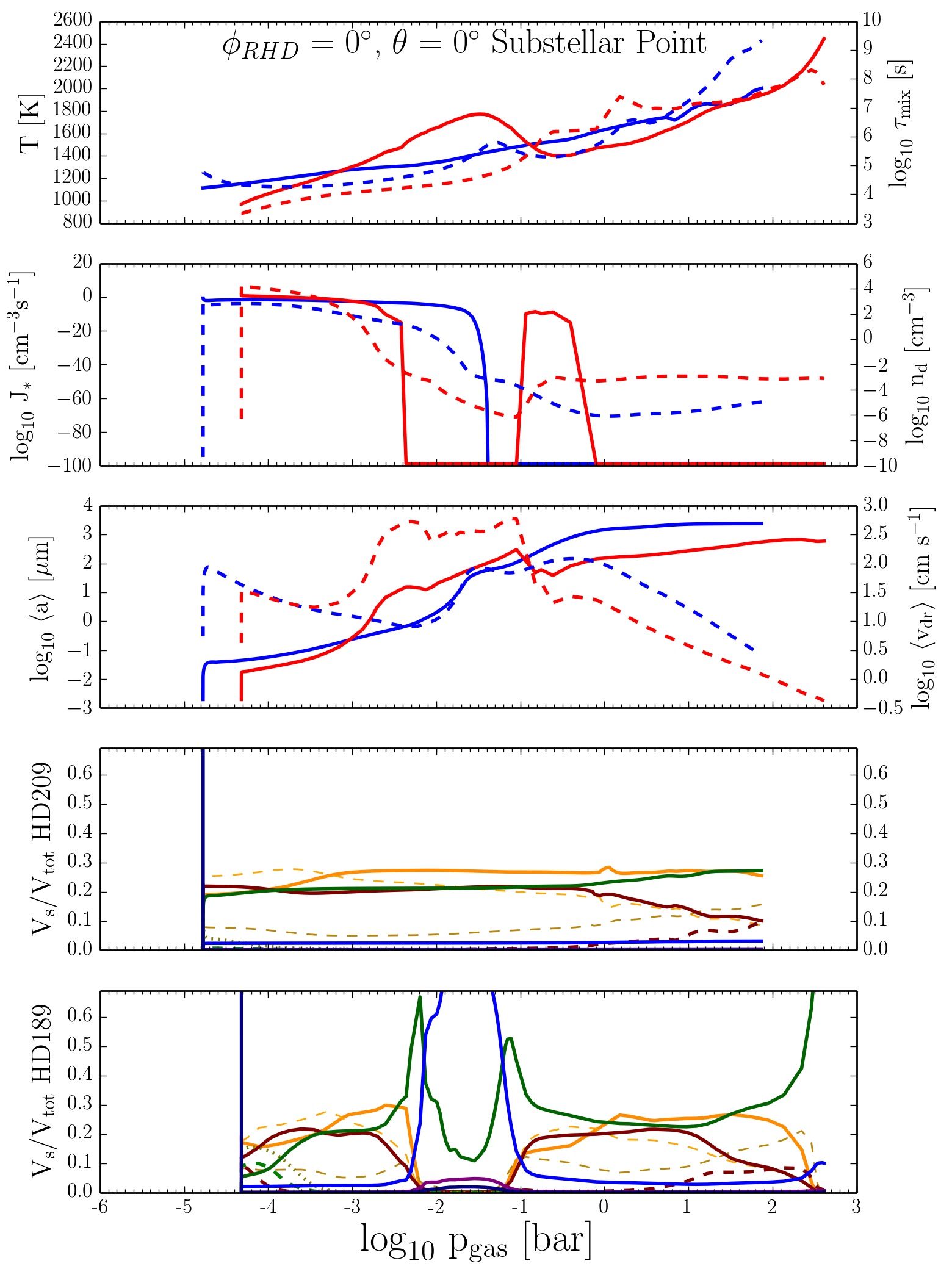}
\caption{The cloud structure at the substellar point
  ($\theta=0^{\circ}$, $\phi=0^{\circ}$) for the two giant gas planets
  HD\,189\,733b (red) and HD\,209\,458b (blue). In the first three
  panels, the dashed lines indicate the properties at the right axis,
  the solid lines to the left axis.  {\bf 1st panel:} gas temperature
  T$_{\rm gas}$ [K] (left), mixing time scale $\tau_{\rm mix}$
  [s$^{-1}$] (right); {\bf 2nd panel:} nucleation rate $J_*$
  [cm$^{-3}$s${-1}$] (left), number of dust particles $n_{\rm d}$
  [cm$^{-3}$] (right); {\bf 3rd panel:} mean grain size $\langle a\rangle$ [$\mu$m] (left), drift velocity for mean grain size  $\langle v_{\rm dr}\rangle$ [cm\,s$^{-1}$] (left); {\bf 4th panel:} material volume fraction
  $V_{\rm S}/V_{\rm tot}$ for HD\,209\,458b; {\bf 5th panel:} material
  volume fraction $V_{\rm S}/V_{\rm tot}$ for HD\,189\,733b. The
  colour coding for the 3rd and 4th panel is: TiO$_2$[s] - solid very
  dark blue, Al$_2$O$_3$[s] - solid blue, CaTiO$_2$[s] - solid purple,
  FeS[s] - dotted green, FeO[s] - dashed green, Fe[s] - solid green,
  SiO[s] - dashed brown, SiO$_2$[s] - solid brown, MgO[s] - dashed
  dirty orange, MgSiO$_3$ - dashed orange, Mg$_2$SiO$_4$ - solid
  orange. }
\label{fig:struc1}
\end{figure*}

\begin{figure*}
\centering
\includegraphics[scale=0.43]{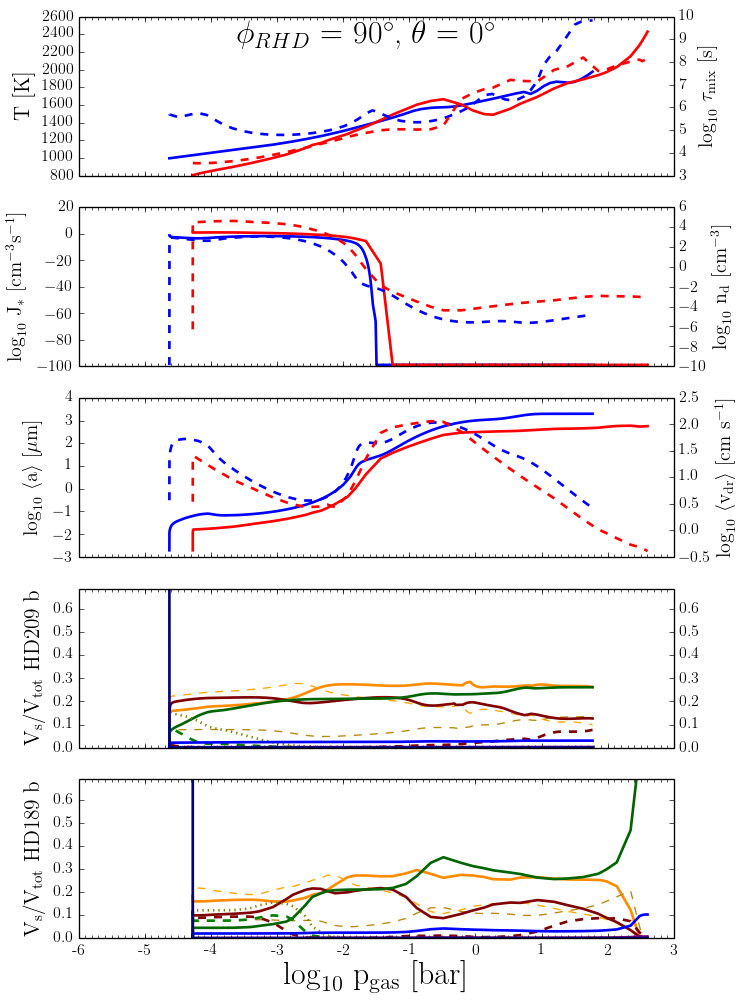}
\includegraphics[scale=0.43]{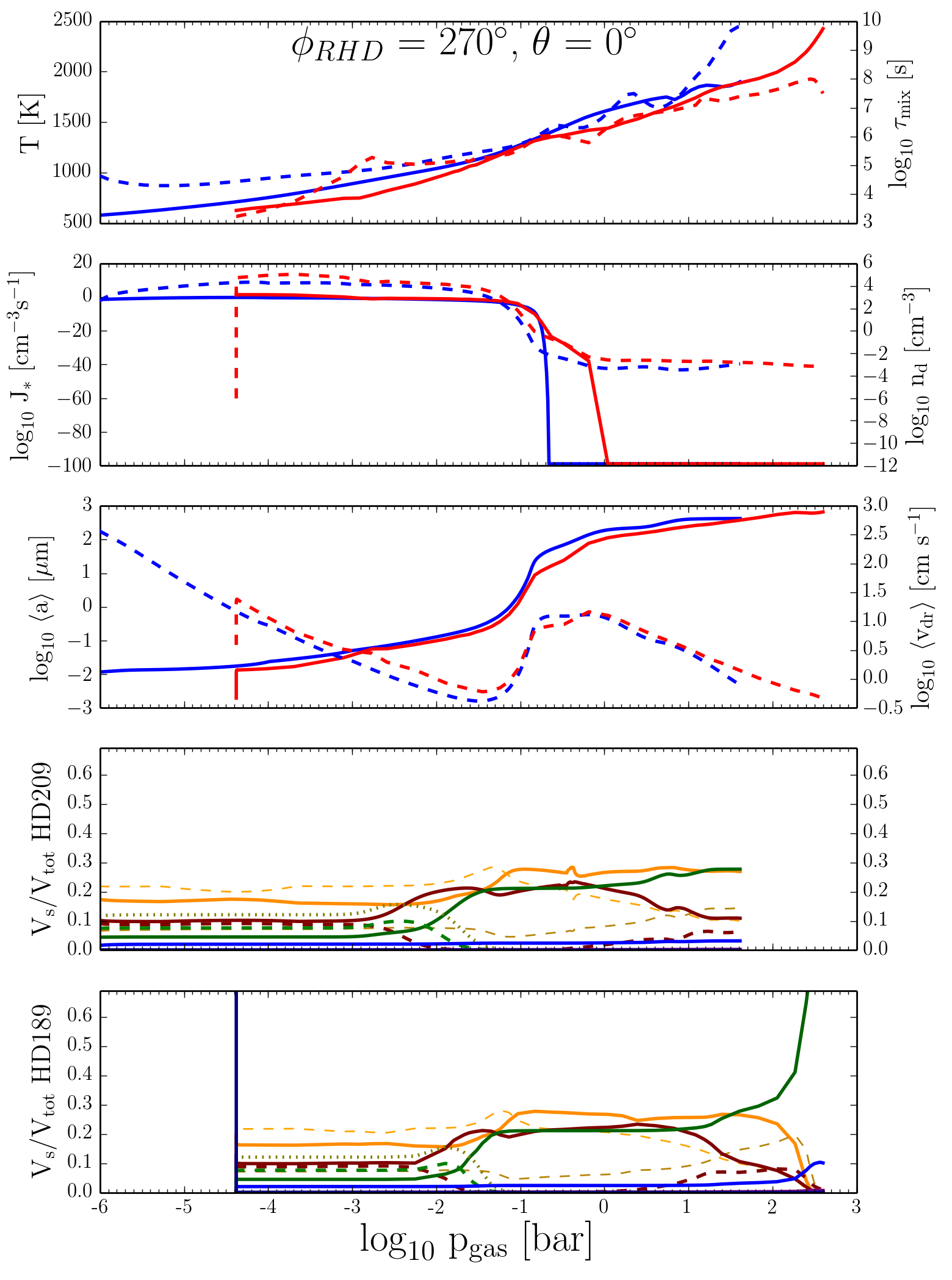}
\caption{The cloud structure at the two terminator points
  ($\theta=0^{\circ}$, $\phi=90^{\circ}$, left) and
  ($\theta=0^{\circ}$, $\phi=270^{\circ}$, right) for the two giant
  gas planets HD\,189\,733b (red) and HD\,209\,458b (blue). The panels
  and line styles have the same meaning like in
  Fig.~\ref{fig:struc1}. The
  colour coding for the 3rd and 4th panel is: TiO$_2$[s] - solid very
  dark blue, Al$_2$O$_3$[s] - solid blue, CaTiO$_2$[s] - solid purple,
  FeS[s] - dotted green, FeO[s] - dashed green, Fe[s] - solid green,
 SiO[s] - dashed brown,  SiO$_2$[s] - solid brown, MgO[s] - dashed dirty
  orange, MgSiO$_3$ - dashed orange, Mg$_2$SiO$_4$ - solid orange.}
\label{fig:struc2}
\end{figure*}

\section{The dust cloud structures of  HD\,189\,733b and  HD\,209\,458b}\label{s:cloudstruc}

HD\,189\,733b and HD\,209\,458b are the two most observed extrasolar
planets (see Sect.~\ref{s:intro}). Observations were analyzed by
applying retrieval methods based on mainly two codes, {\sc Nemesis}
(\citealt{bars2014, lee2014}) and {\sc Scarlet}
(\citealt{benneke2015}), which both assume effective particle sizes,
homogeneous material composition and the number of cloud particles.
The present paper provides more insight into the cloud properties of
HD\,189\,733b and HD\,209\,458b by analyzing the cloud formation
processes based on 3D atmosphere simulations. We demonstrate that the
clouds on HD\,189\,733b and HD\,209\,458b are far richer then
suggested by present observation due to their necessarily
hemispherically averaged nature. 

\subsection{The local view}
The local cloud properties are determined by the rate at which seed
particles can form from the gas-phase (2nd panels
Figs.~\ref{fig:struc1},~\ref{fig:struc2}; and Fig.~\ref{fig:J*}) and
the efficiency with which they can grow (Figs.~\ref{fig:chi}). Both
are determined by the local gas temperature, the local gas density (or
gas pressure, 1st panels Figs.~\ref{fig:struc1},~\ref{fig:struc2}) and
also by the local chemical composition. We discuss how the cloud
structures differ between HD\,189\,733b and HD\,209\,458b in the
following section, the gas-phase composition is discussed in
Sect.~\ref{ss:chemequ}. We note that the local quantities may change
to some extent if the underlying model atmosphere simulation undergoes
updates due to ongoing development work on, for example, opacities
which might affect the deeper atmosphere structure. For the local
view, we chose three different trajectories for each planet
representing the substellar points ($\theta=0^{\circ}$,
$\phi=0^{\circ}$) and the two terminators ($\theta=0^{\circ}$,
$\phi=90^{\circ}$ and $\phi=270^{\circ}$).

\begin{figure*}
\centering
\includegraphics[scale=0.4]{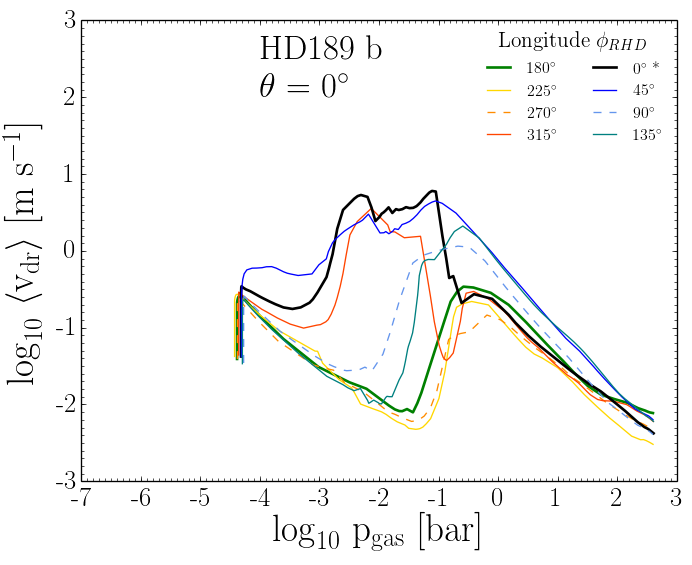}
\includegraphics[scale=0.4]{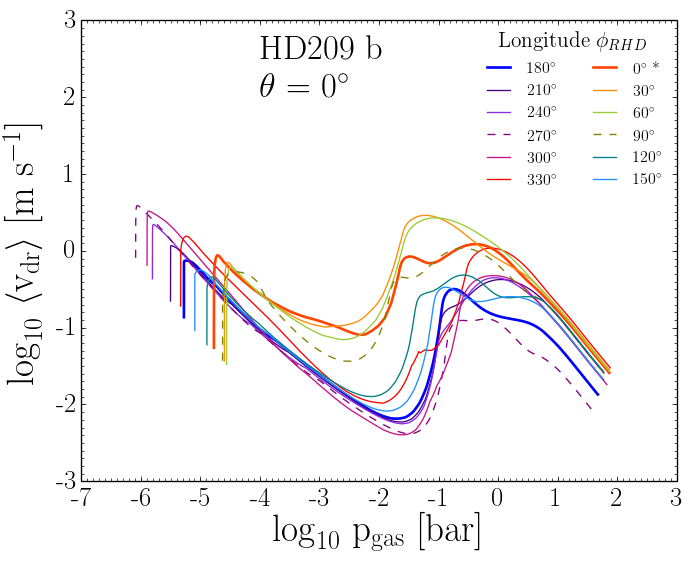}
\caption{Drift velocity for trajectories on the equator (HD\,189\,733b - left,  HD\,209\,458b - right). Large drift velocities in the mid-cloud region relates to large mean grain sizes (see Fig.~\ref{fig:amean}). Small drift velocities near the cloud base are caused by increasing friction due to the increasing local gas density.}
\label{fig:vdr}
\end{figure*}

 HD\,209\,458b clouds form from one single nucleation region that
 spans several orders of magnitude in gas pressure (see also
 Fig.~\ref{fig:J*}).  HD\,189\,733b clouds are characterised by two
 nucleation peaks at low and at high gas pressure near the substellar
 point where the temperature inversion causes a local increase of gas
 temperature, and therewith nucleation ceases until the temperature
 has dropped low enough again. The consequence is a smoother
 distribution of (mean) grain sizes throughout the cloud in
 HD\,209\,458b locally and globally (3rd panels
 Figs.~\ref{fig:struc1},~\ref{fig:struc2}; and Fig.~\ref{fig:amean}).
 HD\,189\,733b clouds do have a middle-region of big grains with
 $\langle a\rangle_{\rm HD\,209\,458b}\approx 1\mu$m $\langle
 a\rangle_{\rm HD\,189\,733b}\approx 10^3\mu$m at $p_{\rm gas}\approx
 10^{-2}$bar. The material composition of the cloud particles are
 considerably different between the two planets at these pressures at
 the substellar point.  In this region, the cloud particles fall
 fastest inside the HD\,189\,733b cloud structure. Similar values of
 gravitational settling are only reached in the lower (inner) cloud in
 HD\,209\,458b (Fig.~\ref{fig:vdr}).

Figures~\ref{fig:struc1},~\ref{fig:struc2} (3rd panel) demonstrate
that the mean grain sizes differ largely between HD\,189\,733b and
HD\,209\,458b at the substellar point, but they are rather similar at
the terminators despite a gas temperature difference of $\approx 500$K
at the upper model boundary. A closer inspection, however, reveals
that for ($\theta=0^{\circ}$, $\phi=90^{\circ}$) the cloud material
composition differs. Both planets show MgSiO$_3$[s] contributing with
20-25\% to the cloud material upto p$_{\rm gas}\approx10^{-2}$bar. The
next most important material is Mg$_2$SiO$_4$[s] (18\%) in
HD\,189\,733b but SiO[s] (20\%) in HD\,209\,458b. HD\,209\,458b has a
larger fraction of Fe[s] contributing to the cloud composition in these
upper regions, while HD\,189\,733b offers a mix of SiO[s], SiO$_2$[s], MgO[s],
FeO[s] contributing with each $\approx 10\%$ but Fe[s]$<$1\%.

We note that HD209\,458b has a large pressure scale height at
($\theta=0^{\circ}$, $\phi=270^{\circ}$) spreading the upper cloud
into regions of lower densities compared to HD\,189\,733b.  If only
the terminators were used to compare the atmospheric structure of
HD\,189\,733b and HD\,209\,458b, one would conclude that they should
be similar as also the (T$_{\rm gas}(z)$, p$_{\rm gas}(z)$) are rather
similar and so is the material composition of the cloud particles. A
more global view, however, suggests that the mean grain sizes are
mostly larger in HD\,209\,458b at any given pressure level
(Fig.~\ref{fig:amean}) unless affected by a temperature
inversion. This is a result of the lower nucleation rate
(Fig.~\ref{fig:J*}) which causes fewer seed particles to form, hence
the cloud particles can grow bigger than in a case of more efficient
nucleation as in HD\,189\,733b. The biggest cloud particles reside at
the lower cloud edge in both models. However note, that none of the
models reaches high enough gas temperatures for the cloud particle to
evaporate completely at such high gas pressures.

\noindent

\begin{figure*}
\centering
\includegraphics[scale=0.4]{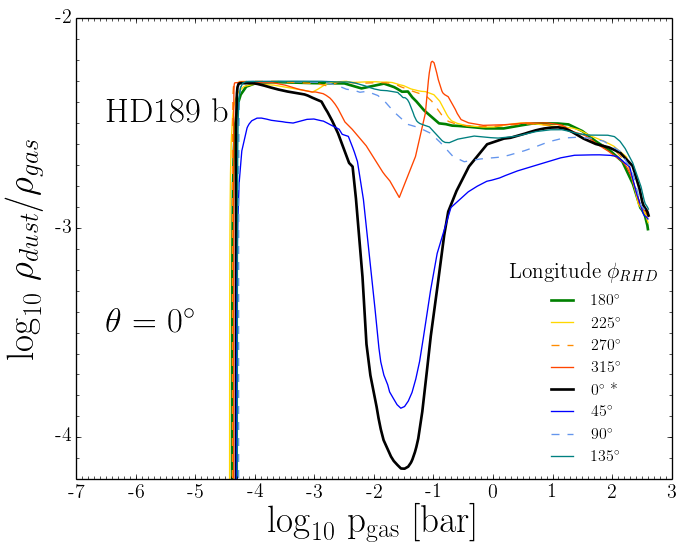}
\includegraphics[scale=0.4]{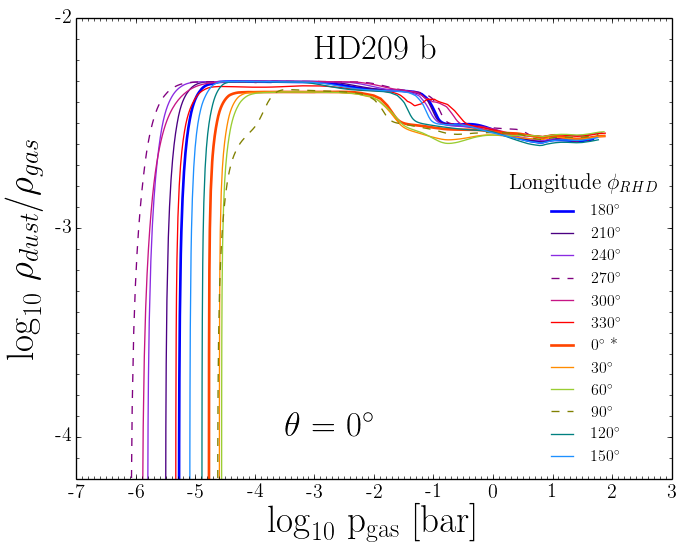}
\caption{The dust loading of the atmosphere in terms of the  dust-to-gas mass density ratio for trajectories on the  equator (HD\,189\,733b - left,  HD\,209\,458b - right). }
\label{fig:rhodrhog}
\end{figure*}

\subsubsection{Dust loading of atmosphere: $\rho_{\rm dust}/\rho_{\rm gas}$ and $\langle v_{\rm dr}\rangle$}
The dynamics of an atmosphere are also determined by the feedback of
the clouds through dynamical processes like friction. The force
equilibrium between (effective) gravity and friction determines how
fast a cloud particle can fall and, hence, to what geometrical
extension (or height) the cloud can grow.  The drift velocity
determines how fast a particular part of the atmosphere is emptied or
filled with cloud material from other regions.  Figure.~\ref{fig:vdr}
shows how the drift velocity (or gravitational settling velocity)
changes with height in the atmosphere at different longitudes for the
two giant gas planets.

Figure~\ref{fig:rhodrhog} show the dust-to-gas ratio, $\rho_{\rm
  dust}/\rho_{\rm gas}$, which can be understood as dust load of the
atmosphere. The comparison of both figures shows that the lowest dust
loading of the atmospheric gas occurs where the drift velocity is
highest.  $\rho_{\rm dust}/\rho_{\rm gas}$ decreases at the cloud bottom
because the cloud particles evaporate.

\noindent
\begin{figure*}
\centering
\includegraphics[scale=0.6]{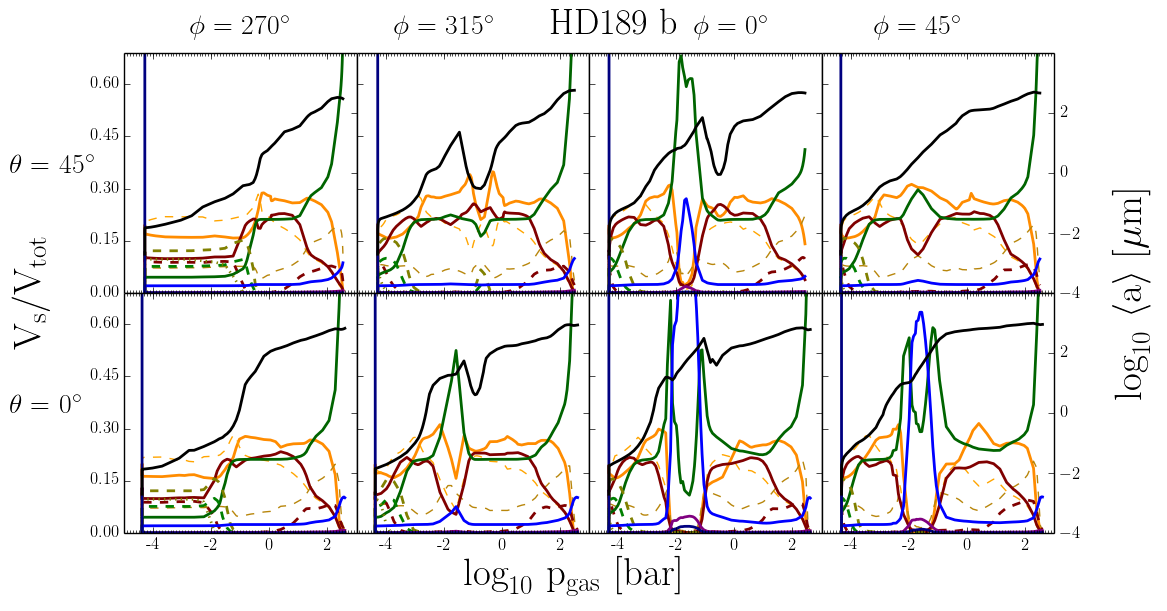}
\includegraphics[scale=0.6]{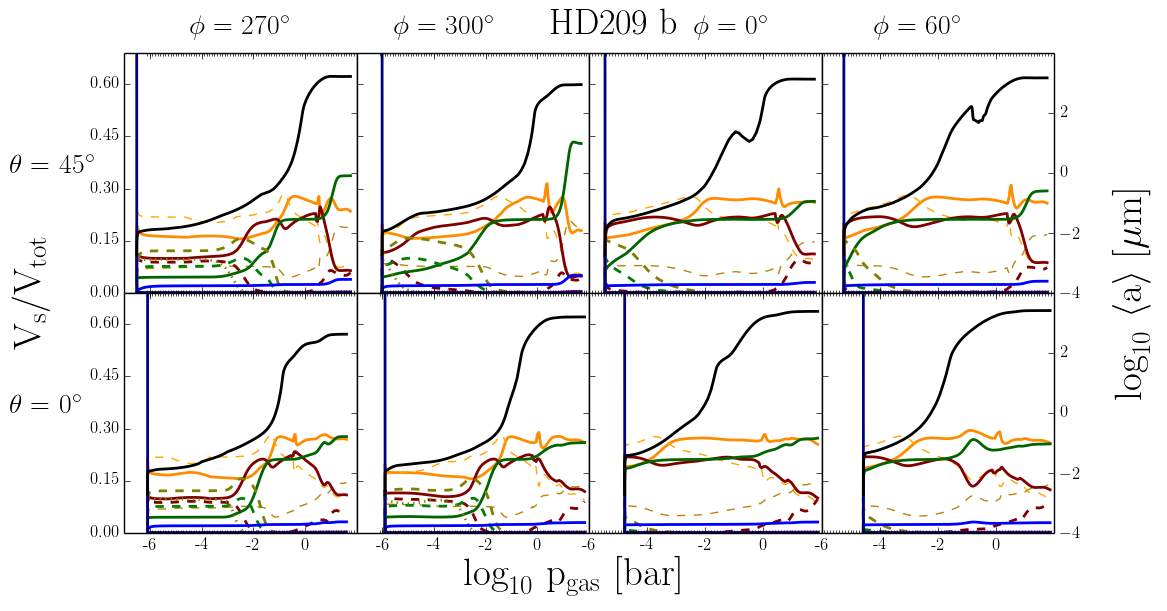}
\caption{The {\bf day-side} cloud particle material composition (color
  coded, left axis) and mean grain sizes (black, right axis) for
  HD\,189\,733b (top) and HD\,209\,458b (bottom) for
  $\theta=45^{\circ}$ and $\theta=0^{\circ}$ (equator). The
  $\phi$-values are somewhat offset for both planets due to the use of
  different grids. The colour code for the material is the same like
  in previous figures and as given in Fig.~\ref{fig:struc1}.  The
  colour coding is: TiO$_2$[s] - solid very
  dark blue, Al$_2$O$_3$[s] - solid blue, CaTiO$_2$[s] - solid purple,
  Fe$_2$O$_3$[s] - dashed light green, FeS[s] - dotted green, FeO[s] - dashed green,
  Fe[s] - solid green, SiO[s] - dashed brown, SiO$_2$[s] - solid brown, MgO[s] - dashed dirty orange, MgSiO$_3$ - dashed orange,
  Mg$_2$SiO$_4$ - solid orange.}
\label{fig:VdaHD189}
\end{figure*}

\noindent
\begin{figure*}
\centering
\includegraphics[scale=0.6]{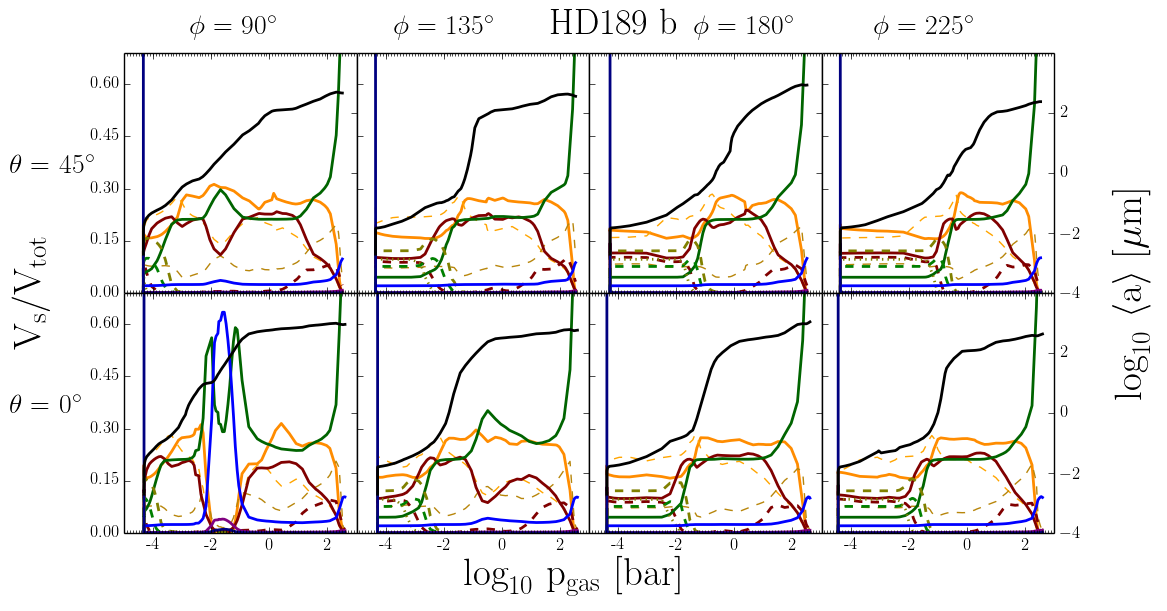}
\includegraphics[scale=0.6]{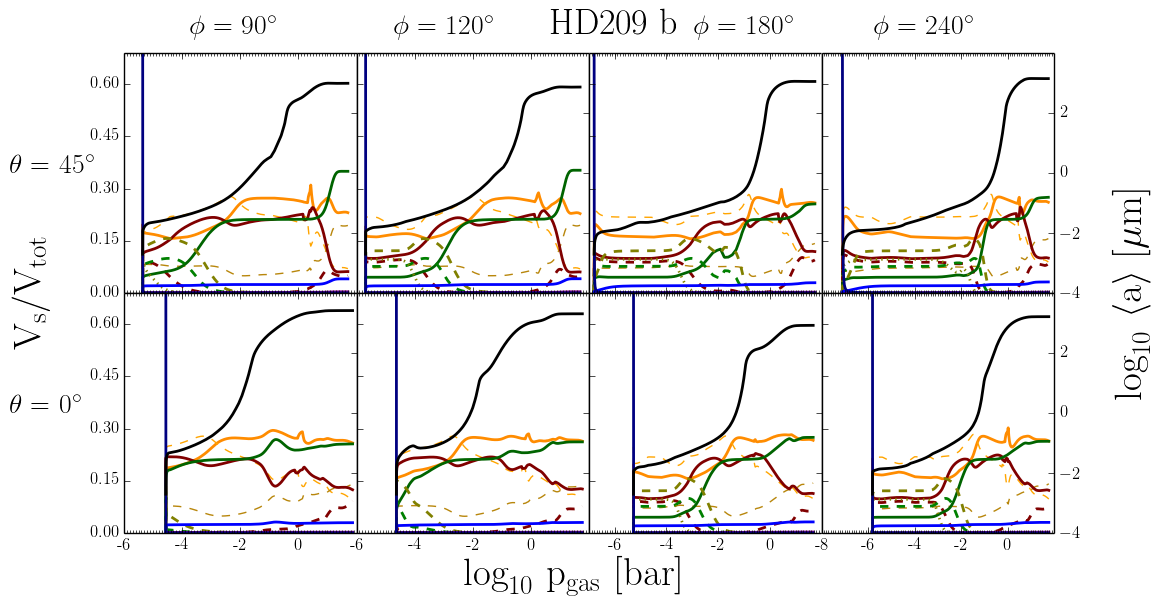}
\caption{The {\bf night-side} cloud particle material composition
  (color coded, left axis) and mean grain sizes (black, right axis)
  for HD\,189\,733b (top) and HD\,209\,458b (bottom) for
  $\theta=45^{\circ}$ and $\theta=0^{\circ}$ (equator). The
  $\phi$-values are somewhat offset for both planets due to the use of
  different grids.  The colour code for the material is the same like
  in previous figures and as given in Fig.~\ref{fig:struc1}. The
  colour coding is: TiO$_2$[s] - solid very
  dark blue, Al$_2$O$_3$[s] - solid blue, CaTiO$_2$[s] - solid purple,
  Fe$_2$O$_3$[s] - dashed light green, FeS[s] - dotted green, FeO[s] -
  dashed green, Fe[s] - solid green, SiO[s] - dashed brown, SiO$_2$[s] - solid brown,
  MgO[s] - dashed dirty orange, MgSiO$_3$ -
  dashed orange, Mg$_2$SiO$_4$ - solid orange. }
\label{fig:VdaHD209}
\end{figure*}

\subsection{The global view}

The thermodynamic properties of the atmospheres of the giant gas
planets HD\,189\,733b and HD\,209\,458b differ considerably due to the
different global parameters of the planets like radii, bulk density
and equilibrium temperature. Therefore, we now discus how the global
cloud structures change which impact the global opacity pattern and
with that observational global quantities like albedo and equilibrium
temperatures. We expect the global cloud results to be less affected
by changes of the underlying 3D atmosphere simulation in the present
two-model approach.

Figures~\ref{fig:VdaHD189} and~\ref{fig:VdaHD209}
provide a quasi-three-dimensional view on the cloud material composition,
$V_{\rm s}/V_{\rm tot}$, and the mean size of the cloud particles,
$\langle a\rangle$, for two different longitudes ($\theta=45^{\circ}$,
$\theta=0^{\circ}$ (equator)), various longitudes at day
(Fig.~\ref{fig:VdaHD189}) and at night (Fig~\ref{fig:VdaHD209}) side
versus the local gas pressure.

Figures~\ref{fig:VdaHD189} and \ref{fig:VdaHD209} offer a 3D
impression of how the cloud material composition and the mean grain
sizes change for the atmospheric trajectories probed in HD\,189\,733b
(top) and HD\,209\,458b (bottom). The cloud material composition
changes with height irrespective of longitude, latitude and planet
considered. The clouds resulting for the 3D thermodynamic structures
used here are predominantly made of a mix of MgSiO$_3$[s] and
Mg$_2$SiO$_4$[s] ($>$30\%), and Fe$_2$O$_3$[s]/Fe[s]/MgO[s] and
SiO$_2$[s] in the upper atmospheric regions where the particles are
small ($10^{-4}\ldots10^{-2}\mu$m). Depending on the local conditions
in the upper atmosphere, SiO$_2$[s] ($\approx 20$\%) can become the second most
abundant material in these cloud particles after MgSiO$_3$[s]. This
occurs at the substellar point ($\phi=0^o$) and westward ($\phi=45^o,
60^o$) of it for both planets. Generally, the cloud particles that
form the upper cloud layers are the most chemically mixed. When
FeS[s], Fe$_2$O$_3$[s], FeO[s], and SiO[s] have evaporated, the middle
cloud is made of basically four materials only: MgSiO$_3$[s] and
Mg$_2$SiO$_4$[s], Fe[s], SiO$_2$[s]. Small amount of Al$_2$O$_3$[s],
and a less than 1\% CaTiO$_3$[s] and TiO$_2$[s] contribute to the
material composition throughout the whole cloud. Slightly more
CaTiO$_3$[s] is present in HD\,189\,733b than in HD\,209\,458b but it
appears in almost negligible amounts in both planets for the (T$_{\rm
  gas}(z)$, p$_{\rm gas}(z)$) part of the atmospheric environment
considered here. 

\subsubsection{Day-night-side difference}
For both planets, the day side (Fig.~\ref{fig:VdaHD189}) shows a
stronger inward-increase of the cloud particles sizes than the night
side. The particles grow more efficiently in HD\,189\,733b than in
HD\,209\,458b because HD\,189\,733b has a more compact atmosphere,
hence higher local gas densities than HD\,209\,458b. 

%\begin{table}
%\caption{Summery of cloud properties}
%\begin{tabular}{p{0.7cm}p{1.2cm}|p{2.2cm}p{2.2cm}}
%\hline\hline
% cloud      &  &  HD 189 &  HD\,209\,458b \\
% \hline
%top & $\langle a\rangle$ [$\mu$m]&  &\\ 
%      & materials & & \\
%middle & $\langle a\rangle$ [$\mu$m]& & \\ 
%      & materials & {\small MgSiO$_3$[s],  Mg$_2$SiO$_4$[s]} & {\small MgSiO$_3$[s], Mg$_2$SiO$_4$[s]}\\
%      &                 &  {\small Fe[s], SiO$_2$[s]}  & {\small  Fe[s], SiO$_2$[s]} \\
%bottom & $\langle a\rangle$ [$\mu$m]& \\ 
%      & materials &\\
%\hline
%\end{tabular}
%\end{table}

\noindent
\begin{figure*}
\centering
\includegraphics[scale=0.35]{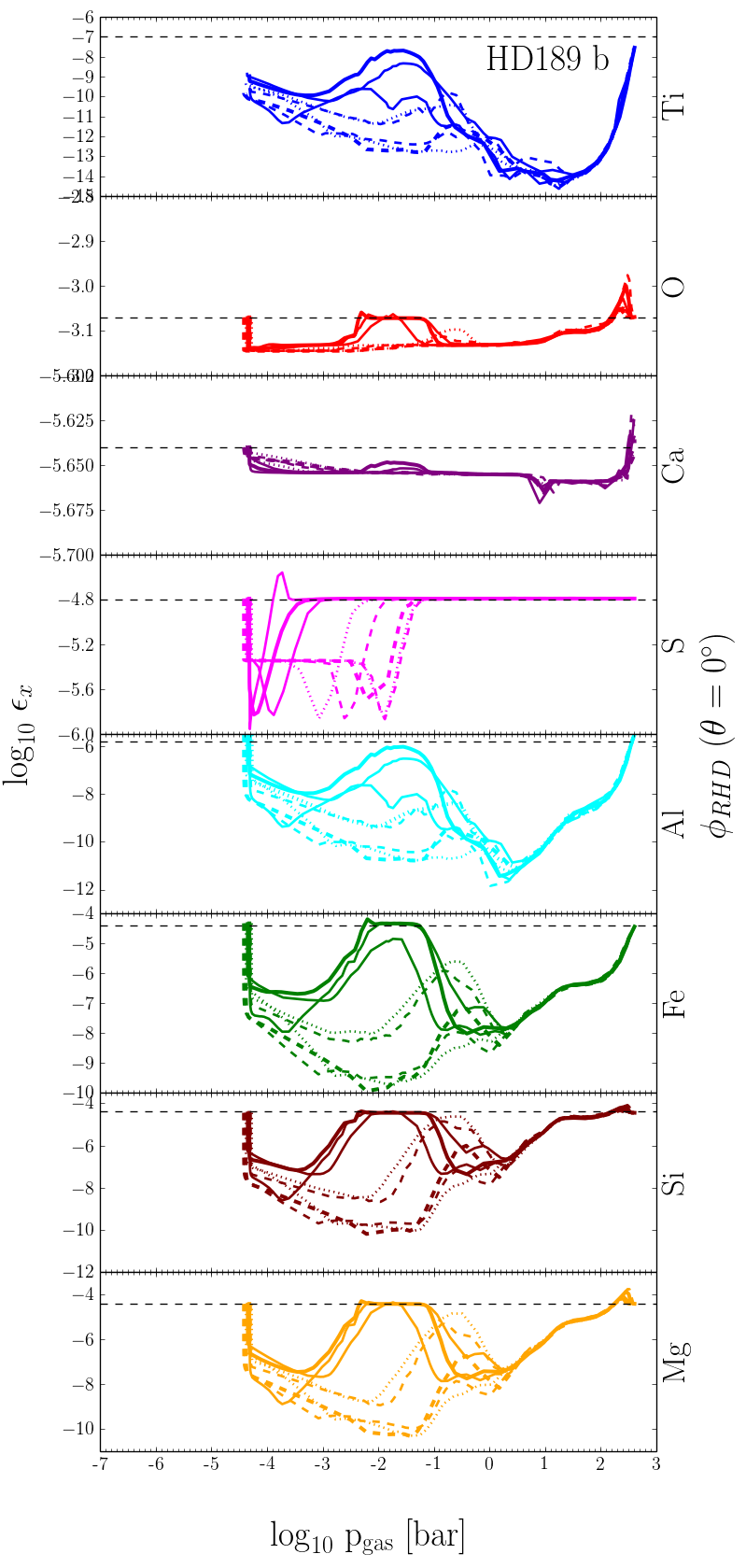}
\includegraphics[scale=0.35]{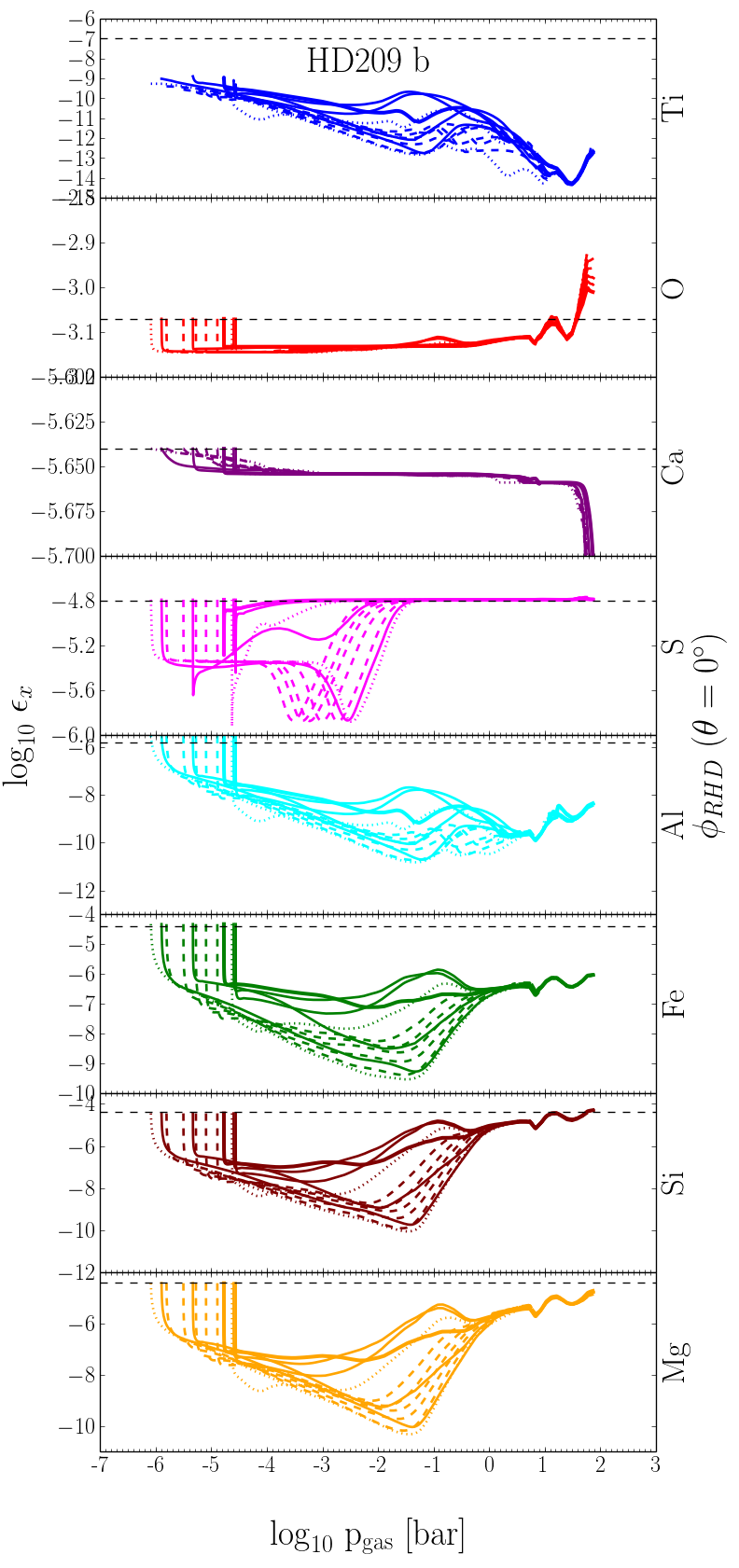}
\caption{The changing element composition of the cloud forming region
  in the atmospheres of the giant gas planets HD\,189\,733b (left) and
  HD\,209\,458b (right). The patterned lines indicate different
  longitudes at the equator, the horizontal dashed black lines
  indicates the initial, undepleted abundance values for each
  element (\citealt{AndGre1989}) with an initial C/O=0.427.}
\label{fig:eabun}
\end{figure*}

\section{The impact of dust cloud formation on gas-phase chemistry in  HD\,189\,733b and  HD\,209\,458b }\label{s:gaschem}

The composition of the atmospheric gas in chemical equilibrium is
determined by the local gas temperature and gas pressure (T$_{\rm
  gas}(z)$, p$_{\rm gas}(z)$) and by the element abundances. All
element abundances that are involved in cloud formation will deviate
from an initial (e.g. solar) value by how much is locked up in the cloud
particles or by how much is released back into the gas phase. This is
of particular interest if the cloud particle formation took place at a
different location compared (e.g. cloud top) to where it evaporates
(e.g. cloud bottom).

\subsection{Atmospheric element abundances}\label{ss:elmab}
Element abundances are an important set of input quantities for every
chemical model, and hence, for every radiative transfer atmosphere
simulation. It is therefore a long-standing challenge to determine
atmospheric element abundances to a high precision for stellar
atmospheres (e.g. \citealt{Grevesse2007,Amarsi2015}). Element
abundances for planetary atmospheres are even harder to determine and
the host star's set of element abundances is usually applied assuming
that planets form from gravitationally unstable protoplanetary disks
keeping the host star's element abundances. This belief has recently
been challenged by chemical disk evolution models
(\citealt{2014Life}). Another challenge for observing the element
abundances from planetary atmospheres arises from the fact that the
formation of clouds causes a selective depletion of elements depending
on the atmopshere's local gas temperature and gas pressure. It results
in a height-dependent element abundance pattern that changes across the
planet's globe as we demonstrate here for two extrasolar
planets. The issue of element abundance determination is also highly
relevant for retrieval methods and the discussion of the C/O
ratio. While no one C/O ratio will exist for a cloud forming
atmosphere (Sect.~\ref{ss:ctoo}), to search for one (planetary) value
for the metallicity is simply asking the wrong question. Both, the C/O
ratio and the metallicity depend on the cloud formation efficiency and
can therefore not be treated as independent from any of the cloud
parameters invoked in retrieval methods.  We demonstrate how the
element abundances change due to cloud formation in the two giant gas
planets HD\,189\,733b and HD\,209\,458b, and how the abundance pattern
may differ between these two planets.

Figure~\ref{fig:eabun} summarizes the element abundances for the 8
cloud forming elements (O, Ca, S, Al, Fe, Si, Mg) forming 12 solid
species (MgSiO$_3$[s], Mg$_2$SiO$_4$[s], SiO[s], SiO$_2$[s], MgO[s],
Fe[s], FeO[s], Fe$_2$O$_3$[s], Al$_2$O$_3$[s], TiO$_2$[s],
CaTiO$_3$[s], FeS[s]) for various test trajectories across the equator
($\theta=0^{\circ}$) of HD\,189\,733b (left) and HD\,209\,458b
(right).  The element abundances follow the same depletion hierarchy
in both planets: Ti is the most depleted, followed by Al, then Fe, Si
and Mg by about the same order of magnitudes. The least overall
depleted elements are O, Ca and S. Individual trajectories that probe
the atmosphere differ widely in their depletion pattern. The most
prominent example is the substellar point where in the case of
HD\,189\,733b all Si, Mg and Fe is returned to the gas phase through
the evaporation of the Si/Mg-binding materials (MgSiO$_3$[s],
Mg$_2$SiO$_4$[s], SiO[s], SiO$_2$[s], Fe[s], FeO[s], Fe$_2$O$_3$[s])
as result of the temperature inversion. A similar trend occurs in
HD\,209\,458b only at the cloud base.

\begin{figure}
\hspace{0.5cm}
\includegraphics[scale=0.45]{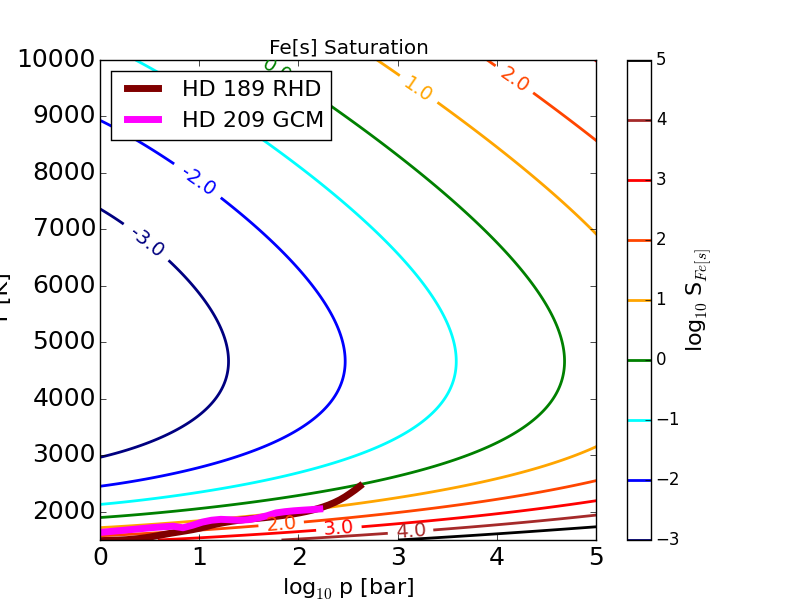}
\caption{The thermal stability in terms of supersaturation ($S=p_{\rm
    Fe}/p_{\rm sat,Fe}$; $p_{\rm Fe}$ - Fe vapour pressure, $p_{\rm
    sat,Fe}$ - Fe[s] saturation vapour pressure) for solid iron,
  Fe[s], in the (T$_{\rm gas}$, p$_{\rm gas}$) plane. Both vertical
  atmosphere (T$_{\rm gas}$, p$_{\rm gas}$) trajectories for
  HD\,189\,733b and HD\,209\,458b do not cross the line of thermal
  stability ($S=1$), hence, Fe[s] will not evaporate at the cloud
  base. The solar value for the Fe-element abundance is used for this
  plot.}
\label{fig:Fethstab}
\end{figure}

The element depletion at the cloud top is somewhat model dependent as
the location of the cloud top (e.g. where element depletion starts)
depends on the setting of the upper model boundary. The location of
the cloud base can in principle be defined as where all cloud
particles have evaporated and the bound elements have been
returned to the gas phase. But this is only possible if the models
extend to high enough gas temperatures that the cloud particles can
evaporate completely. Both, HD\,189\,733b and HD\,209\,458b, reach
such high local gas pressures in the lower atmosphere that the thermal
stability of all materials considered here is extended to
substantially higher temperatures compared to, for example, brown
dwarfs. This is demonstrated in Fig.~\ref{fig:Fethstab} for
Fe[s]. However, some of the elements show an element enrichment that
results from cloud particles being formed elsewhere and having rained
into these regions where some of their solids become thermally
unstable. This is the case for HD\,189\,733b on the day and the night
side (compare top of Figs.~\ref{fig:VdaHD189}, ~\ref{fig:VdaHD209})
where all the Mg/Si/O minerals have evaporated except MgO[s]. As
MgSiO$_3$[s], Mg$_2$SiO$_4$[s] and SiO[s]/Fe$_2$O$_3$[s] make up $>$60\%
of the cloud particle volume, its evaporation produces a substantial
enrichment at the cloud base in oxygen, and to a lesser extent in Si
and Mg in HD\,189\,733b. Inspecting Fig.~\ref{fig:eabun} shows that
HD\,209\,458b receives an even stronger oxygen enrichment at the cloud
base which, however, results mainly from MgSiO$_3$[s] and SiO$_2$[s]
evaporation (compare bottom of Figs.~\ref{fig:VdaHD189},
~\ref{fig:VdaHD209}). The changing element abundance of Ca does not
affect the cloud material volume remarkably.

\noindent

\begin{figure*}
%\hspace*{-1.5cm}\includegraphics[scale=0.5]{plotsChem/HD189_0_0_depleted}
\hspace*{-1.5cm}\includegraphics[scale=0.45]{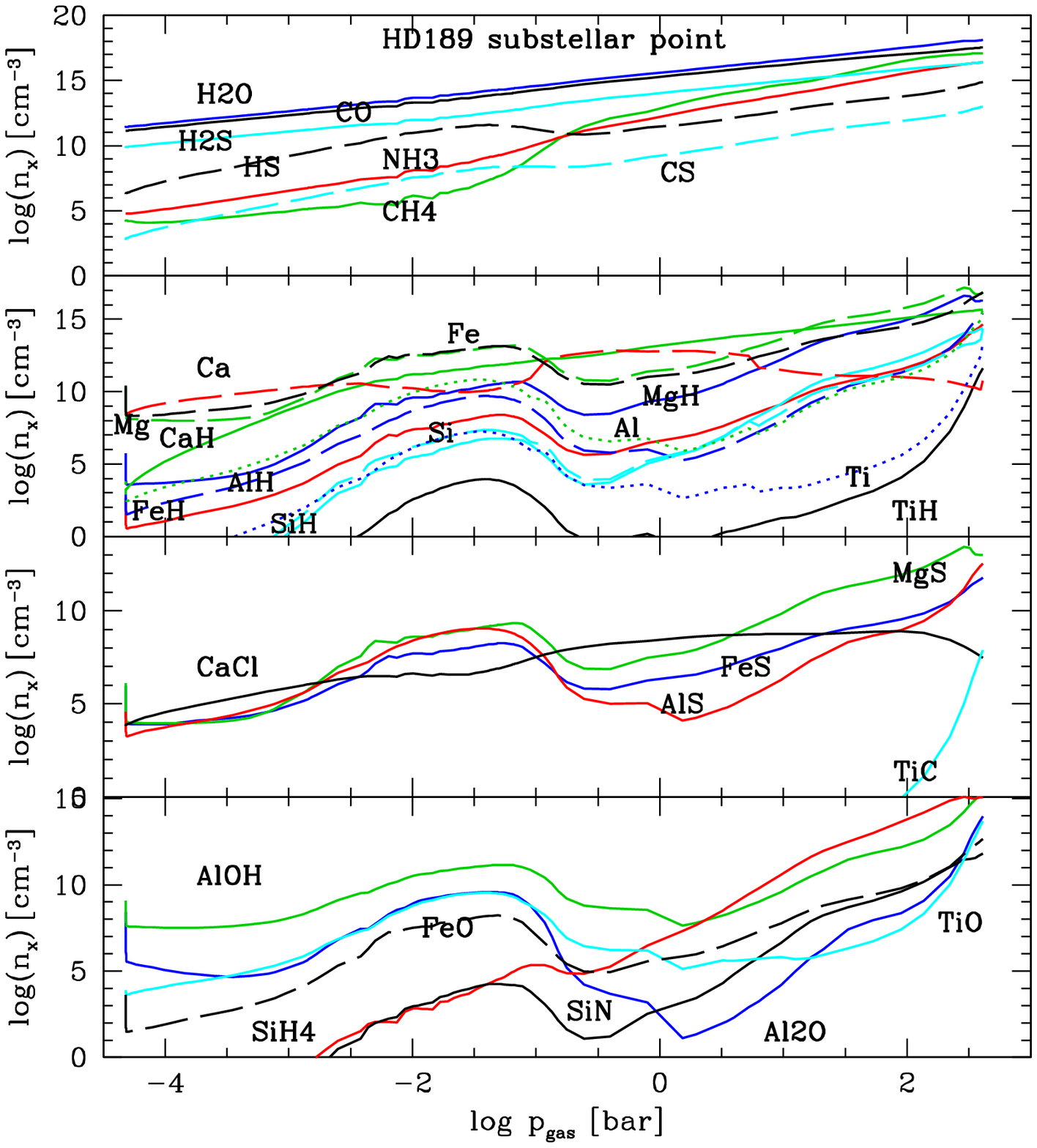}
\hspace*{-1.5cm}\includegraphics[scale=0.45]{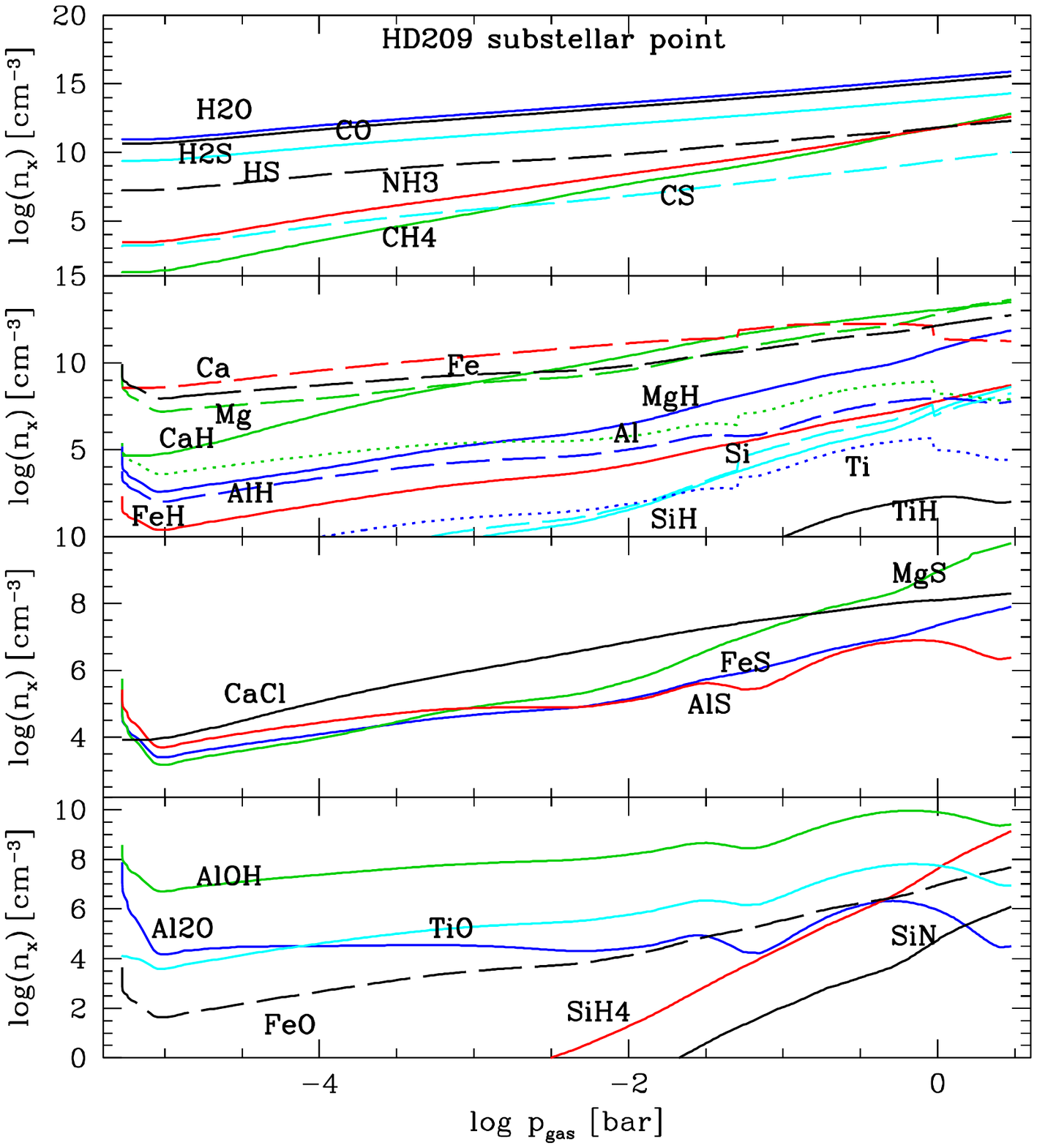}\\*[-3.0cm]
%\hspace*{-1.5cm}\includegraphics[scale=0.5]{plotsChem/HD189_substellar_ratio.eps}
%\hspace*{-1.5cm}\includegraphics[scale=0.5]{plotsChem/HD209_substellar_ratio.eps}\\*[-1.0cm]
\caption{Chemical gas composition at the substellar point
  ($\theta=0^{\circ}$, $\phi=0^{\circ}$) HD\,189\,733b (left) and
  HD\,209\,458b (right) in a collisionally dominated, inner (deeper)
  atmosphere. Shown are the number densities of atoms and molecules
  linked to the elements (O, Al, S, Mg, Ti Fe, Si Ca) involved in
  cloud formation. 
%{\bf Bottom:} The impact of cloud formation on the
%  gas-phase composition.  The undepleted case is calculated for solar
%  abundances and the depleted case for the element abundances after
%  cloud formation (Fig~\ref{fig:eabun}).
 }
\label{fig:molabun18920900}
\end{figure*}

%\begin{figure*}
%\centering
%\includegraphics[scale=0.45]{plotsChem/HD209_substellar_depleted}
%\includegraphics[scale=0.45]{plotsChem/HD209_substellar_undepleted}\\*[-1.0cm]
%\caption{Gas-phase composition  HD\,209\,458b at the substellar point ($\theta=0^{\circ}$, $\phi=0^{\circ}$): O% Al, S, Mg, Ti Fe, Si Ca.  {\bf Left:}  dust-depleted elements according to left of Fig~\ref{fig:eabun}, {\bf R%ight:} un-depleted gas-phase (solar abundance)}
%\label{fig:molabun20900}
%\end{figure*}

\subsection{Abundances of atoms, molecules and electrons}\label{ss:chemequ}

\begin{figure}
\centering
\includegraphics[scale=0.42]{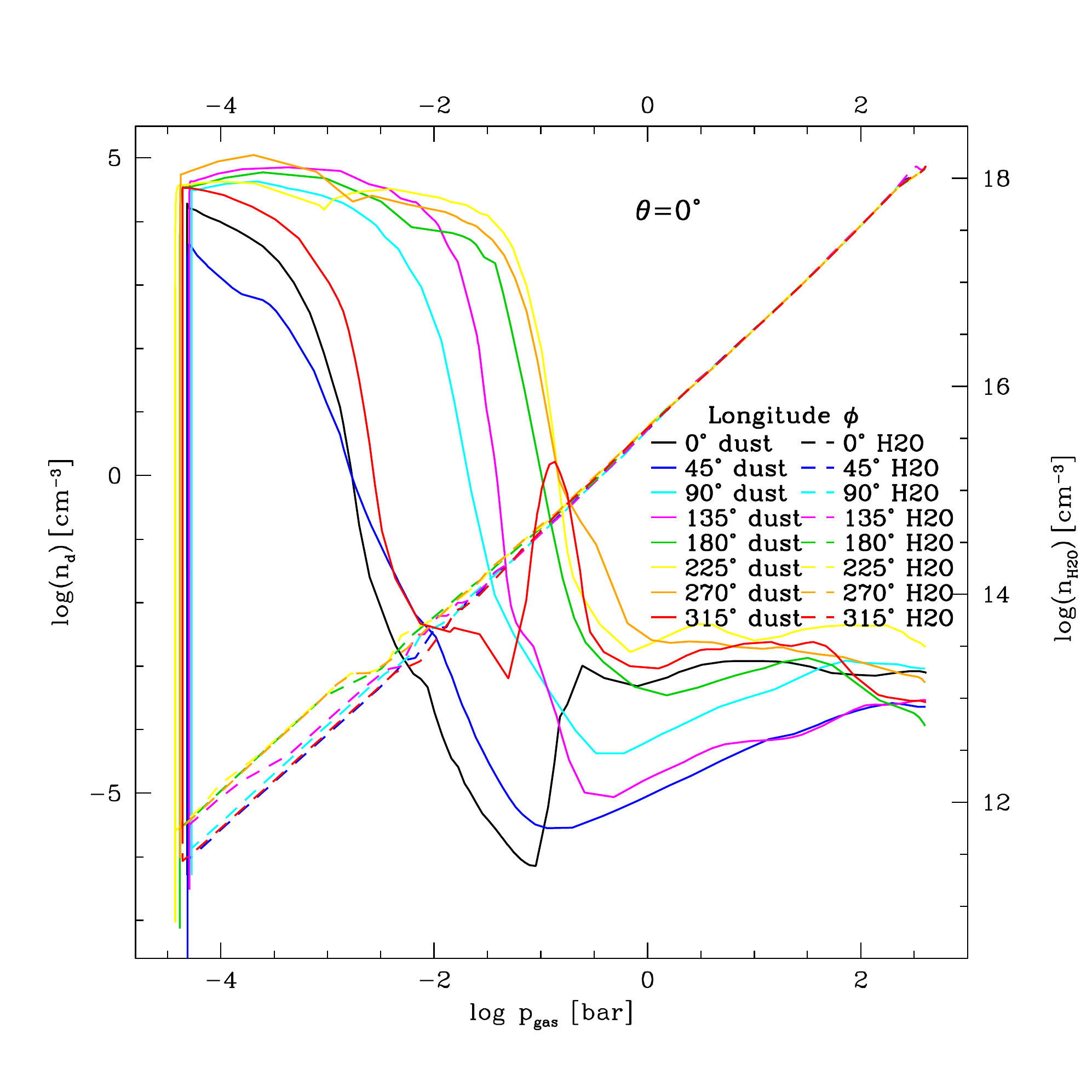}\\*[-1.2cm]
\includegraphics[scale=0.42]{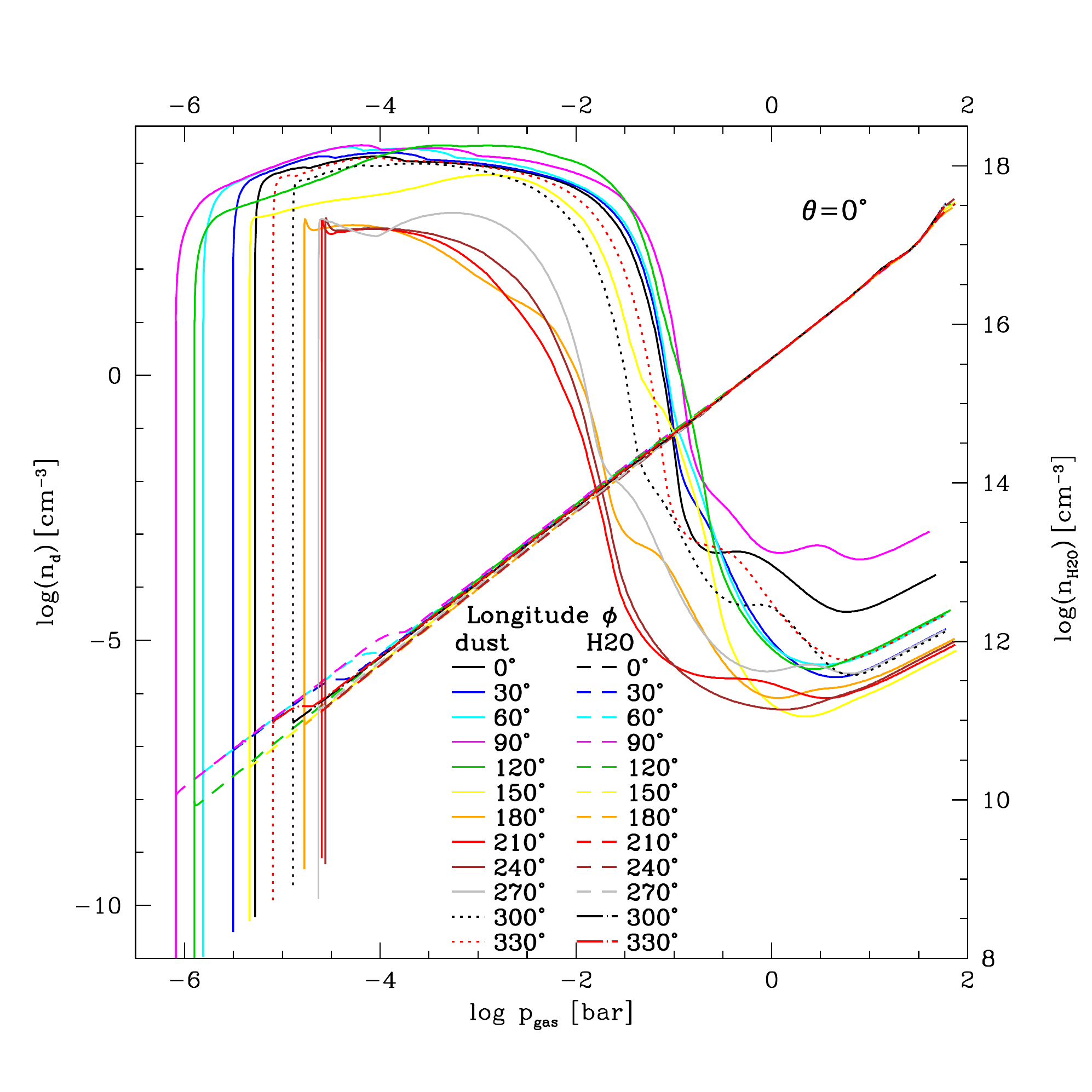}
\caption{The water abundance (dashed lines) compared to the location of the cloud top (in terms of $n_{\rm d}$) for HD\,189\,733b (top) and HD\,209\,459b (bottom).}
\label{fig:H2ondHD209}
\end{figure}

We consider the collisional dominated inner (deeper) part of an
atmosphere where kinetic gas-phase modelling is not required, but
which composes the inner boundary for kinetic gas-phase rate networks
(e.g. \citealt{agu2012,rim2015}).  We consider the part of the
atmosphere where the formation of dust clouds influences the local
gas-phase chemistry.  The cloud formation causes a depletion of those
elements which take part in the condensation process (e.g. Fe, Mg, O,
Ti; Fig.~\ref{fig:eabun}) resulting in reduced abundances of
respective molecules.  We neglect any external radiation field effect
on the gas phase chemistry.  We are interested in
\begin{itemize}
\item by how much the gas-phase composition changes due to the element
  depletion by cloud formation and how this effect might differ between
  the two planets,  HD\,189\,733b and  HD\,209\,458b (Figs.\ref{fig:molabun18920900})
\item TiO abundance in a cloud forming atmosphere (Figs.\ref{fig:molabun18920900})
\item how the C/O-ratio changes across the globe and between the two planets (Figs.\ref{fig:CtoOHD189})
\item how abundant PAH and other large carbon-hydrate molecules could be in  HD\,189\,733b and  HD\,209\,458b (Figs.~\ref{fig:PAHsHD189}, ~\ref{fig:C2H6HD189})
\item how the water abundance compares to the cloud location (Fig.~\ref{fig:H2ondHD209}).
\end{itemize}

\subsubsection{Chemical gas-phase abundance as a result of element depletion by cloud formation}
The first observation from our results is that n(CO)$\gg$ n(CH$_4$) at
the substellar point and at both terminators in HD\,189\,733b and
HD\,209\,458b in the case of oxygen depletion by cloud formation
(Fig~\ref{fig:molabun18920900}). CH$_4$ approach CO in abundance
in our test case if no oxygen-depletion occurs (see
Fig.~\ref{fig:molabun1899020},~\ref{fig:molabun20990270}). CO has been
suggested to exist in the upper atmosphere of HD\,189\,733b
(\citealt{des2009,dek2013,rodler2013}), and \cite{knu2012} suggest
that more CO is required than suggested by chemical equilibrium
calculation. More CO is required in order to suppress the Spitzer
4.5$\mu$m flux to the observed level with the models applied in
\cite{knu2012}.

Figure~\ref{fig:molabun18920900} demonstrates how the molecular
abundances change as a result of element depletion.
%: $\log(n_{\rm
% undepleted}/n_{\rm depleted})<0$ indicating a higher abundance of
%the respective molecule after depletion.  Most of the molecules were
%more abundant for an undepleted gas (solar abundances), hence
%$\log(n_{\rm undepleted}/n_{\rm depleted})>0$. However, CO$_2$,
%H$_2$O, HS and H$_2$S increase in abundances.
% due to the decrease of available oxygen in both planets. 
All molecules containing Fe/Al/Mg/Si/O or a combination of those follow element depletion pattern.
%The same effect occurs for CO and MgS to a lesser degree in the high
%density part of HD\,189\,733b. 
Molecules like NH$_3$ and N$_2$ are not
visibly affected by cloud formation because N is not part of any
condensing species considered here. CaH and CaCl are also not affected
by the cloud formation despite Ca being part of CaTiO$_3$ as
condensing species.

Figure~\ref{fig:molabun18920900} further demonstrates that the
abundance of individual molecules is reduced by many orders of
magnitudes which change throughout the atmosphere. It is further
important to note that this effect occurs also for CH$_4$ and CS which
are molecules that are not affected by depletion of one of their
constituent elements. Instead, they are influenced by element
depletion through other molecules like CO$_2$ for CH$_4$ and AlS, FeS,
MgS for CS.

Comparing the abundances of molecules of major interest for
observations, CO, H$_2$O and CH$_4$, for HD\,189\,733b and
HD\,209\,459b shows that the abundance of CO and H$_2$O are very
similar in both atmospheres at the substellar point
(Fig.~\ref{fig:molabun18920900}) in a gas that is affected by
element depletion due to cloud formation. HS differs by one order of
magnitude between the two planets. Figure~\ref{fig:molabun1899020} and
~\ref{fig:molabun20990270} (left columns) demonstrate that differences
increase at the terminator points which will affect transit
spectroscopy.  H$_2$O remains the most abundant of these molecules and
of similar abundance in both atmospheres, but CO and CH$_4$ strongly
decrease in the upper-most layers at the east terminator ($\phi=90^o$)
in HD\,209\,459b. The differences in the local (T$_{\rm gas}$(z),
p$_{\rm gas}$(z)) structures between the terminator and substellar
trajectories are imprinted into the local molecular abundances: all
trajectories show distinct chemical differences in both planets. We
refer for more details to
Figs.~\ref{fig:molabun18920900},~\ref{fig:molabun1899020} and
~\ref{fig:molabun20990270}.  Cloud free models will overestimate
differences in molecular abundances between both planets (compare
right columns of Figs.~\ref{fig:molabun1899020} and
~\ref{fig:molabun20990270}).

Given the interest in additional opacity species to possibly account
for temperature inversions in atmospheres of irradiated planets, we
discuss our findings about the TiO abundance. First, Ti is one of the
most depleted elements in an oxygen rich atmosphere where cloud
formation takes place. This is a result rather independent of the
cloud modelling approach simply because of the low solar abundance
value for Ti. 
%Figure~\ref{fig:molabun18920900} (bottom) shows that 
TiO is depleted by 5-6 orders of magnitudes in both planets inside the
cloud region. Only TiH, AlH, AlS, and Al$_2$O appear more depleted
than TiO. Considering now that the upper boundaries of our present
simulations for HD\,189\,733b (left) and HD\,209\,458b are
artificially fixed, it is reasonable to expect that the cloud expands to
lower pressures and with that the TiO depletion would continue to higher
atmospheric layers.

Figure~\ref{fig:molabun18920900}  suggest that a temperature
inversion might be detectable by atomic emission lines as Fe and Mg
appear very abundant and even enhanced to a small extent. Temperature
inversions are related to a local positive temperature gradient at the
high-pressure part of it.  Molecules like TiO and CO seem not too well
suited to search for a temperature inversion because TiO has a very
low abundance in the cloud region and CO is simply not affected by the
temperature inversion, at least according to our equilibrium chemistry
calculation inside the irradiation-shielded and collisional dominated
atmosphere. However,  strong molecular lines may be used to derive local temperatures, for example from high-resolution spectra as in \cite{schw2015}.

\subsubsection{Water abundance}\label{ss:water}
Figure~\ref{fig:H2ondHD209} compares the location of the cloud in
terms of cloud particle number density, $n_{\rm d}$, along the equator
to the H$_2$O gas abundance (dashed lines). Water is the most abundant
gas-phase absorber in ultra-cool atmospheres, and
Fig.~\ref{fig:H2ondHD209} shows where the molecule is most abundant
compared to the cloud location.  The detection of atmospheric molecular water in
HD\,189\,733b has been reported by \cite{tod2014,cro2014} near $6\mu$m
from secondary eclipse emission spectrum and \cite{Mcc2014} report the
detection of water in the transit absorption spectrum of
HD\,189\,733b. \cite{evans2015} initial results from Spitzer
observations suggest no water detection for HD\,209\,458b but the
spectrum would suggest the detection of haze and depleted CO. However,
\cite{deming2013} provide evidence for molecular water at shorter wavelengths,
1.4$\mu$m, from Hubble observations, which recently was confirmed by \cite{sing2016}. While molecular water is omnipresent in
both atmospheres, in HD\,189\,733b and HD\,209\,459b, modelled here,
we have an indication for a drop in CO abundance at the east
terminator ($\phi=90^o$). This result, however, would need to be
confirmed by kinetic gas-phase calculations to test the effect of the
stellar radiation field.  Returning to Fig.~\ref{fig:H2ondHD209}
suggests that a considerably larger portion of the atmosphere of
HD\,209\,458b is affected by the dust cloud. The lower mass of
HD\,209\,458b leads to a smaller surface gravity which leads to a
larger atmospheric scale height resulting in a larger geometrical
extension of the clouds. Our results concerning the presence and
location of gaseous water in relation to the clouds may explain some of the
observed differences between the two planets. In HD\,189\,733b,
strong molecular water features have been reported while HD\,209\,458b
H$_2$O-absorption appears much shallower.
%seems to have much less.  
However, by considering the location of the clouds in
the two atmospheres, we see that obscuring clouds exist high in the
atmosphere of HD\,209\,458b, but much deeper in HD\,189\,733b. Cloud
opacity calculations suggest that gaseous water could be observable on both
planets, on HD\,189\,733b and on HD\,209\,458b
(Fig.~\ref{fig:opac90detailed},~\ref{fig:opac00detailed},~\ref{fig:opac90detailed})
as the cloud extinction is low at wavelength related to Spitzer bands.
A closer inspection of the 3.6$\mu$m, 4.5$\mu$m, and 5.8$\mu$m opacity
in Figs.~\ref{fig:opacHD189} and ~\ref{fig:opacHD209} as indicative of
the IRAC 1,2 and 3 bands, shows that the opacity peaks at $\approx$0.1
bars. 
%Comparing to Fig.~\ref{fig:H2ondHD209}, the water content at 0.1
%bars is higher on HD\,189\,733b then in HD\,209\,458b in agreement
%with observations.
%However, the atmosphere models used in this study may not cover
%a large enough pressure range.

\subsubsection{Changing carbon-to-oxygen ratio}\label{ss:ctoo}
\begin{figure}
{\ }\\*[-1.6cm]
\centering
\includegraphics[scale=0.39]{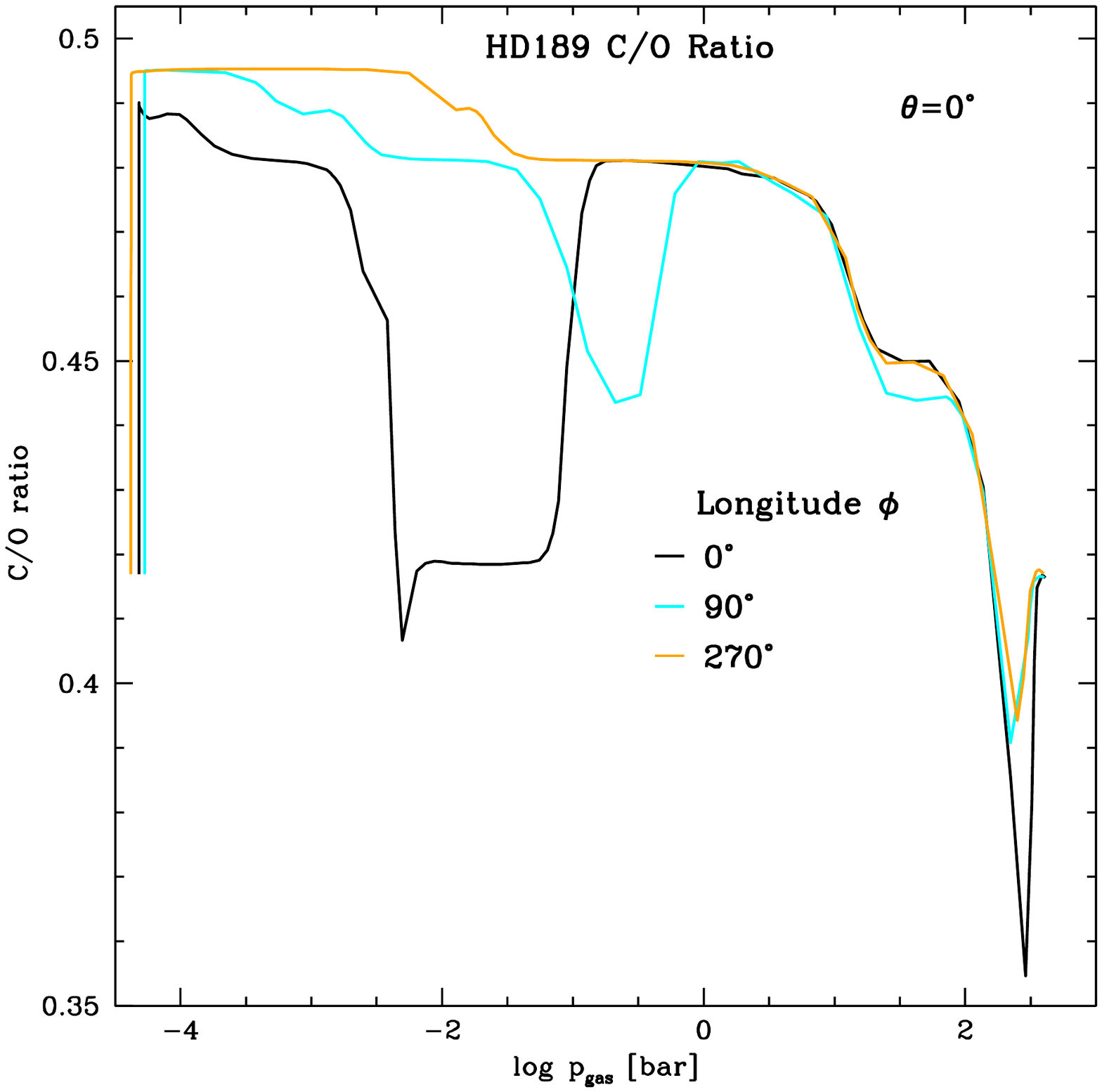}\\*[-2.7cm]
\includegraphics[scale=0.41]{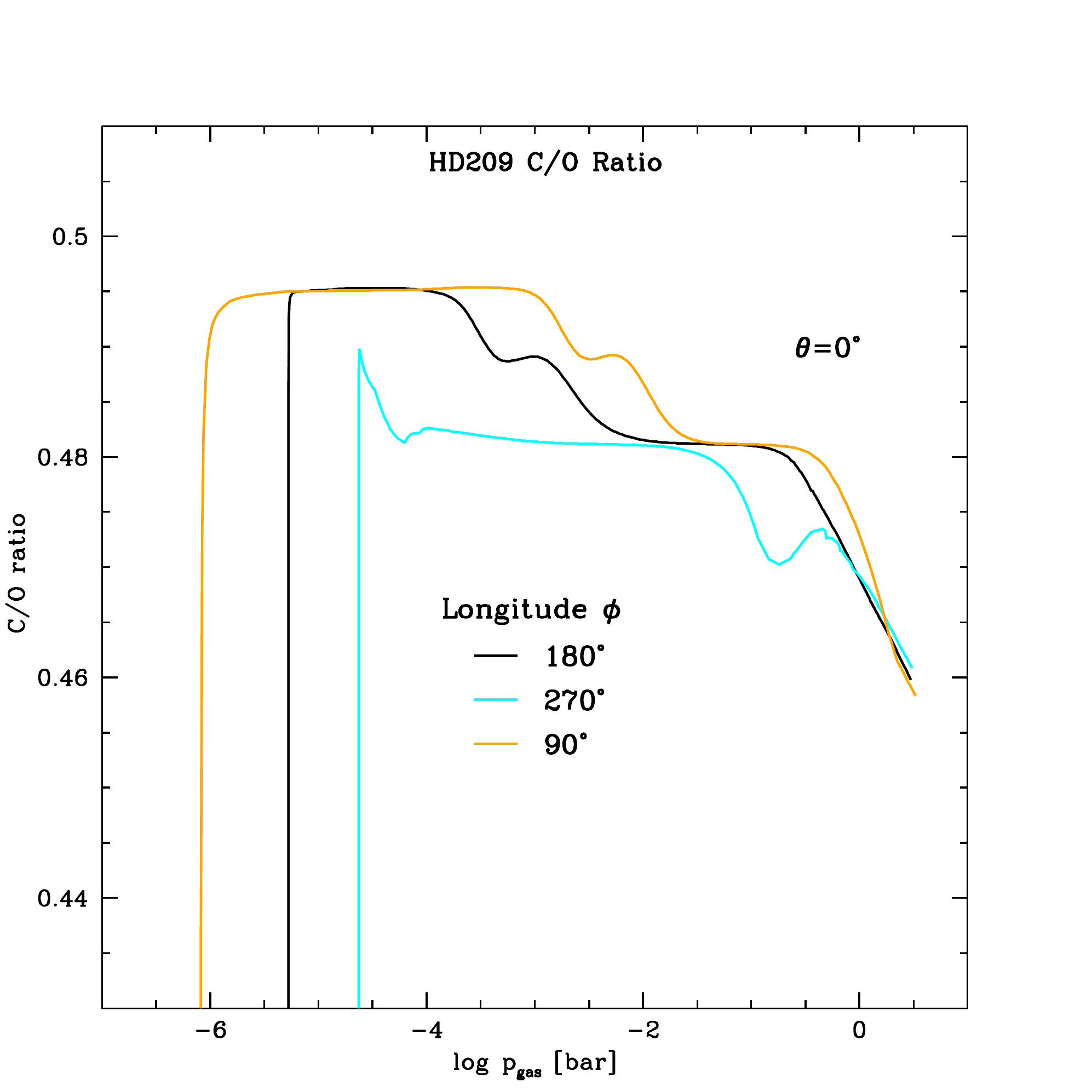}
\caption{The changing carbon-to-oxygen ratio, C/O, in the atmospheres
  of HD\,189\,733b (top), HD\,209\,458b (bottom) plotted  at the substellar point (black solid) and the two terminators ($\phi=90^o$ - cyan,
  $\phi=270^o$ - brown). The initial, solar value is C/O=0.427
  (\citealt{AndGre1989}) for both planets. (Note that in these plots
  the HD\,209\,458b-$\phi$ has a 180$^o$-offset compared to
  HD\,189\,733b.)}
\label{fig:CtoOHD189}
\end{figure}

The carbon-to-oxygen ratio (C/O$=\log\epsilon_{\rm C}/\log\epsilon_{\rm
  O}$, $\epsilon_{\rm C}$ - carbon element abundance, $\epsilon_{\rm
  O}$ - oxygen element abundance) is often used as indicator for
potential carbon-binding molecules and it has recently attracted
attention in the exoplanetary community with respect to the study of
pre-biotic carbohydrate molecules. The C/O ratio has been used a
global parameter (similar to log(g) or T$_{\rm eff}$) to argue for the
existence of carbon planets where C/O$>$1 (\citealt{mad2012}).
\cite{2014Life} demonstrated that a considerable overabundance of
carbon over oxygen is very unlikely to occur during disk evolution,
and that cloud formation is one mechanism to tip an atmosphere into a
carbon-rich regime if C/O is already close to one. \cite{benneke2015}
concludes from his analysis that HD209458b, WASP-12b, WASP-19b,
HAT-P-1b, and XO-1b but also WASP-17b, WASP-43b and HD\,189\,733b have
atmospheres with a global C/O$<$0.9, hence these atmosphere appear
oxygen-rich. We summarize our findings about the C/O ratio for
HD\,189\,733b and HD\,209\,458b based on the combined modelling
approach of this paper in
Fig.~\ref{fig:CtoOHD189}. Figure~\ref{fig:CtoOHD189} shows that the
C/O ratio changes with height throughout the cloud. It first increases
due to the massive depletion of oxygen from an initial value of
C/O=0.427 to C/O=0.495 for both planets near the top of the
cloud. Depending on longitude, it remains at the this maximum level or
starts to drop due to oxygen being released to the gas through
silicate evaporation. In the case of HD\,189\,733b, the oxygen
abundance is enriched to values larger than the initial solar value
which is a clear sign for cloud particles moving through the
atmosphere and, hence, transporting elements into different
regions. This is particularly prominent at the cloud base of
HD\,189\,733b where C/O has decreased to almost 0.35.  Note that these
changes are solely caused by changes in the $\epsilon_{\rm O}$ as
$\epsilon_{\rm C}=\epsilon_{\rm C}^{\rm solar}$ in the present
models. Such a changing C/O ratio has implications for synthetic
spectrum calculation, for example for young objects where the accreted
dust from the protoplanetary disk serves as condensation seed and
clears out the atmosphere until the supersaturation ($S=p_{\rm
  X}/p_{\rm sat,X}$, $X$ any key species of a condensing material)
drops below one for the key species of a condensing material,
i.e. until $S=1$.  Our investigations show again that it remains an
illusion to attribute one C/O ratio value or one metallicity value to a
planet that forms clouds inside its atmosphere. With that we confirm
our previous findings in (\citealt{Bilger2013, 2014Life}).

\begin{figure}
\includegraphics[scale=0.4]{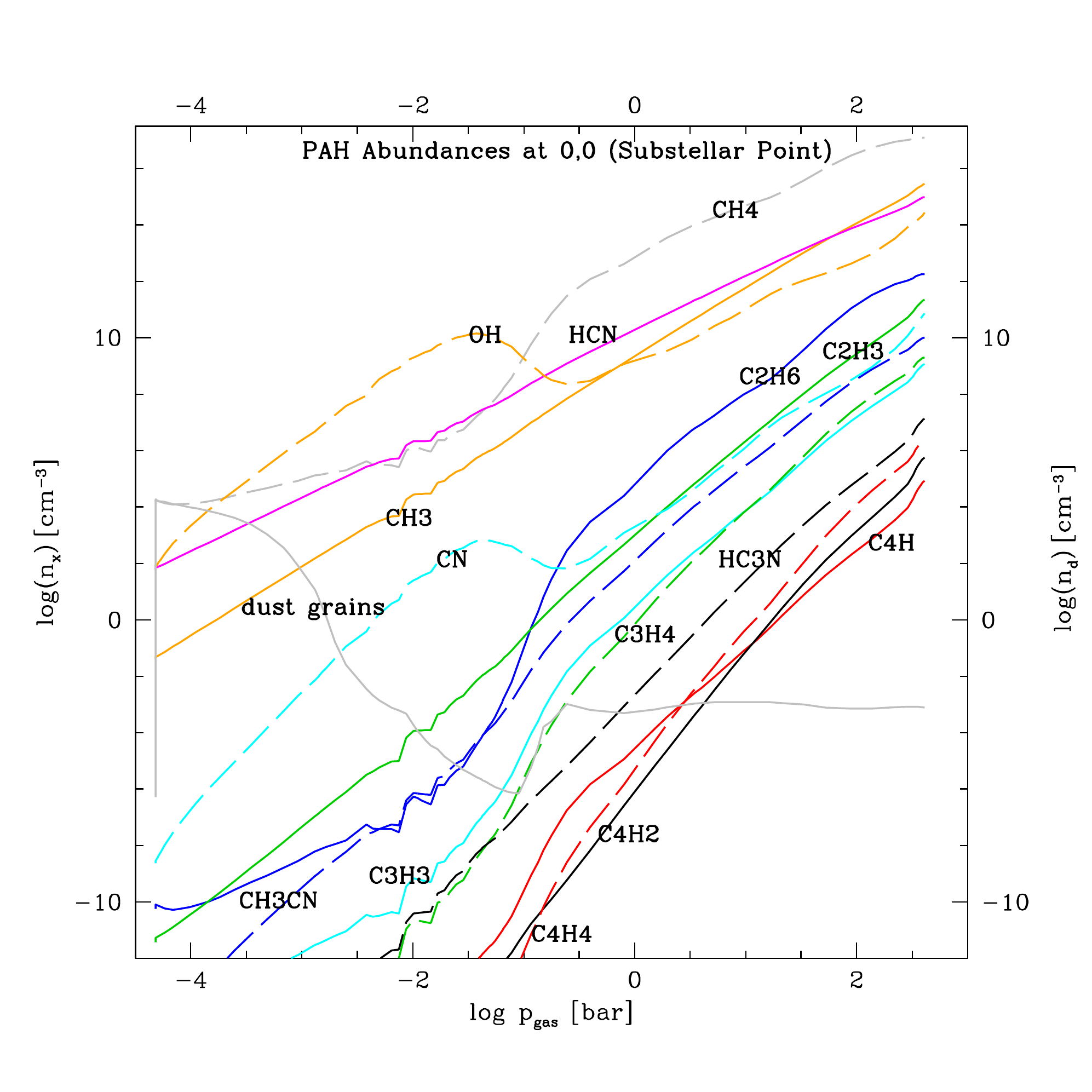}\\*[-1.2cm]
\includegraphics[scale=0.4]{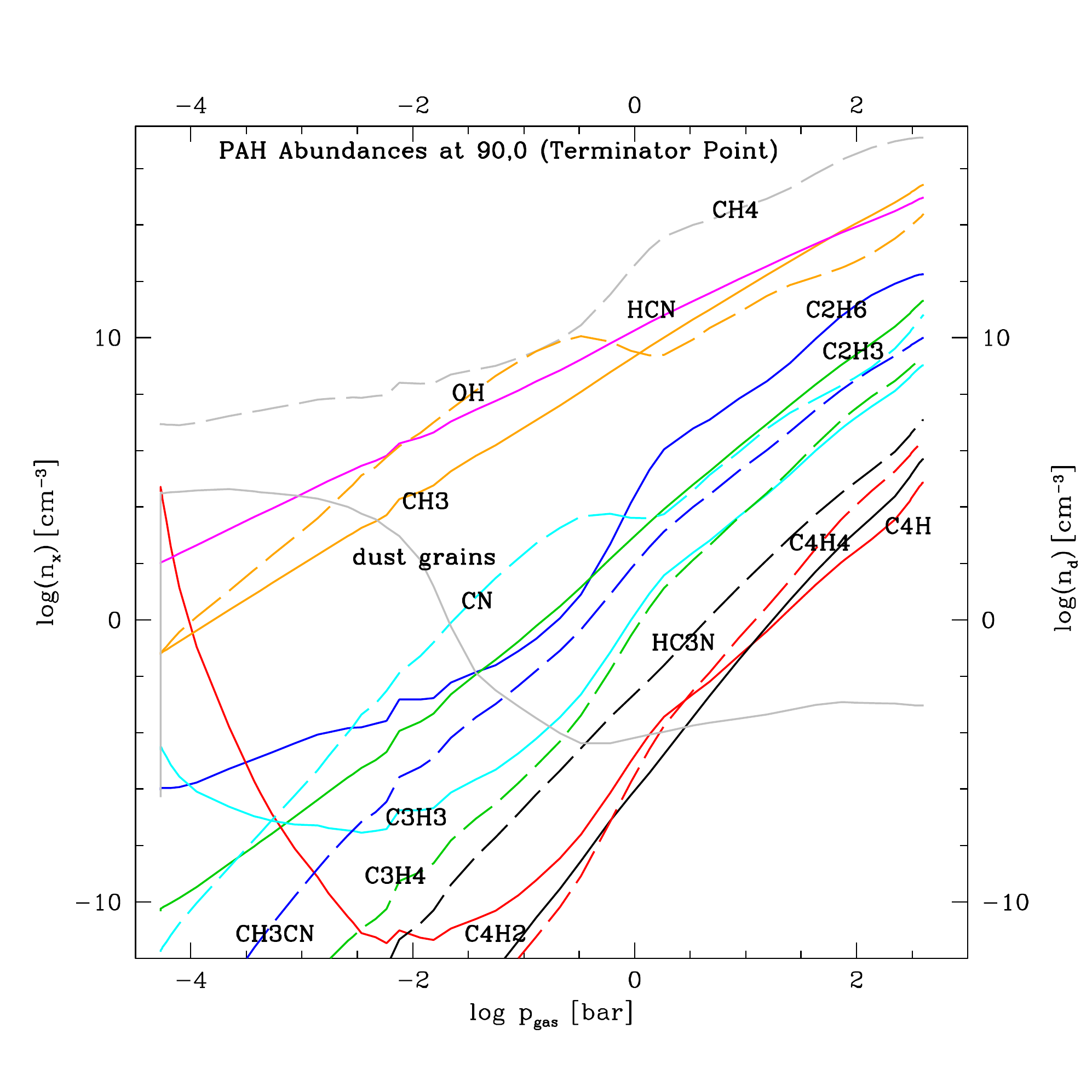}\\*[-1.2cm]
\includegraphics[scale=0.4]{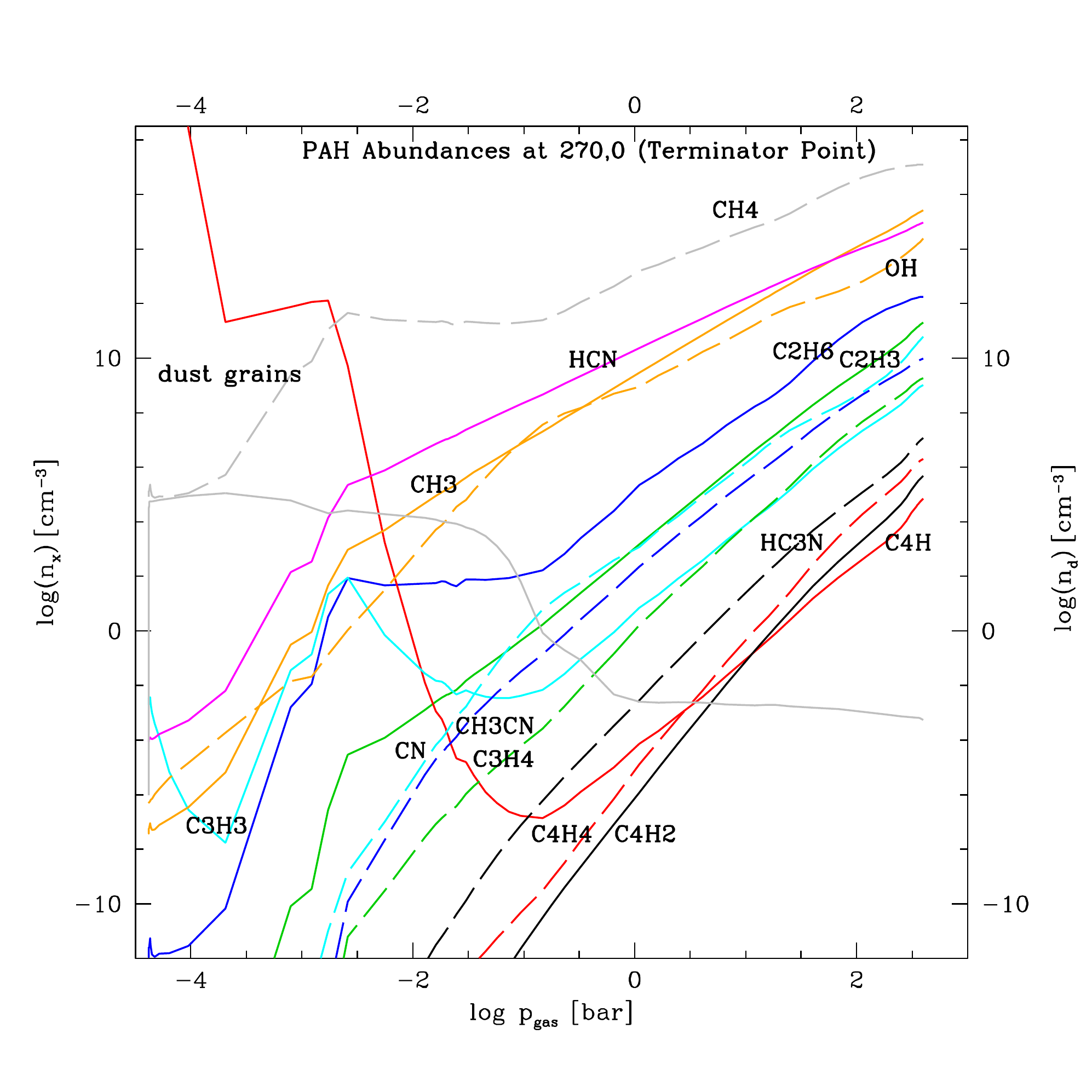}
\caption{PAH abundances (left axis) in HD\,189\,733b. The number of cloud particles is over-plotted (right axis) to compare to the cloud location.}
\label{fig:PAHsHD189}
\end{figure}

\begin{figure}
\includegraphics[scale=0.4]{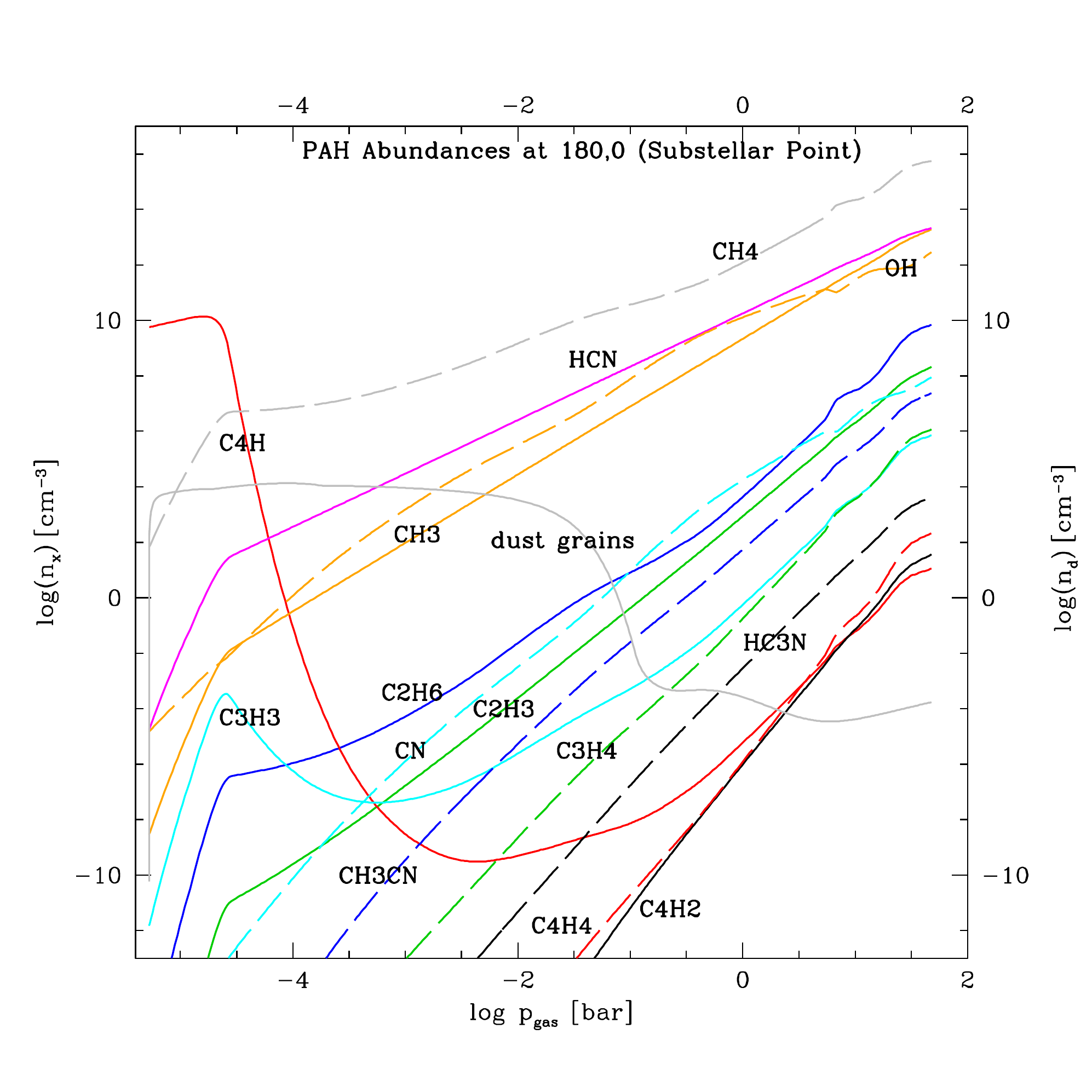}\\*[-1.2cm]
\includegraphics[scale=0.4]{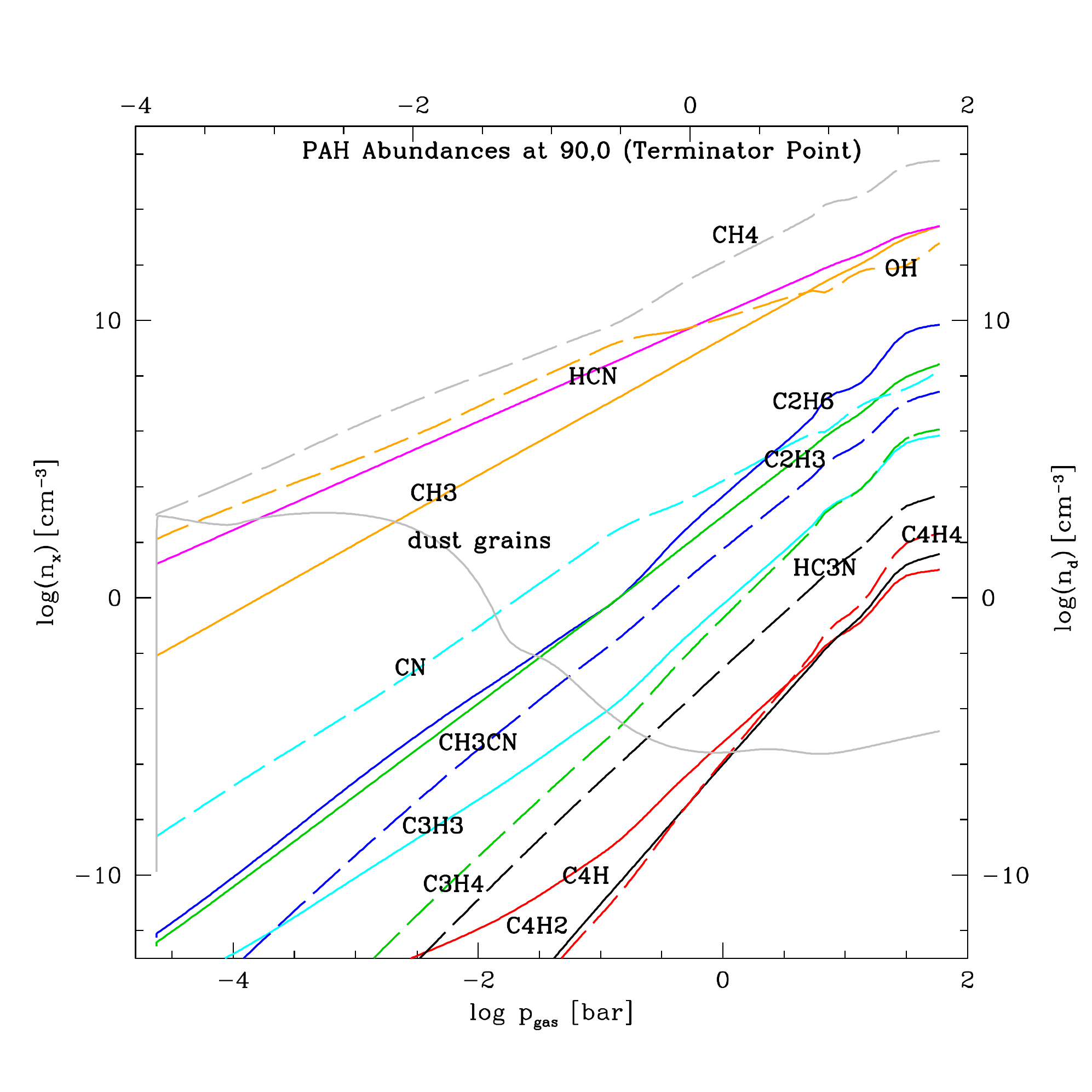}\\*[-1.2cm]
\includegraphics[scale=0.4]{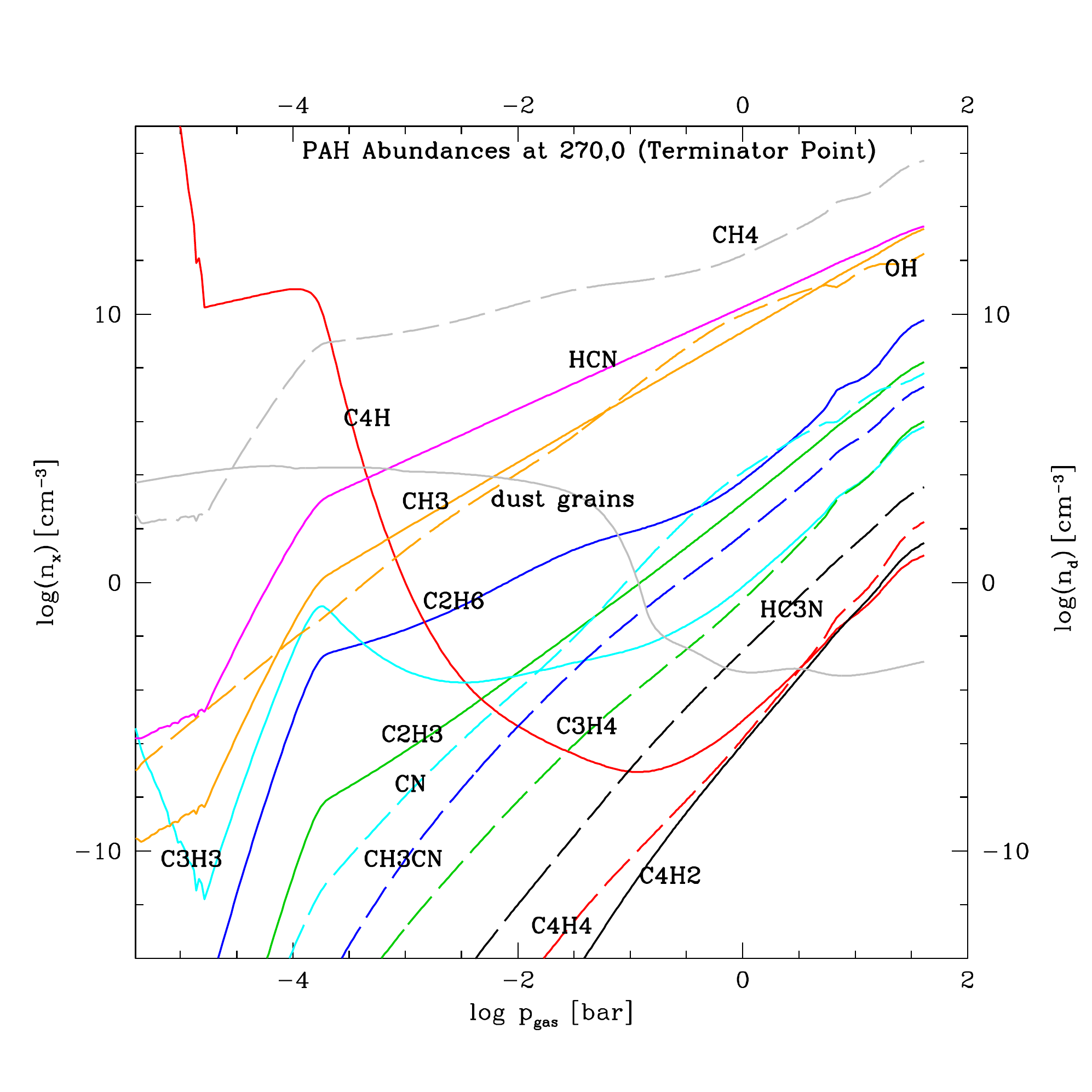}
\caption{PAH abundances (left axis) in HD\,209\,458b. The number of cloud particles is over-plotted (right axis) to compare to the cloud location. }
\label{fig:PAHsHD209}
\end{figure}

\subsection{The abundances of hydrocarbon and cyanopolyyne molecules}

Hydrocarbon and cyanopolyyne molecules are of growing interest for
extrasolar planets due to their potential link to biomolecules
(e.g. \citealt{rim2015} and references therein). The oxygen
depletion caused by the formation of clouds has a strong impact on the
abundance of the dominant molecules as re-iterated in
Sect.~\ref{ss:chemequ}. We are now in the position to investigate what
affect this would have on the abundance of carbohydrate molecules
including large N-bearing species HC$_{\rm x}$N, complex hydrocarbons
C$_{\rm n}$H$_{\rm 2n\pm 2}$, C$_2$H$_{\rm 2n}$, CH-bearing radical
C$_{\rm x}$H, CH$_{\rm x}$ and C$_{\rm x}$.  \cite{Bilger2013} have
demonstrated that cloud formation leads to a increase in the C/O ratio
such that several hydrocarbon molecules and cyanopolyyne
molecules can be present in brown dwarf and giant gas planet
atmospheres. A large surface gravity and/or a decrease in metallicity
lead to an increased abundance of these species.  \cite{Bilger2013}
show that CO, CO$_2$, CH$_4$ and HCN contain the largest fraction of
carbon.

Figures~\ref{fig:PAHsHD189} and ~\ref{fig:C2H6HD189} show the
abundance of hydrocarbon molecules and cyanopolyyne molecules in
HD\,189\,733b and HD\,209\,458b, respectively. We note that these
abundances are for the collisionally dominated part of the atmosphere
only, where the assumption of chemical equilibrium is valid. 

The most abundant hydrocarbon molecule is CH$_4$ in both planets,
followed by HCN, OH and CH$_3$ with changing importance depending on
the local thermodynamic conditions. These molecules reach abundances
comparable to MgH, MgS, CaH or Fe at p$_{\rm gas}>$1bar
(Fig.~\ref{fig:molabun18920900}). The larger hydrocarbon and
cyanopolyyne molecules abundance decrease strongly towards the upper
atmosphere but are rather substantial in the high-density part of the
atmosphere where p$_{\rm gas}>$1bar. Large molecules like C$_2$H$_6$
and C$_2$H$_3$ followed by CH$_3$CN are the most abundant molecules
after CH$_4$, HCN, OH and CH$_3$.  The number density of the most
abundant hydrocarbon molecule (C$_2$H$_6$) changes considerably in
particular in the low-density part of the atmosphere of both planets
(Fig.~\ref{fig:C2H6HD189},~\ref{fig:C2H6HD189}). The C$_2$H$_3$
abundances, in contrast, do not change very much with longitude.

We therefore conclude that hydrocarbon and cyanopolyyne molecules can
be rather abundant in the inner, dense part of the atmospheres of
HD\,189\,733b and HD\,209\,458b but non-equilibrium processes will
need to be studied to see if these molecules could reach an
observational abundance also in the upper part of the atmosphere. It
is, however, likely that these molecules would be hidden beneath cloud
layers as the comparison with the cloud particle number density in
Figures~\ref{fig:PAHsHD189} and ~\ref{fig:C2H6HD189} suggests.

These findings are also of interest for future kinetic gas phase
studies that take into account the influence of high-energy radiation
like for example cosmic rays or the host star's radiation field.

\section{Different cloud opacity for different planets}\label{s:opacity}

The cloud extinction is relevant for understanding observations of hot
spots or molecular abundances as discussed in Sect.~\ref{ss:water}
for water.  Many authors very approximatively describe the chemically
complex clouds on extrasolar planets like HD\,189\,733b and
HD\,209\,459b. We therefore present the cloud extinctions for
HD\,189\,733b and HD\,209\,459b for their day-side and the night-side
in conjunction with the $\langle a(z)\rangle$ (height-dependent mean
cloud particle size). The material composition that relates to an
$\langle a(z)\rangle$ for the atmospheric trajectories considered here
can be retrieved from Figs.~\ref{fig:VdaHD189}
and~\ref{fig:VdaHD209}. We demonstrate how the cloud opacities differ
between the two planets in the following section.

\noindent
\begin{figure*}
\centering
\includegraphics[scale=0.6]{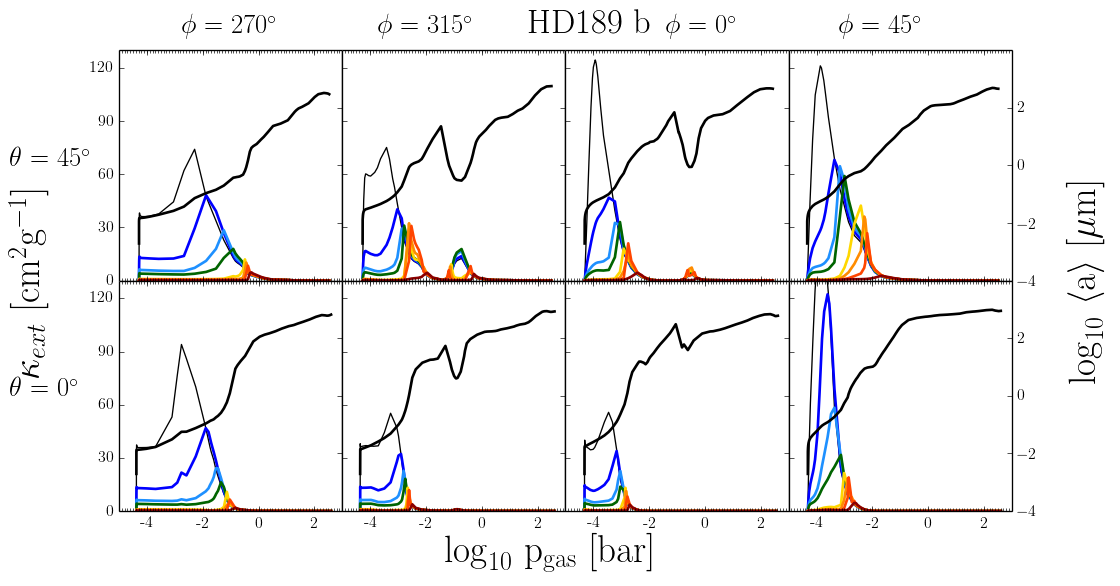}\\
\includegraphics[scale=0.6]{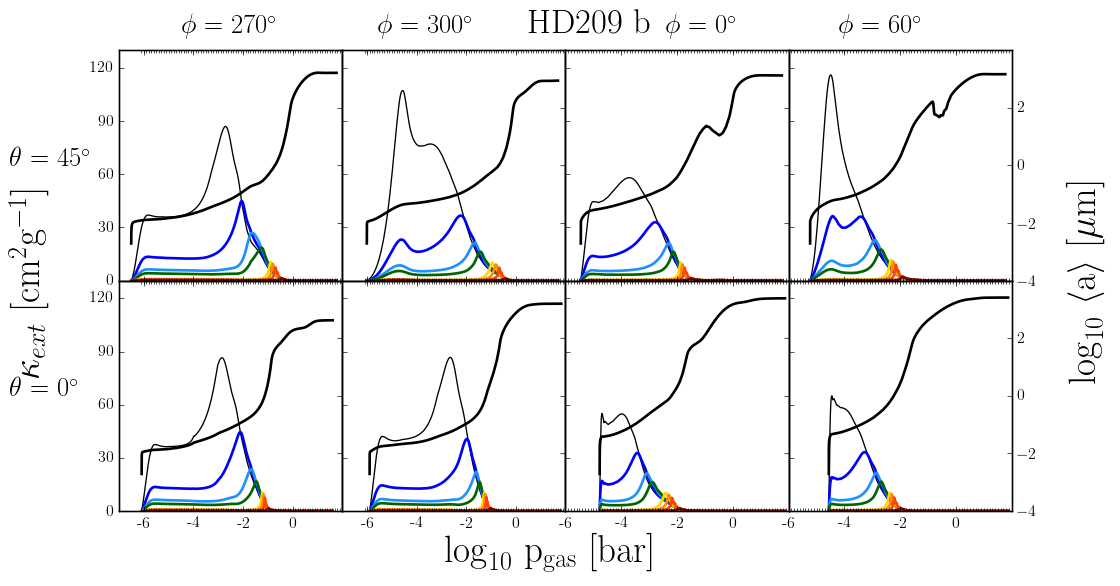}\\
\includegraphics[scale=0.6]{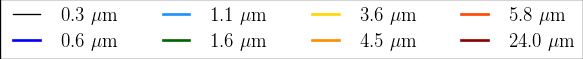}
\caption{The {\bf day-side} cloud extinction, $\kappa_{\rm ext}$
  [cm$^2$g$^{-1}$] at the equator ($\theta=0^o$) and in the northern
  hemisphere ($\theta=45^o$) for different wavelengths $0.3\ldots
  24\mu$m (absorption + scattering, colour coded).  The right axis
  depicts the mean grain size, $\langle a\rangle$ [$\mu$m] for each
  longitude. Top panels - HD\,189\,733b, bottom panels -
  HD\,209\,458b).}
\label{fig:opacHD189}
\end{figure*}

\noindent
\begin{figure*}
\centering
\includegraphics[scale=0.6]{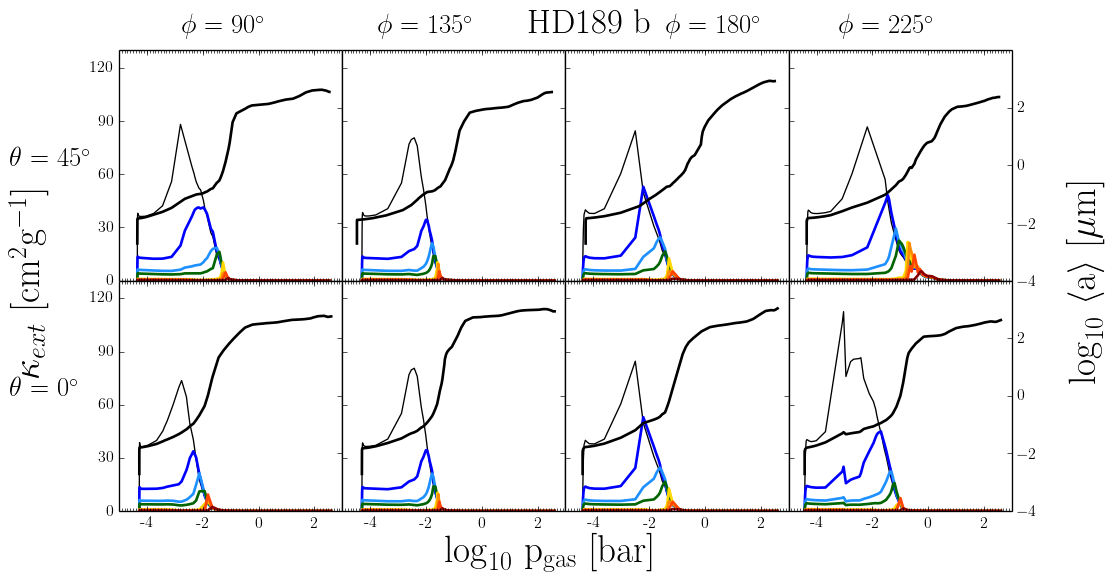}\\
\includegraphics[scale=0.6]{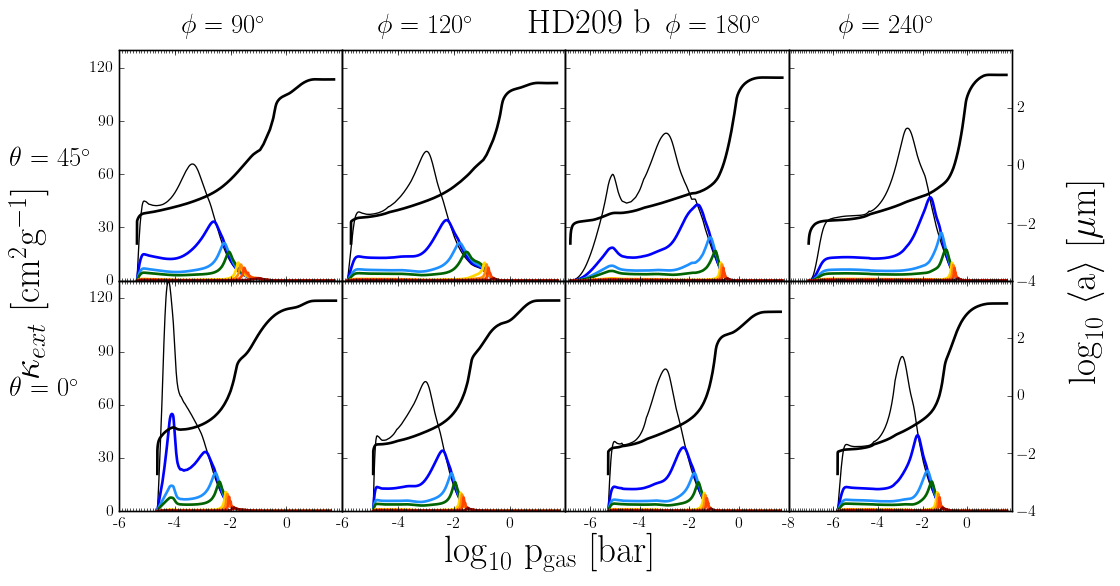}\\
\includegraphics[scale=0.6]{plots/legendv3.png}
\caption{The {\bf night-side} cloud extinction, $\kappa_{\rm ext}$
  [cm$^2$g$^{-1}$] at the equator ($\theta=0^o$) and in the northern
  hemisphere ($\theta=45^o$) for different wavelengths $0.3\ldots
  24\mu$m (absorption + scattering, colour coded).  The right axis
  depicts the mean grain size, $\langle a\rangle$ [$\mu$m] for each
  longitude. Top panels - HD\,189\,733b, bottom panels -
  HD\,209\,458b).}
\label{fig:opacHD209}
\end{figure*}

\subsection{Cloud extinction from  cloud formation}
Figures~\ref{fig:opacHD189} shows the day-side cloud extinction,
$\kappa_{\rm ext}$ [cm$^2$g$^{-1}$], and Fig.~\ref{fig:opacHD209} the
night-side cloud extinction as sum of the cloud absorption and
scattering. Figures~\ref{fig:opac00detailed} (substellar point) and
~\ref{fig:opac90detailed} (terminator) provide more details of the
individual contributions of scattering and absorption of the cloud
layers.

Generally, the cloud opacity is individual to each of the two
planets. In both planets, however, the cloud opacity is largest in the
upper cloud layers which are dominated by particles of $\langle
a\rangle \le 0.1\mu$m. Both clouds are darkest if observed at
$0.3\mu$m and appear successively lighter if observed at longer
wavelengths as the extinction decreases with increasing
wavelengths. The extinction drops rapidly at a certain pressure level
in both planets and this is associated with the strong rise in
particles sizes. Once the mean particles sizes (panel 3 in
Figs.~\ref{fig:struc1},~\ref{fig:struc2}) has reached a maximum value,
the cloud extinction drops to its minimum. HD\,209\,458b shows
generally a smother change in cloud particle size
(Fig.~\ref{fig:amean}, bottom) as result of the formation of seed
particles occurring rather homogeneously along those trajectories
studied here.  Our results for HD\,189\,733b demonstrates that a
temperature inversion in the atmosphere has a distinct effect on the
cloud properties that will also affect the cloud extinction: Seed
formation occurs intermittently along a certain trajectory and in
comparison between different trajectories (Fig.~\ref{fig:J*},
$\phi=0^o, 45^o, 315^o$). As consequence, the mean grain size changes
vividly along such a trajectory, hence the cloud particle sizes change
strongly with atmospheric height, but also with longitude as
Fig.~\ref{fig:amean} (top) demonstrates. Following now the cloud
extinction at $\phi=0^o, 45^o$ and $\phi=315^o$ in HD\,189\,733b in
Fig.~\ref{fig:opacHD189} shows, that the cloud opacity shows the most
unpredictable changes as function of wavelength and also with
atmospheric height at these longitudes. To put these results in
perspective, most of the transit spectroscopy observations are fitted
with one particle size for the whole cloud, retrieval methods assume
2-4 particle sizes (e.g. \citealt{kok2011,bars2014,lee2014}).
However, it is interesting to note that both planets have similar
cloud opacity at the 90$^o$-terminator despite HD\,209\,458b having
lower gas pressure at the cloud top. The differences between the two
planets in cloud opacity at the 270$^o$-terminator originates from a
difference in number of cloud particles at the cloud top as the
r.h.s. of Fig.~\ref{fig:struc2} indicates, but not from a difference
in particle sizes or materials. 

While the material composition is important for the element depletion
of the gas phase, it is less important for the cloud extinction near
the top of the atmosphere. Here, the number of particles and the mean
grain size have the largest impact on the cloud extinction. It is
therefore not surprising that observations for transitting planets can
be fit by almost every material for which optical data are available
(e.g. \citealt{wak2015,nik2015}). This also means, that the cloud
extinction is the result of the cloud formation process as the mean
grain size will be determined by the rate at which condensation seeds
can form and where. We have discussed this above for the example of a
temperature inversion. In conclusion, we demonstrated that only a
detailed cloud model lifts the $n_{\rm d}$/$\langle a\rangle$/ $V_{\rm
  s}/V_{\rm tot}$ degeneracy in retrieval methods. This degeneracy is
artificially generated by assuming independence of cloud parameters
which is simply incorrect.

\section{Discussion}\label{s:disc}

\subsection{Cloud properties in perspective}
The desire to extract meaningful information from observations (or to
decide what a meaningful information is) led to the development of
retrieval methods which mainly solve the radiative transfer problem
for prescribed parameters. The challenge is now to statistically
evaluate and re-set these initial parameters such that a best fit to a
set of observational information is achieved. Therefore, observations
were fitted by a selected number of cloud size bins, a prescribed
material composition and number of cloud particles, prescribed
pressure levels for cloud deck and the cloud base and other
assumptions for the gas composition.

\smallskip
\noindent
{\bf HD\,189\,733b:}\\ \cite{leca2008} identifed MgSiO$_3$[s]
as a possible abundant condensate with particle size $\approx
10^{-2}\,\ldots\,0.1\mu$m. \cite{bars2014} used STISS observations of
HD\,189\,733b to retrieve a monodisperse cloud layer made of
MgSiO$_3$[s] (or MnS[s]). Best fits were achieved for small particles $<0.3 \mu$m or 10
$\mu$m particles, but a degeneracy with the Na abundance occurred. The
authors settled on a best fit by a homodisperse cloud layer made of
$<0.1 \mu$m-sized MnS[s] particles and 50 ppmv of
Na.\\ \cite{lee2014}, for the same object, chose MgSiO$_3$[s] as
material for their chemically homogeneous cloud particles. They also
derive a vertically homogeneous layer for 0.1$\mu$m-sized particles.
Both groups treat their cloud parameters and gas phase abundances as
independent parameters and find a degeneracy between gas composition
and cloud particle parameters. \cite{lee2014} emphasis that this
degeneracy does not affect H$_2$O.\\ 
\cite{pont2013} imply that various cloud particle sizes would be
required to fit to their transit spectra of HD\,189\,733b.

\smallskip
\noindent
{\it Our results} indeed show
that the upper part of the cloud on HD\,189\,733b is dominated by
small cloud particles of $\approx 0.01\,\ldots\,0.1\mu$m and an
abundance of $n_{\rm d}\approx 10^{-4}$ cm$^{-3}$. The material
composition of a mix of materials with the most abundant materials
being $\approx 22$ \% MgSiO$_3$[s], $\approx 20$ \% SiO$_2$[s],
$\approx 18$ \% Mg$_2$SiO$_4$[s], $\approx 16$ FeS[s] which quickly is
replaced by the same amount of Fe[s], and $\approx 9$ \%
MgO[s]. These detailed volume fractions are for the $90^o$
terminator. We have demonstrated that the relative abundance of the
materials does change across the globe and that MgSiO$_3$[s] is the
dominating material at the cloud top with $\approx 20$ \% volume
fraction but with other silicates only marginally less abundant.

\bigskip\bigskip
\noindent
{\bf HD\,209\,458b:}\\ Not much seems known from observation about
cloud properties on HD\,209\,458b, except the suggestion that the
observation of Si$^{2+}$ in the planet's exosphere indicates that no
clouds involving Si should form (\citealt{kos2013}). \cite{bur2008}
and \cite{rowe2006} conclude that there are no (high-altitude)
reflecting clouds in HD\,209\,458b.  \cite{hood2008} invoked iron and
enstatite (MgSiO$_3$[s]) assuming a homodisperse cloud particle size
of 5$\mu$m for their 3D radiative transfer calculation on fractal
clouds.\\ {\it Our results} show that the mean cloud particle size can
be as small as 0.01$\mu$m at the cloud top which might be accessible
by transit spectroscopy. These particles are made of a mix of
materials with again $\approx 21$ \% volume fraction by MgSiO$_3$[s]
and $\approx 18$ \% Mg$_2$SiO$_4$[s] or SiO$_2$[s] depending on
latitude.

\medskip
{\it Limitations} of these results arise from the confinement of our
simulation to a limited computational domain. One might therefore
wonder if other elements like Na, K, Mn, Cu, or Zn would change our
results if the atmosphere would extent to lower pressures.  Element
abundance determination is an ever evolving topic, but also the newest
literature confirms that Fe(7.47), Mg (7.59), Si (7.51), are
considerably more abundant in the Sun than Na (6.21), Mn (5.42),
K(5.04), Zn (4.56), or Cu (4.18)
(\citealt{scott2015,grev2015,scott2015b}) but considerably less than
oxygen.  Hence, if we use the Sun as a reference, all less abundant
elements can only contribute to the cloud particle growth with a very
small fraction. The growth of TiO$_2$[s] is such an example. While
TiO$_2$ is very important as nucleation seed, it plays a very minor
role for the cloud particle sizes or the material composition. We are
therefore confident that the number of elements will not alter the
cloud particle sizes nor their material composition substantially.

%\subsection{3D atmosphere simulations}

\section{Conclusion}\label{s:concl}

The results of present global  circulation models, coupled with a
non-equilibrium cloud formation model, suggest that HD\,189\,733b and
HD\,209\,458b are covered with mineral clouds throughout the entire
modelling domain. Both giant gas planets, HD\,189\,733b and
HD\,209\,458b, have similarly chemically complex mineral clouds. Their
cloud particles are a mix that is dominated by Si/Fe/Mg/Al-O silicates
and oxides, and the relative abundance of these materials changes
depending on the local thermodynamic properties. Independent of
longitude and latitude, a strong height-dependence of the cloud
particle material composition and their mean size occur in both
planets. The cloud material composition in the upper and intermediate cloud layers is generally more rich in
HD\,189\,733b than in HD\,209\,458b meaning more materials contribute
to each cloud particle. HD\,209\,458b  generally has a small-grain
cloud part region of several pressure magnitudes (cloud top) with a sharp
growth zone leading to big mixed grains ($\langle a\rangle \approx
10^3\mu$m) with an iron content of $\approx 20$\% at the cloud base.  The cloud particles
in HD\,189\,733b grow to a maximum size of $\langle a\rangle \approx
10^{2.2}\mu$m and are made of 90\% Fe[s] and 10\% TiO$_2$[s] at the cloud base.  If only
the terminators were used to compare the atmospheric structure of
HD\,189\,733b and HD\,209\,458b, one would conclude that they should
be similar as their (T$_{\rm gas}(z)$, p$_{\rm gas}(z)$) structures
are rather similar at these points, and so is the material composition
of the cloud particles. A more global view, however, suggests a more
differentiated picture.

 The chemical diversity of the cloud particles is imprinted into the
 remaining element abundances which is also strongly
 height-dependent. Both, the C/O ratio and the metallicity depend on
 the cloud formation efficiency and can therefore not be treated as
 independent from any of the cloud parameters as it is commonly
 assumed in retrieval methods. It remains an illusion to attribute one
 C/O ratio value to any particular cloud forming planet.

The individual cloud results for HD\,189\,733b and HD\,209\,458b allow
to understand the difference in the detected molecular water on HD\,209\,458b: The
H$_2$O content at 0.1 bars is higher on HD189 733b then in HD209 458b
%in agreement with observations. 
Hydrocarbon and cyanopolyyne molecules
can be rather abundant in the inner, dense part of the atmospheres of
HD\,189\,733b and HD\,209\,458b but non-equilibrium processes will
need to be studied to see if these molecules could reach an
observational abundance also in the upper part of the atmosphere. The
presence of large hydrocarbon molecules is not to be confused with the
formation of hydrocarbon clusters which then could be interpreted as
hydrocarbon hazes.

In conclusion, we have demonstrated that only a detailed modelling of
the cloud formation processes allows to disentangle the cloud
parameter degeneracy that is an artificial result from retrieval
methods by their assumption of the independence of parameters.

\section*{Acknowledgments}

%\bigskip
We highlight financial support of the European Community under the FP7
by the ERC starting grant 257431 and by an ERC advanced grant 247060. JK
acknowledges the Rosen fellowship from the Brooklyn College New York,
US. Some of the calculations for this paper were performed on the
DIRAC Facility jointly funded by STFC, the Large Facilities Capital
Fund of BIS, and the University of Exeter.

\footnotesize{
\bibliographystyle{aa}
\bibliography{bib}{}}

%\clearpage
%\newpage

\begin{appendix} \label{s:appendix}
\section{Cloud property for  HD\,189\,733b and  HD\,209\,458b}

\begin{figure}
\includegraphics[scale=0.4]{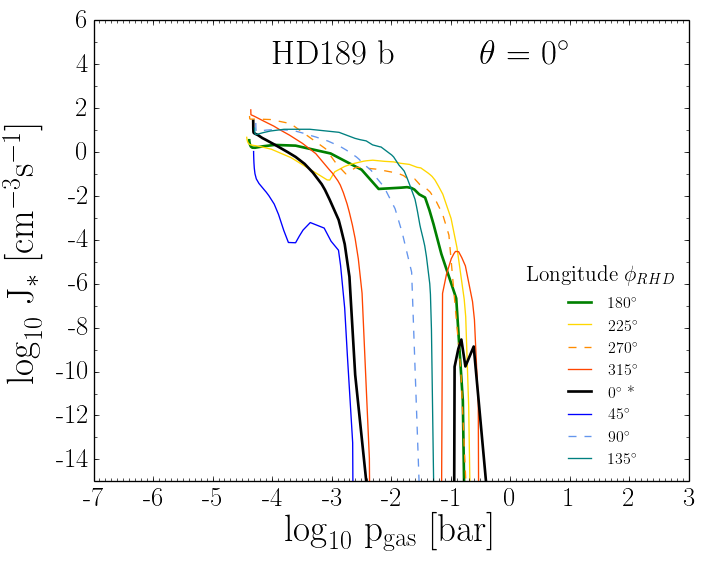}
\includegraphics[scale=0.4]{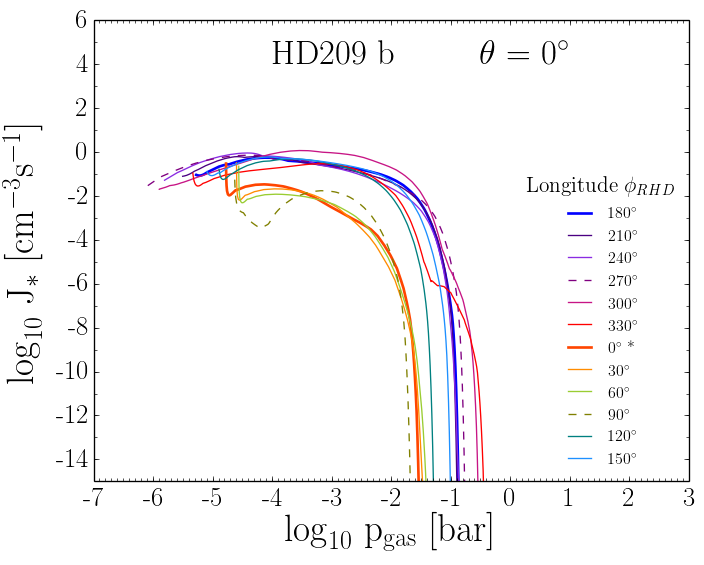}
\caption{The rate of formation of condensation seeds (nucleation
  rate), $J_*$ [cm$^{-3}$ s$^{-1}$], along the equator. HD\,189\,733b -- top, 
  HD\,209\,458b -- bottom. }
\label{fig:J*}
\end{figure}
\begin{figure}
\includegraphics[scale=0.4]{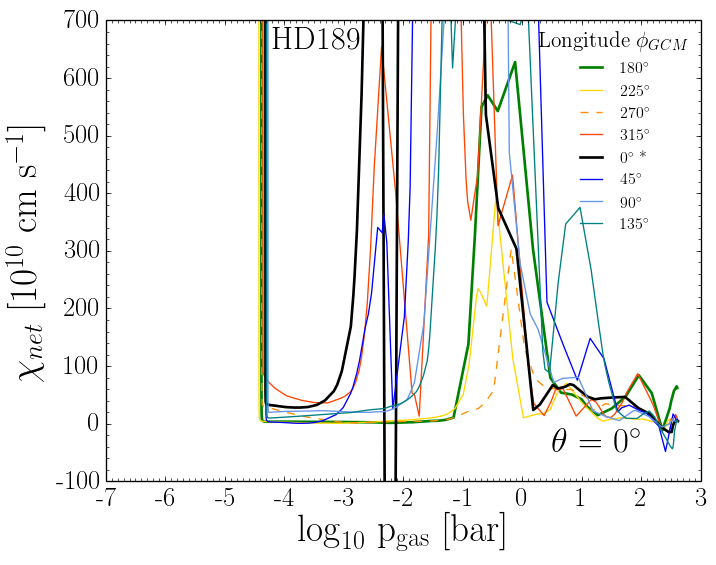}
\includegraphics[scale=0.4]{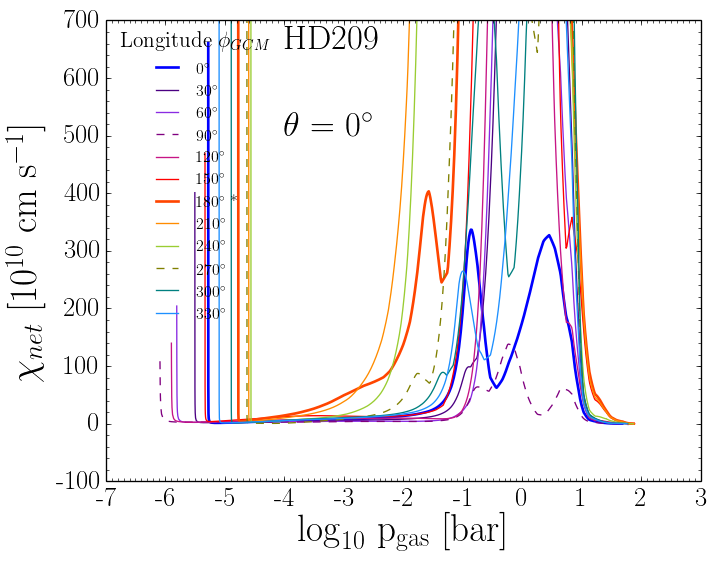}
\caption{The net growth velocity of cloud particles at longitudes along the
  equator.  HD\,189\,733b -- top, HD\,209\,458b -- bottom.}% {\bf Should be $10^{-10}$!!!}}
\label{fig:chi}
\end{figure}
\begin{figure*}
\includegraphics[scale=0.4]{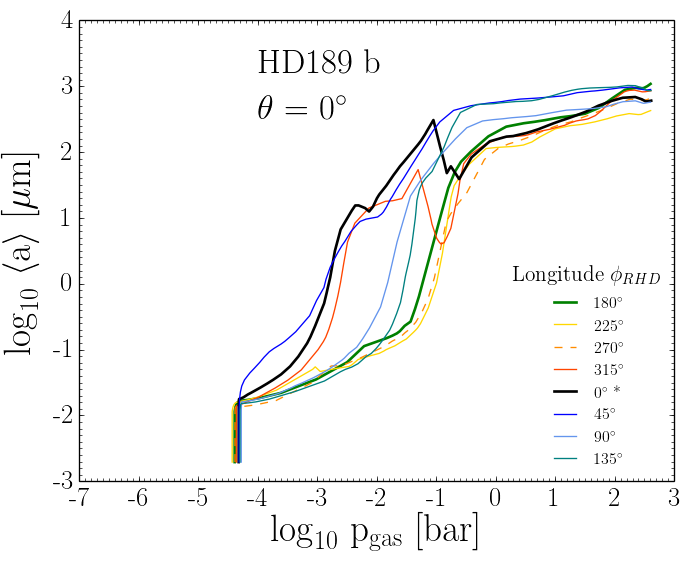}
\includegraphics[scale=0.4]{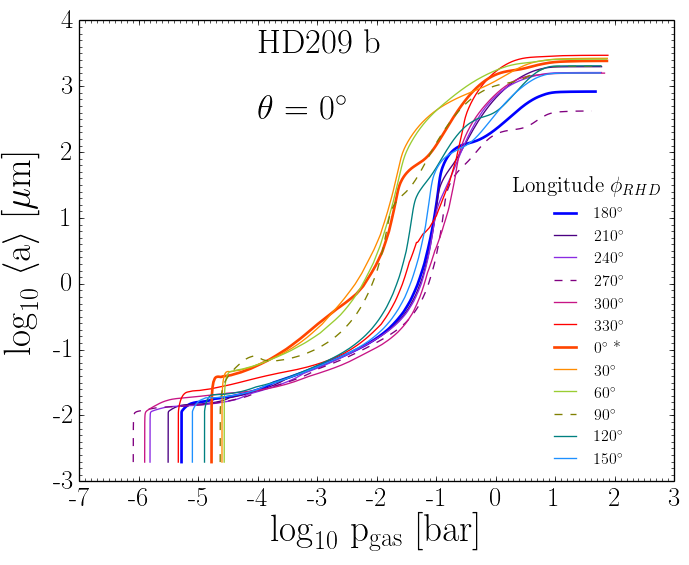}
\caption{Cloud particle mean grain size, $\langle a\rangle$ [$\mu$m]
  along the equator. HD\,189\,733b -- top, HD\,209\,458b -- bottom.}
\label{fig:amean}
\end{figure*}

%\subsection{Time-scale comparison}
%Figure~\ref{fig:taus} compares the dust formation time scales ($\tau_{\rm nuclea}$, $\tau_{\rm gr}$) and the large-scale dynamic time scales ($\tau_{\rm sink}$, $\tau_{\rm mix}$) that determine the cloud formation.

%\noindent
%\begin{figure}
%\includegraphics[scale=0.4]{plots/timescales_0_0_HD189}
%\includegraphics[scale=0.4]{plots/timescales_180_0}
%\caption{Times scales at the Substellar Point: {\bf check with Graham's results; do we need that figure?}}
%\label{fig:taus}
%\end{figure}

\subsection{Equilibrium gas-phase abundance}
\begin{figure*}
\centering
\begin{tabular}{cc}
\hspace*{-1.0cm}\includegraphics[scale=0.5]{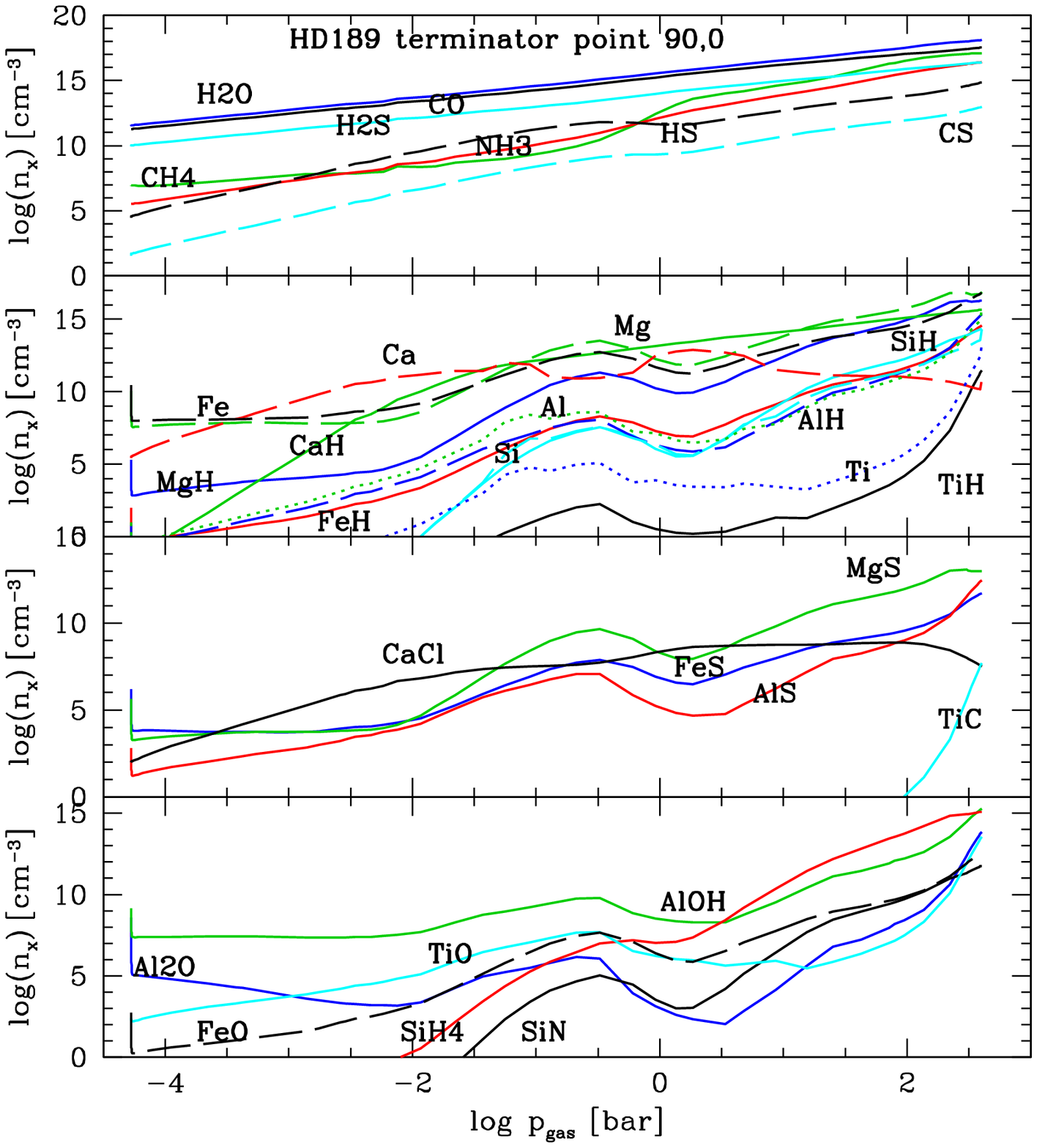} & 
\hspace*{-1.5cm}\includegraphics[scale=0.5]{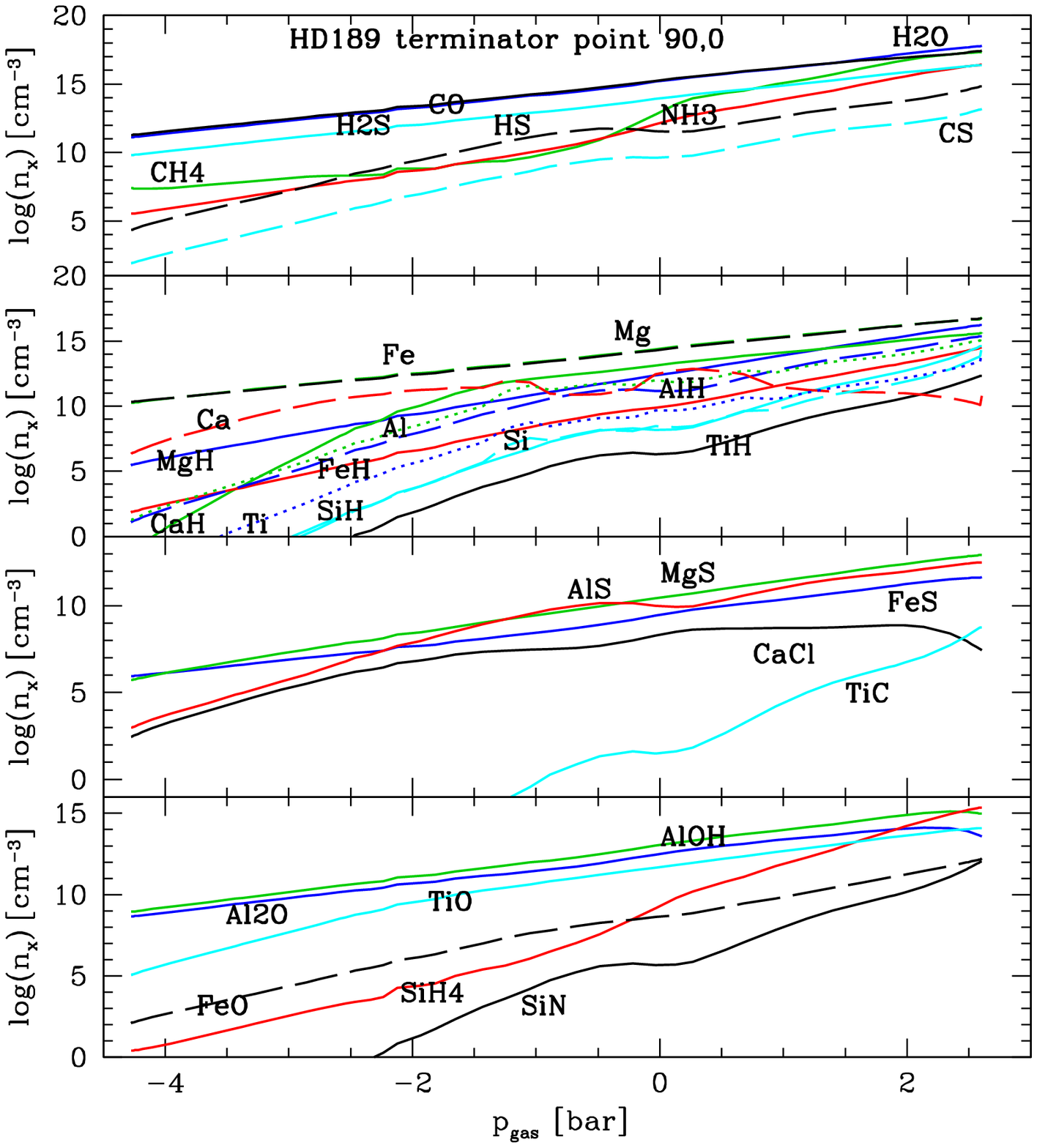}\\*[-4.5cm]
\hspace*{-1.0cm}\includegraphics[scale=0.5]{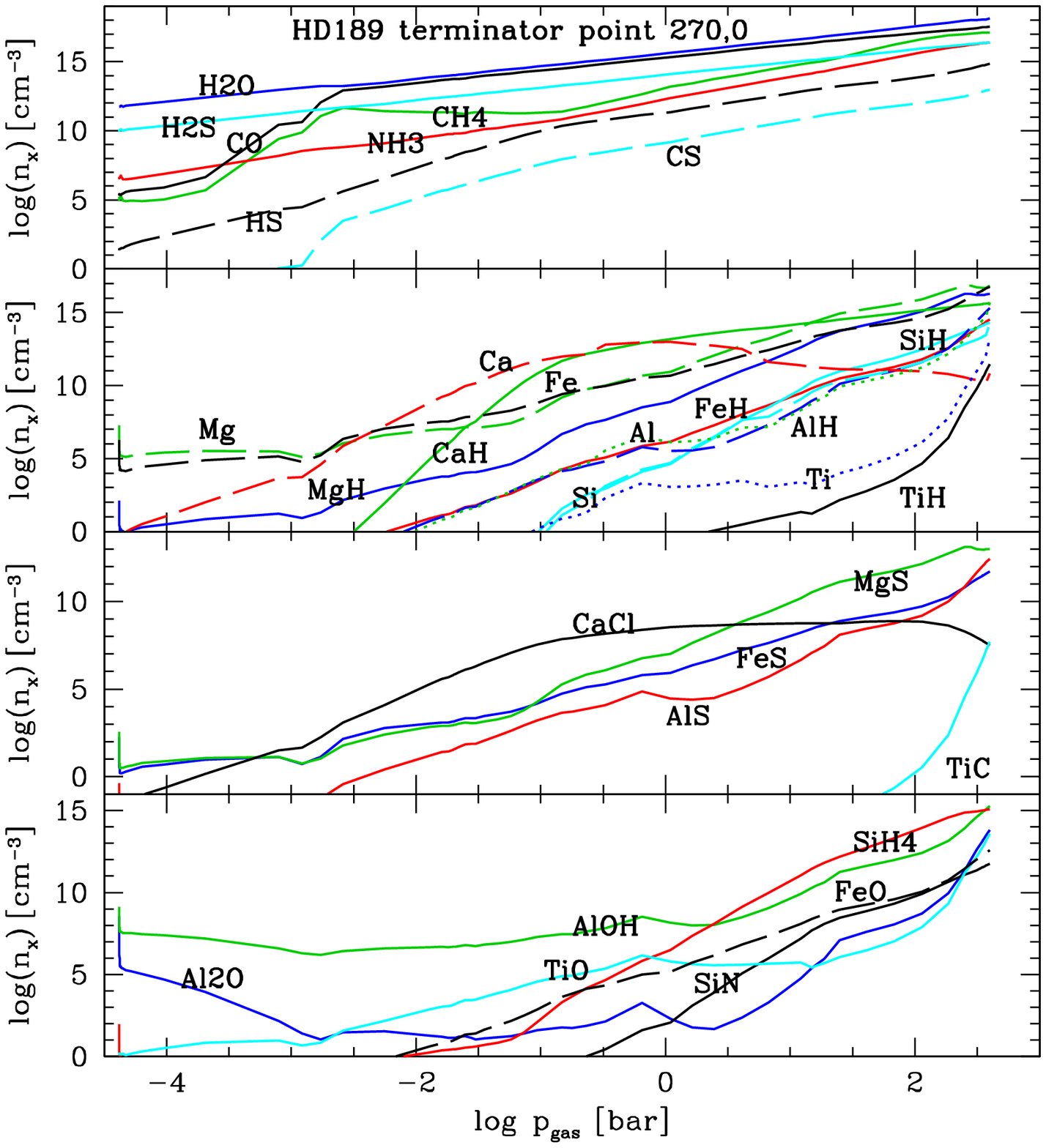} & 
\hspace*{-1.5cm}\includegraphics[scale=0.5]{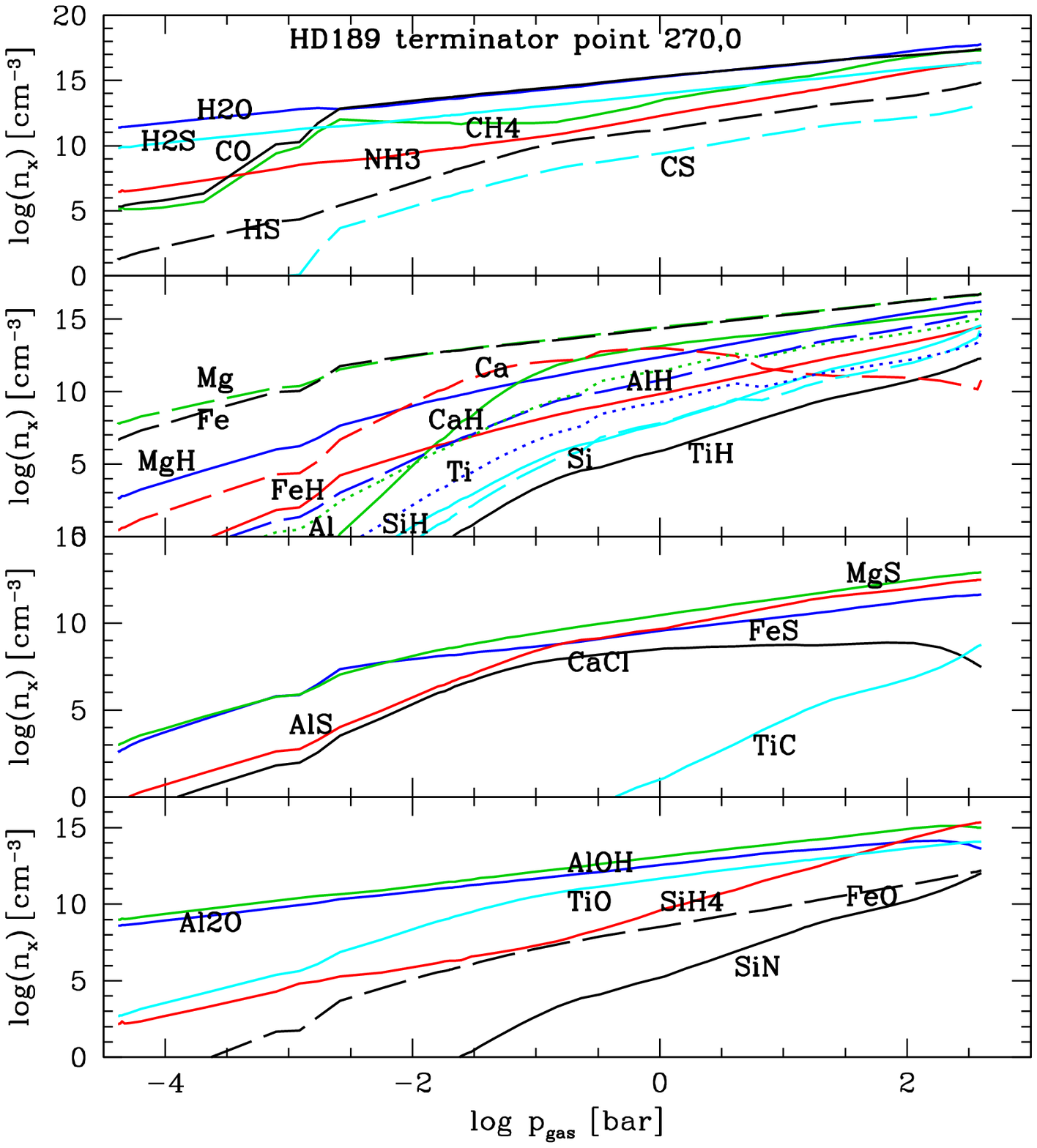}\\*[-3.8cm]
\end{tabular}
\caption{Gas-phase composition  HD\,189\,733b at the  two terminators ($\theta=0^{\circ}$, $\phi=90^{\circ}, 270^{\circ}$): O Al, S, Mg, Ti Fe, Si Ca.  {\bf Left:}  dust-depleted elements according to right of Fig~\ref{fig:eabun}, {\bf Right:} un-depleted (solar abundance) gas-phase.}
\label{fig:molabun1899020}
\end{figure*}

\begin{figure*}
\centering
\begin{tabular}{cc}
\hspace*{-1.0cm}\includegraphics[scale=0.5]{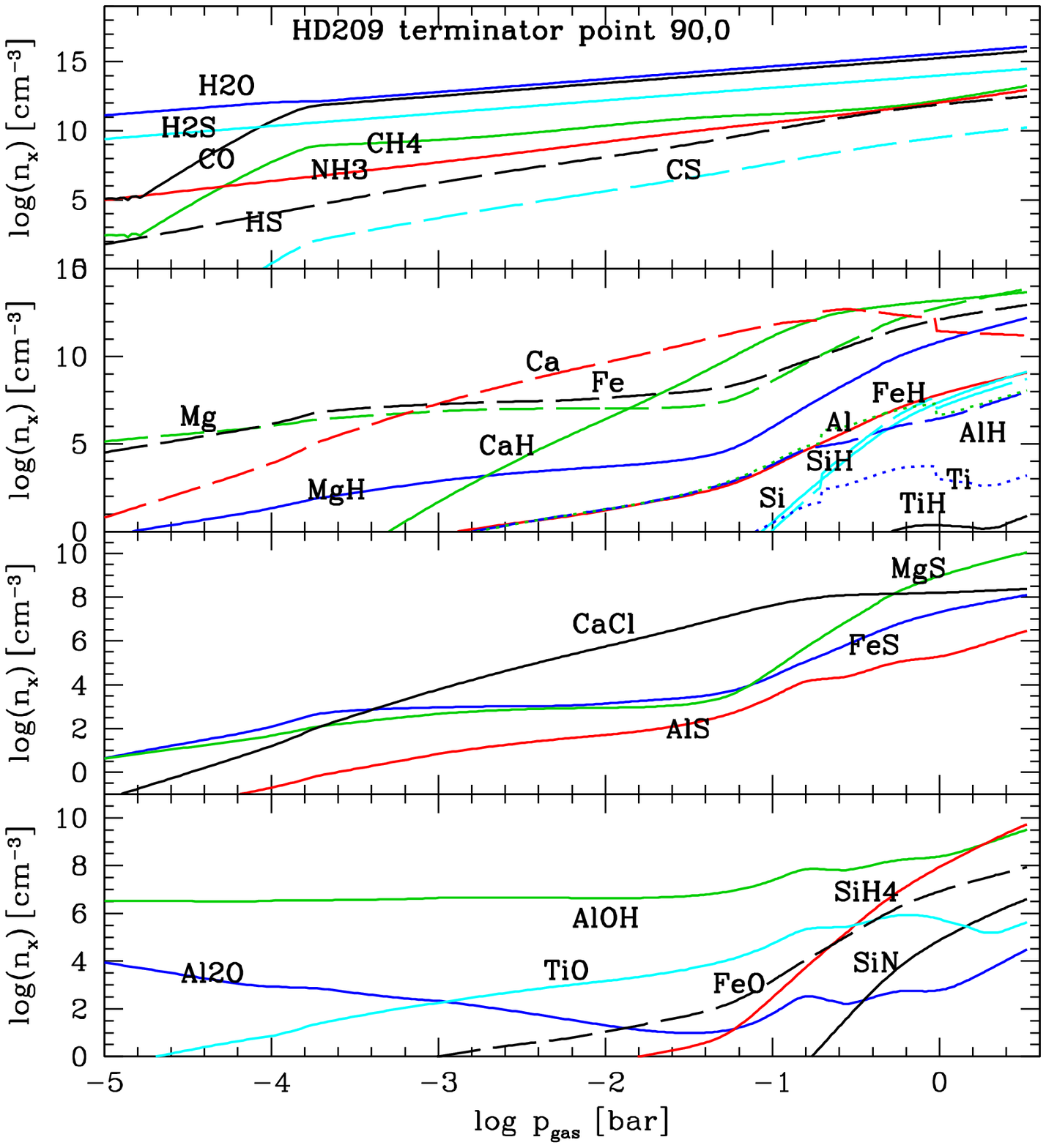} & 
\hspace*{-1.5cm}\includegraphics[scale=0.5]{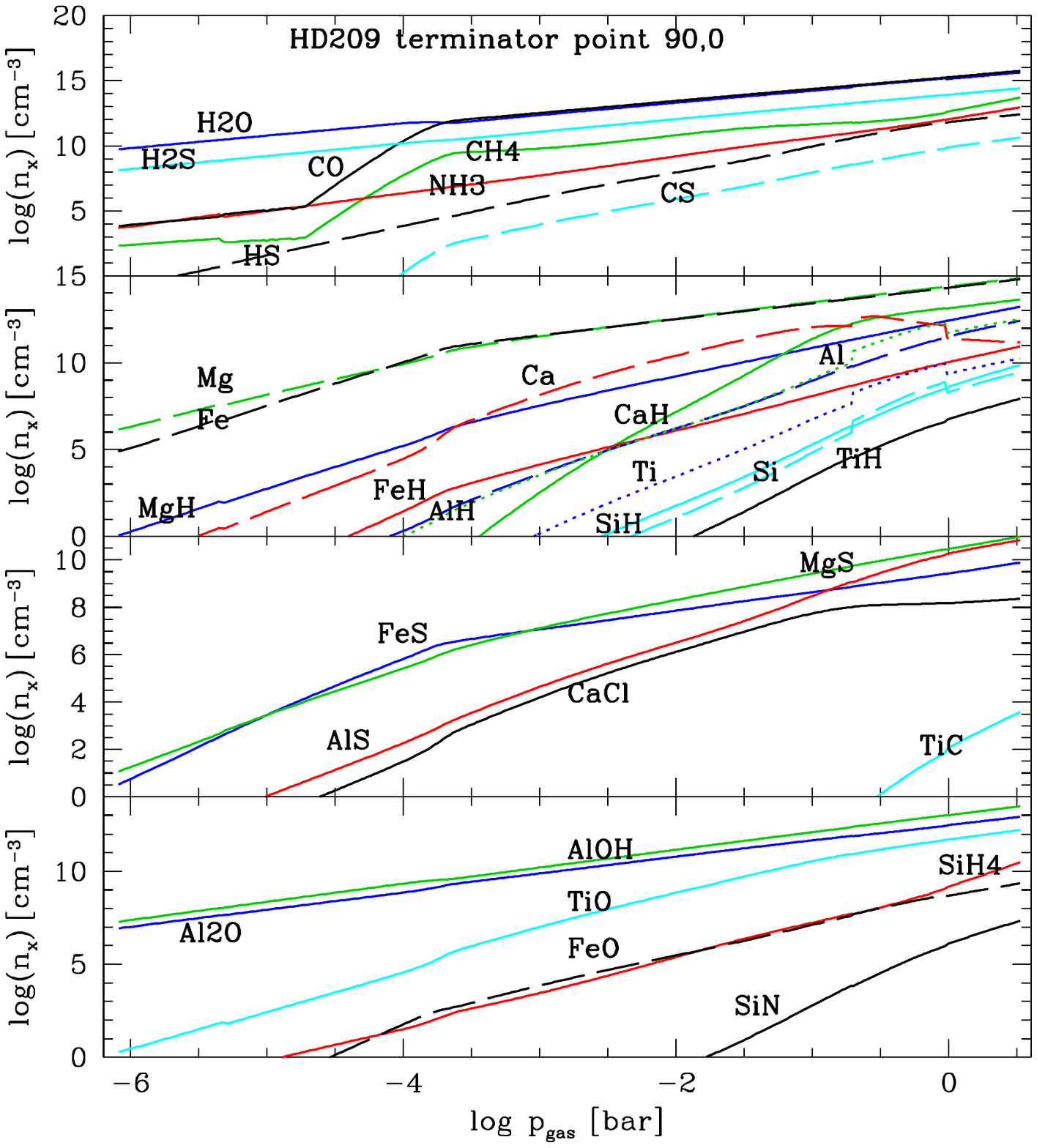}\\*[-4.5cm]
\hspace*{-1.0cm}\includegraphics[scale=0.5]{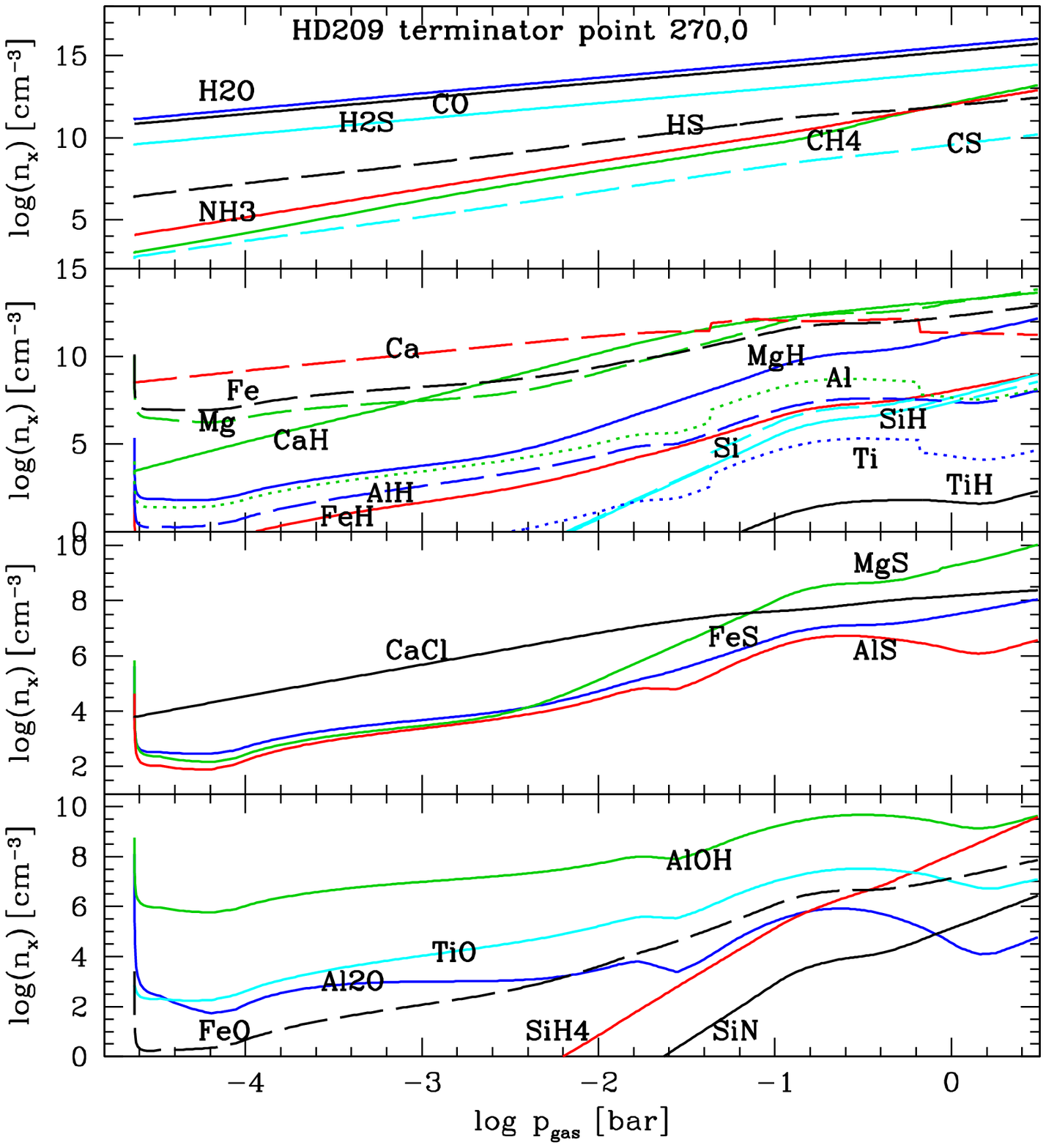} & 
\hspace*{-1.5cm}\includegraphics[scale=0.5]{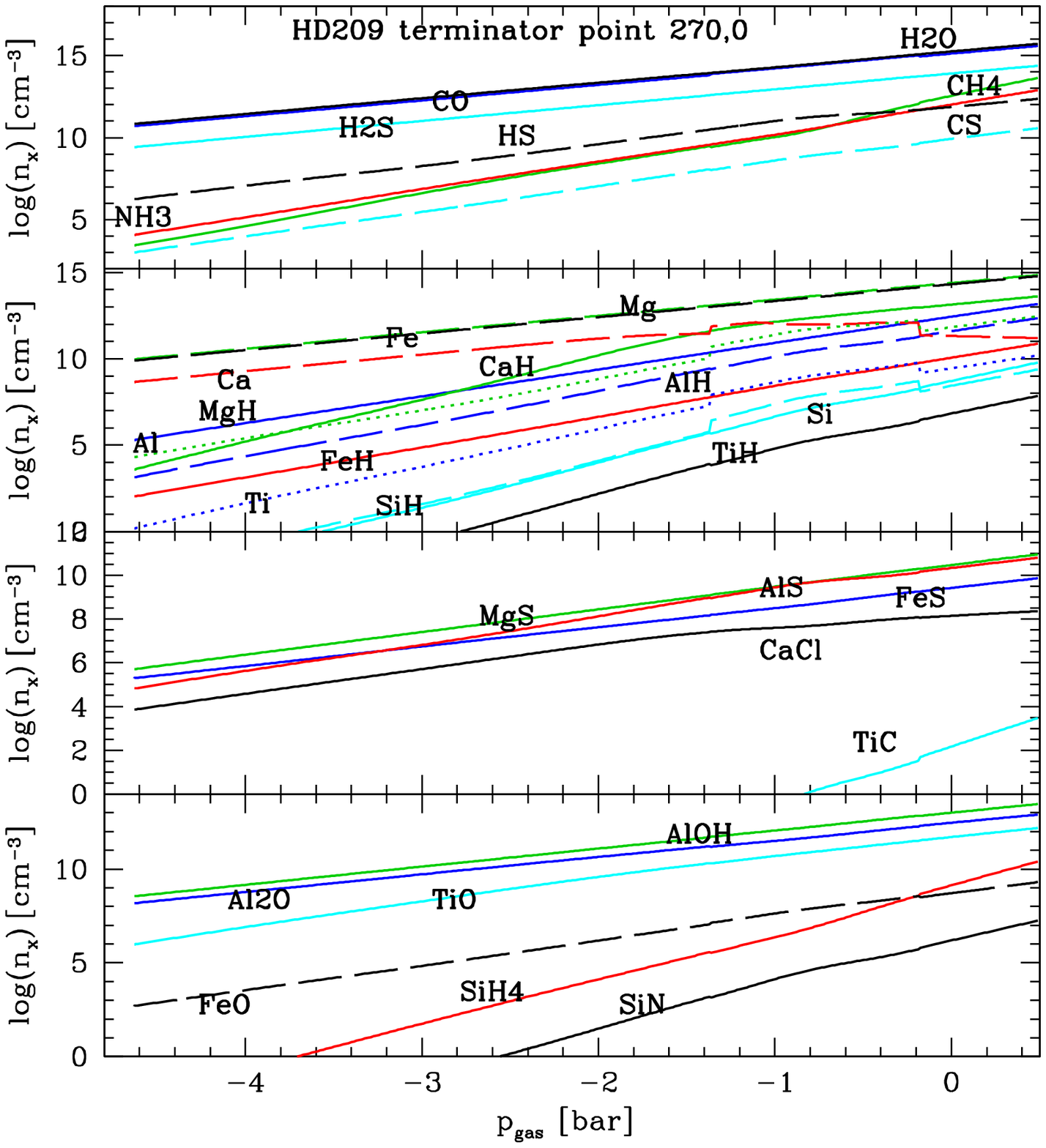}\\*[-3.8cm]
\end{tabular}
\caption{Gas-phase composition  HD\,209\,458b at the  two terminators ($\theta=0^{\circ}$, $\phi=90^{\circ}, 270^{\circ}$): O Al, S, Mg, Ti Fe, Si Ca.  {\bf Left:}  dust-depleted elements according to right of Fig~\ref{fig:eabun}, {\bf Right:} un-depleted (solar abundance) gas-phase.}
\label{fig:molabun20990270}
\end{figure*}

The equilibrium gas-abundance of the most dominating gas-species at
the terminator points for HD189\,733b (Fig.~\ref{fig:molabun1899020})
and for HD209\,458b (Fig.~\ref{fig:molabun20990270}) as complementary
material to the results shown in the main text for the substellar
points only.

\begin{figure*}
\begin{tabular}{cc}
\includegraphics[scale=0.4]{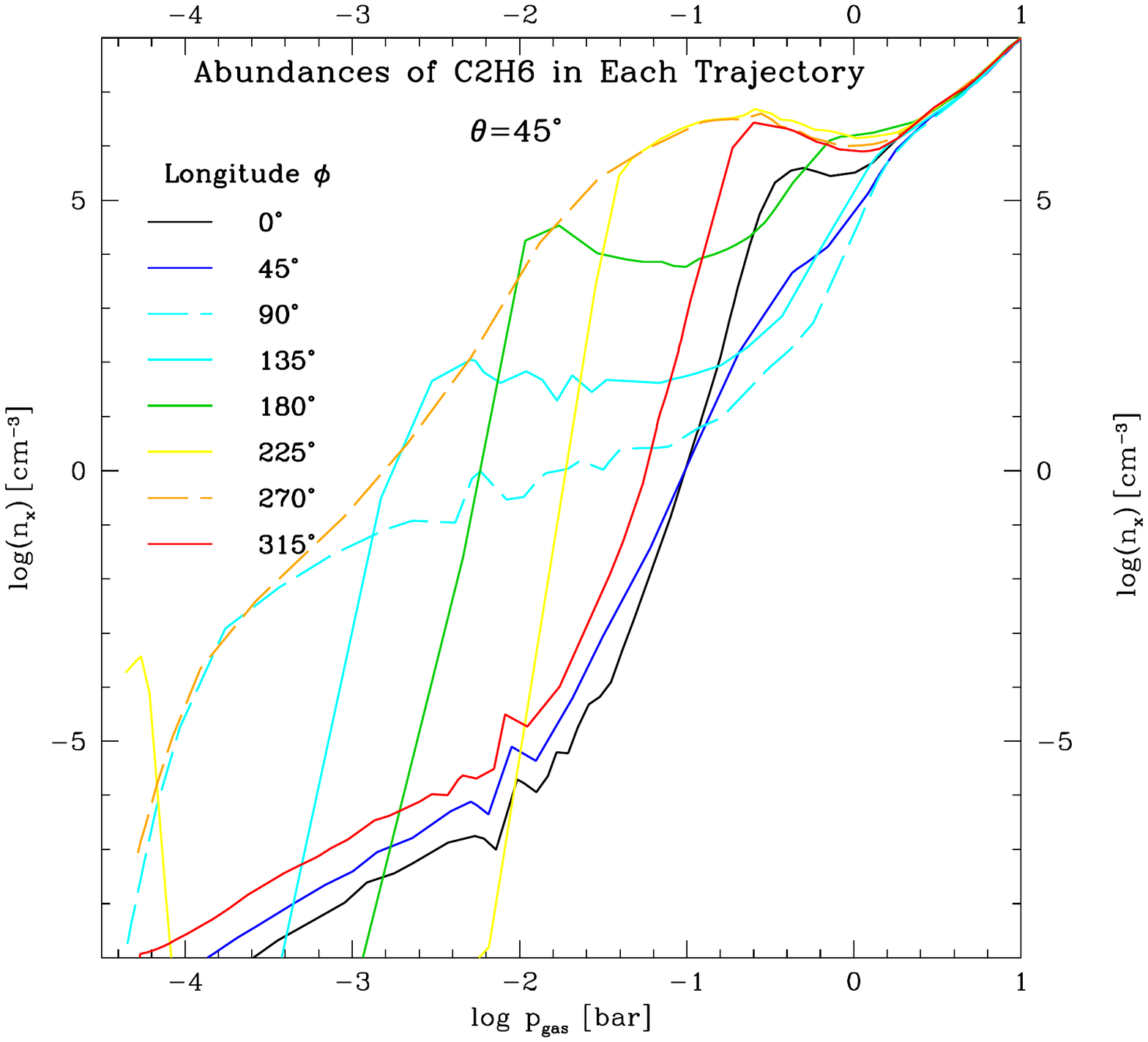} & 
\includegraphics[scale=0.4]{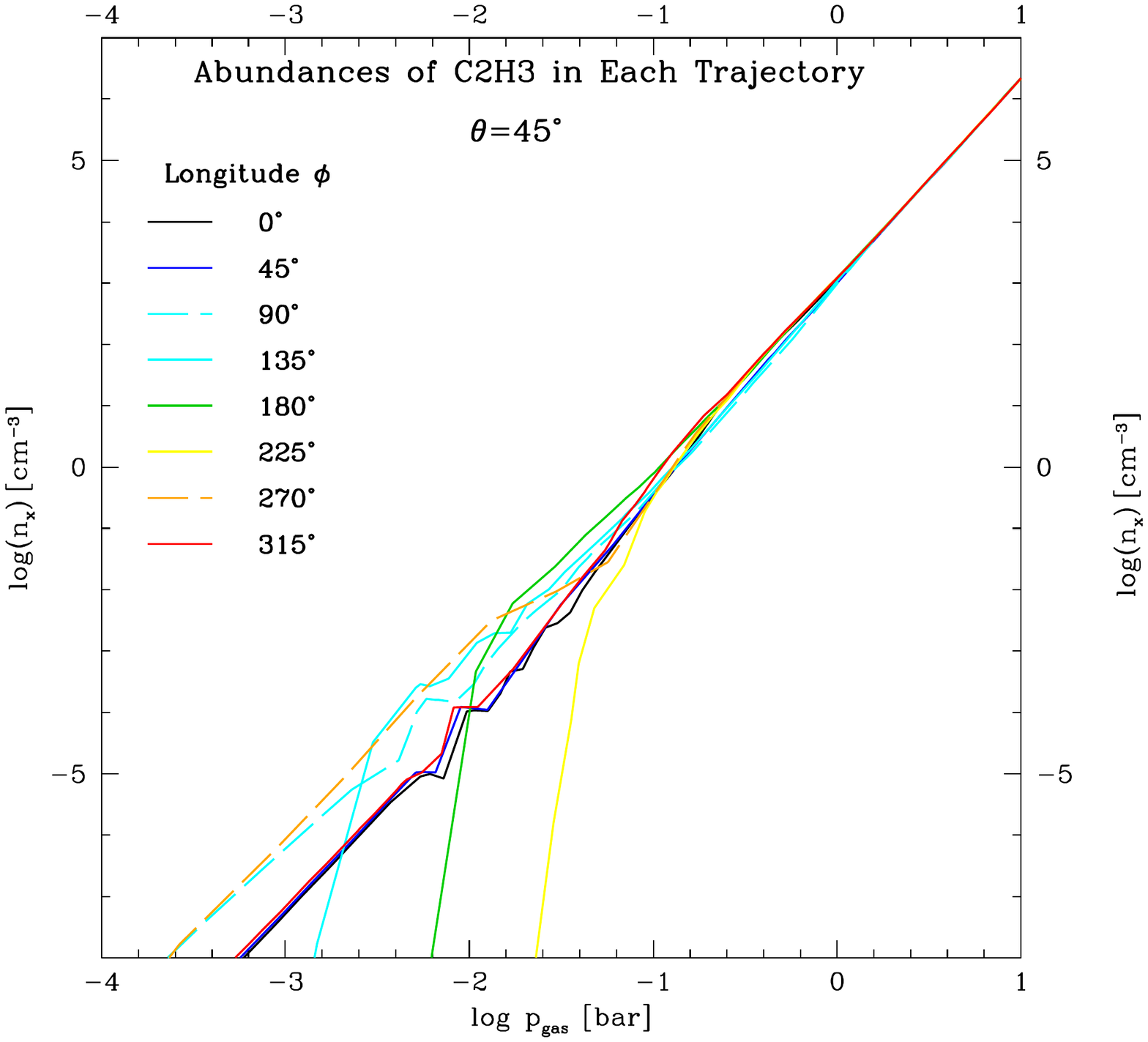}\\*[-4.2cm]
\includegraphics[scale=0.4]{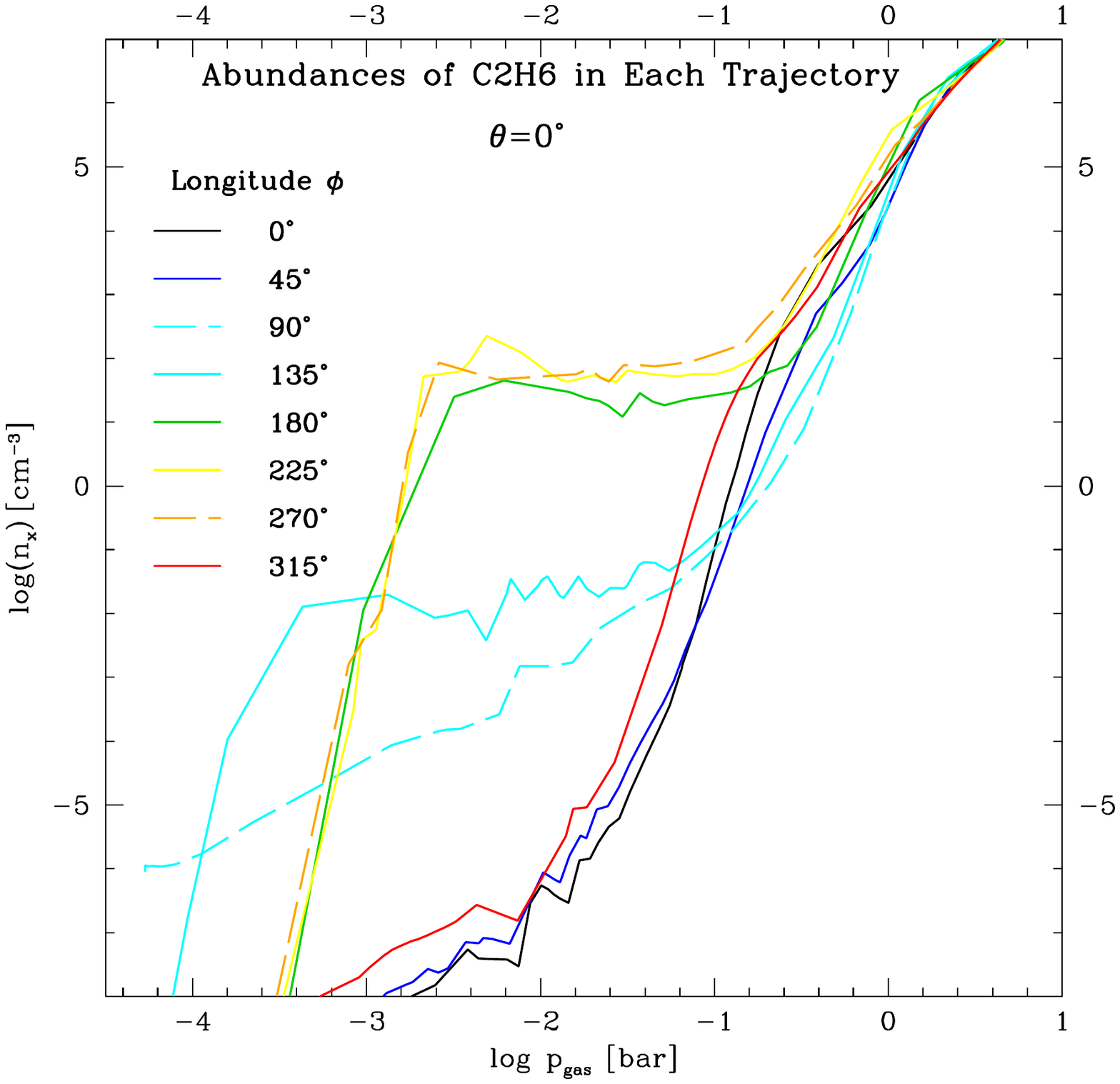} & 
\includegraphics[scale=0.4]{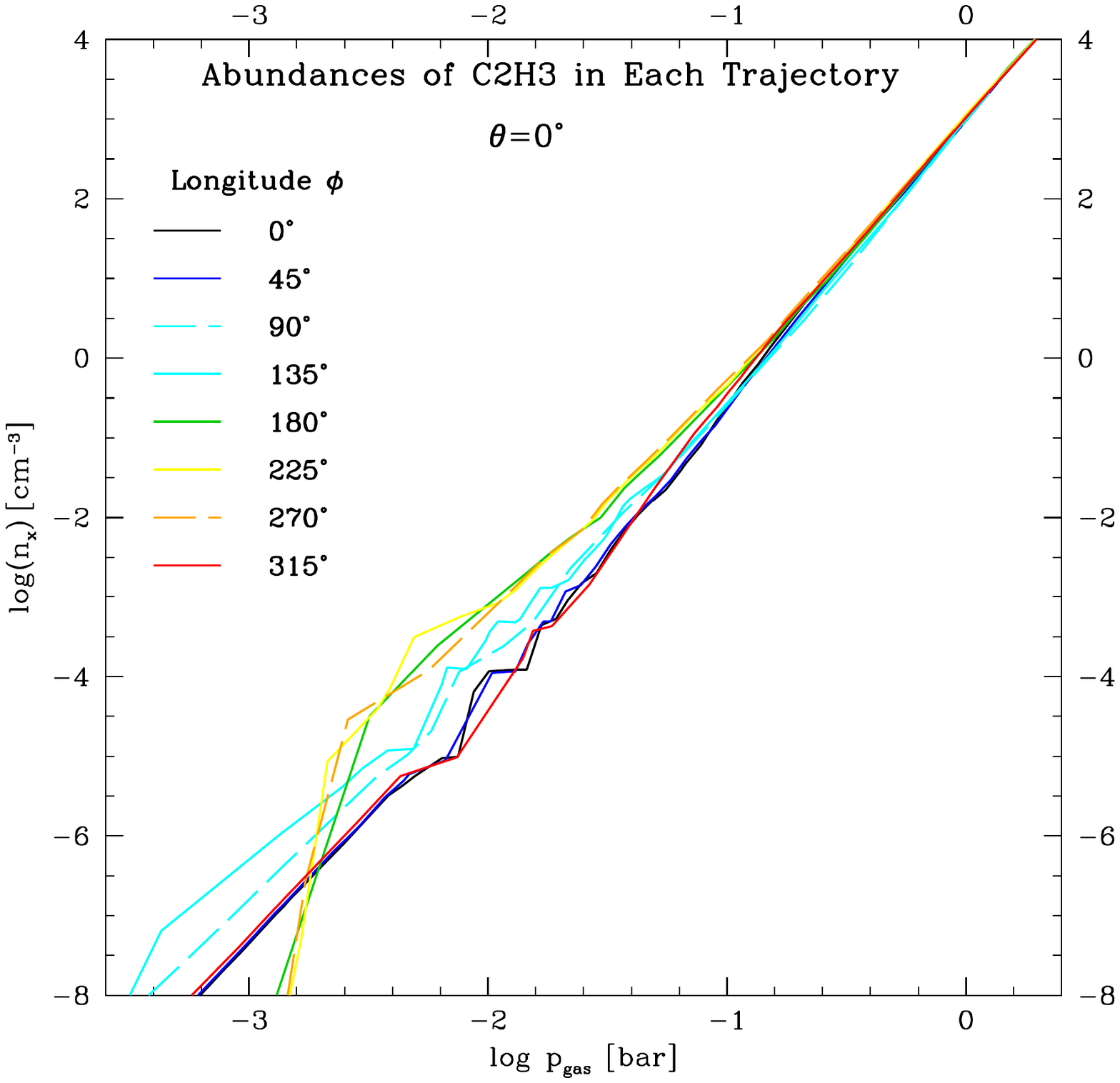}\\*[-2.5cm]
\end{tabular}
\caption{The abundances of the most abundant hydrocarbon molecules in HD\,189\,733b along the equator (bottom) and in the northern hemisphere (top).}
\label{fig:C2H6HD189}
\end{figure*}

\begin{figure*}
\begin{tabular}{cc}
\includegraphics[scale=0.4]{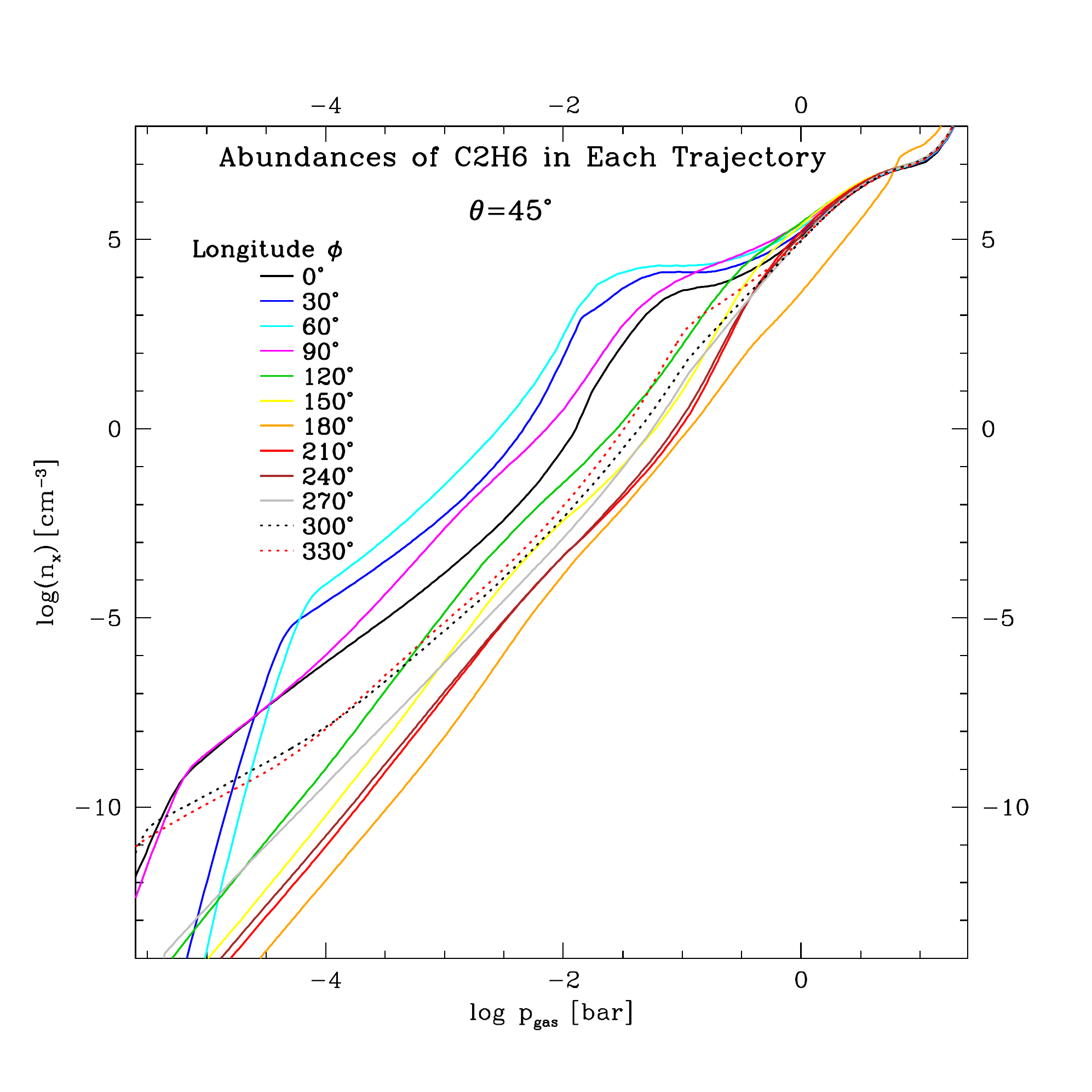} & 
\includegraphics[scale=0.4]{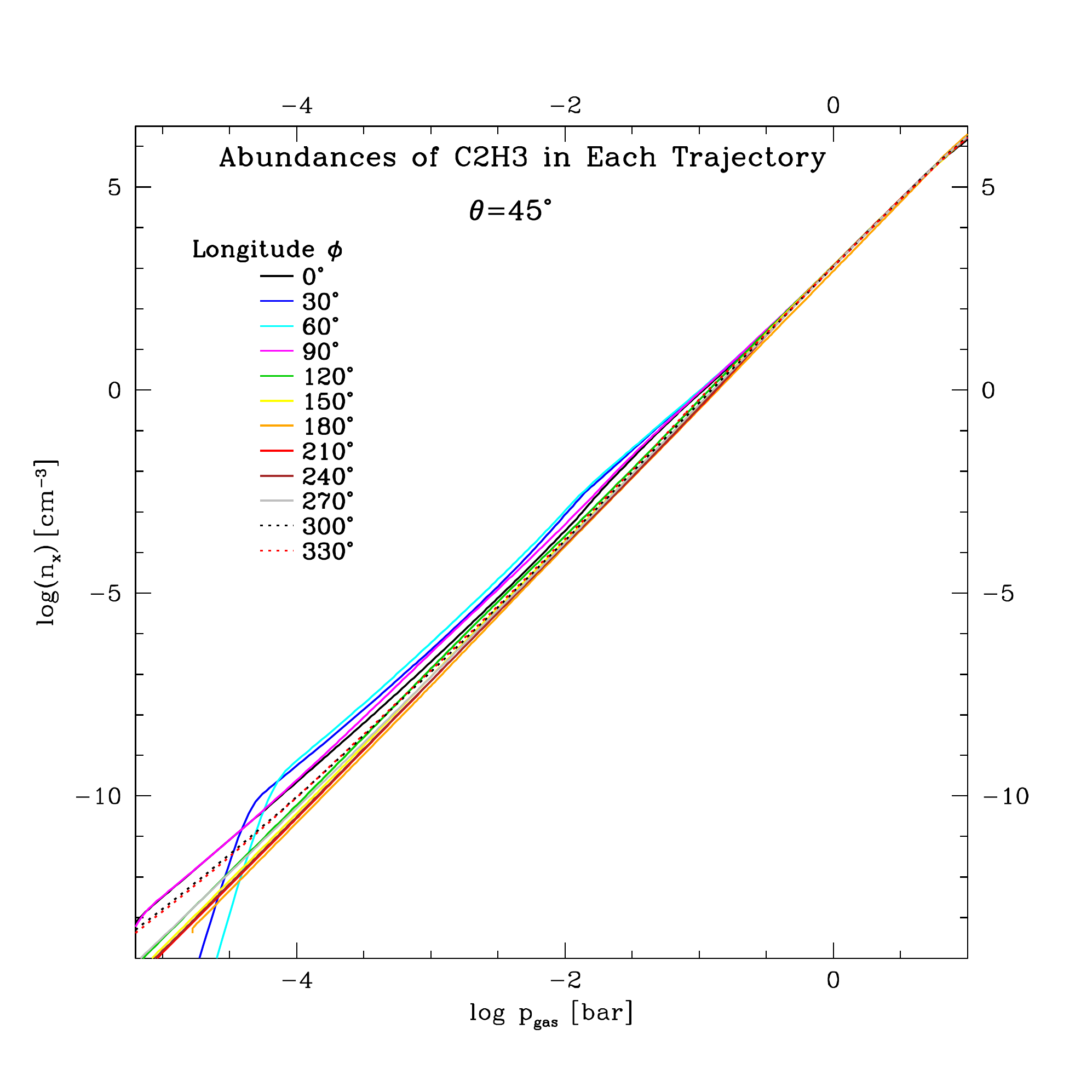}\\*[-1.2cm]
\includegraphics[scale=0.4]{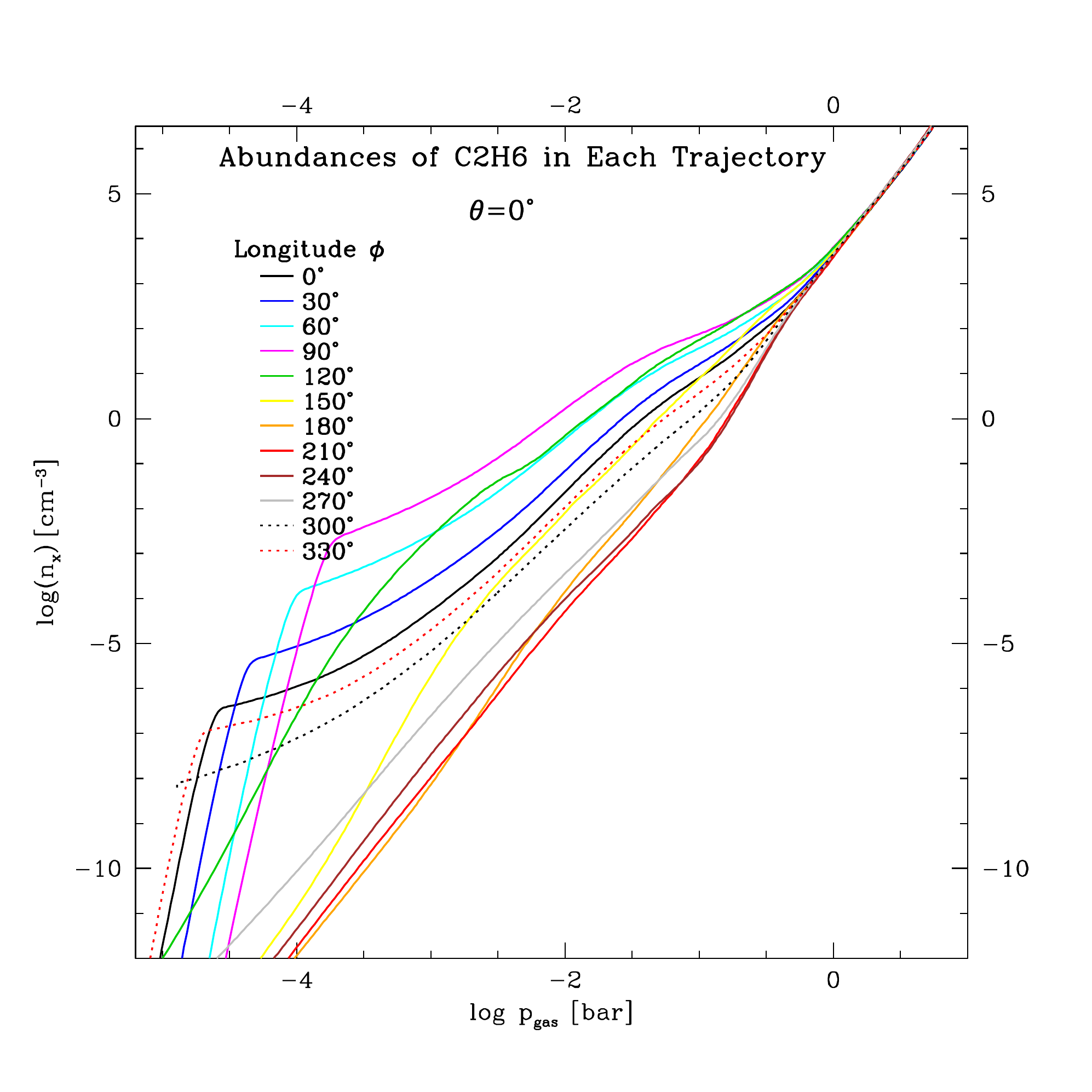} & 
\includegraphics[scale=0.4]{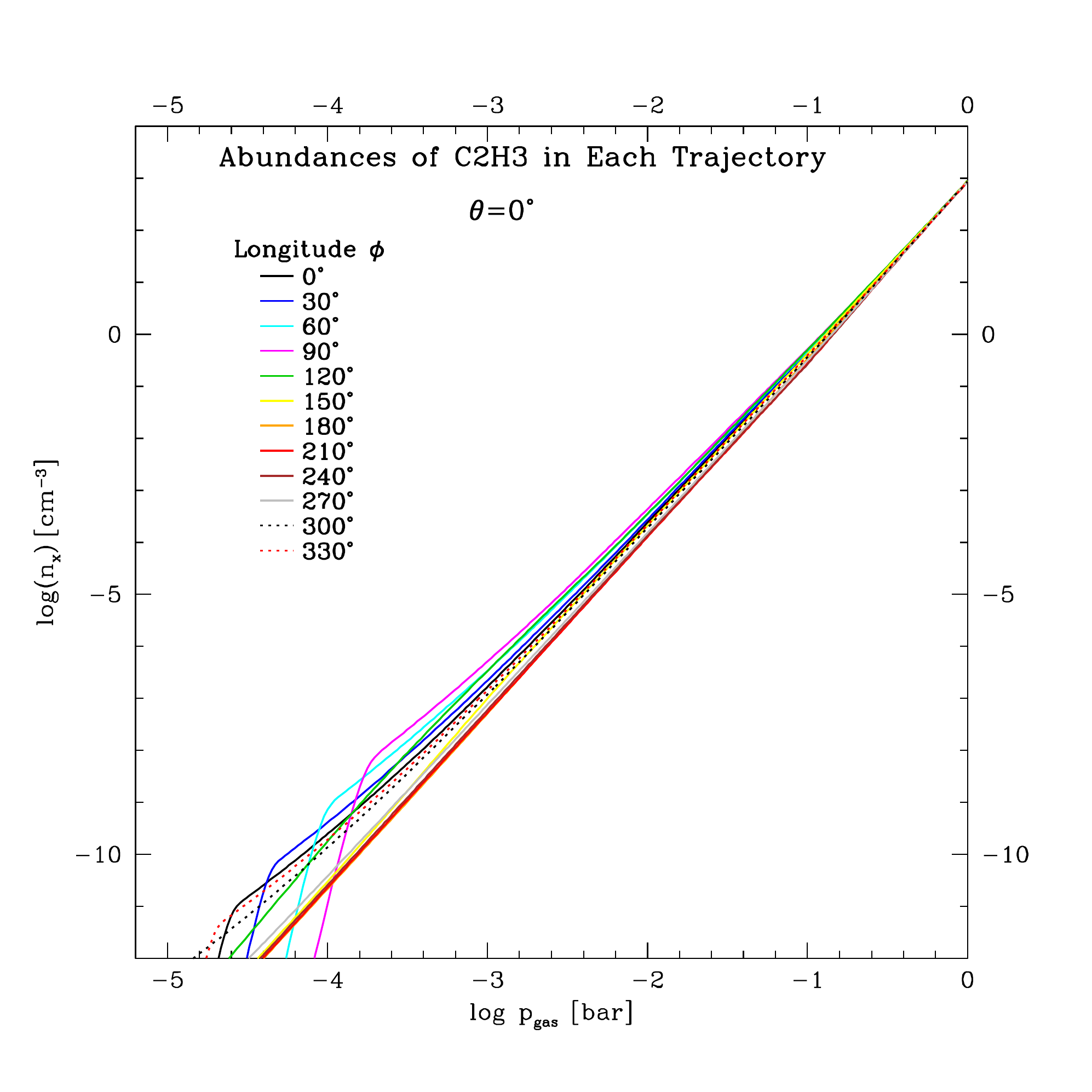}\*[-0.5cm]
\end{tabular}
\caption{The abundances of the most abundant hydrocarbon molecules in HD\,209\,458b along the equator (bottom) and in the northern hemisphere (top).}
\label{fig:C2H6HD189}
\end{figure*}

\clearpage
\newpage
\subsection{Cloud absorption, scattering and extinction}
\begin{figure*}
\begin{tabular}{cc}
\includegraphics[scale=0.35]{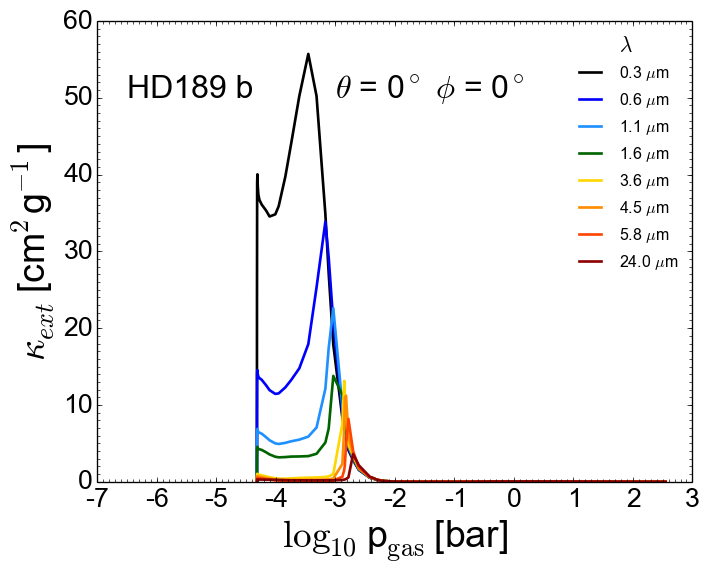} & \includegraphics[scale=0.35]{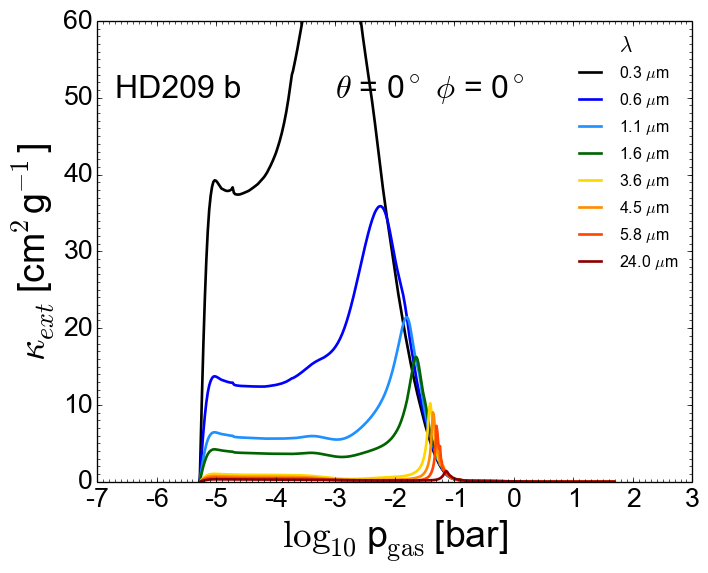}\\
\includegraphics[scale=0.35]{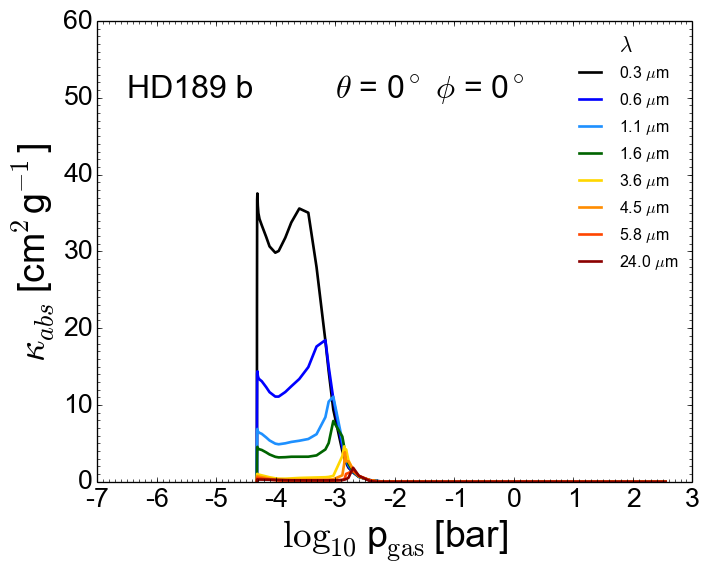} & \includegraphics[scale=0.35]{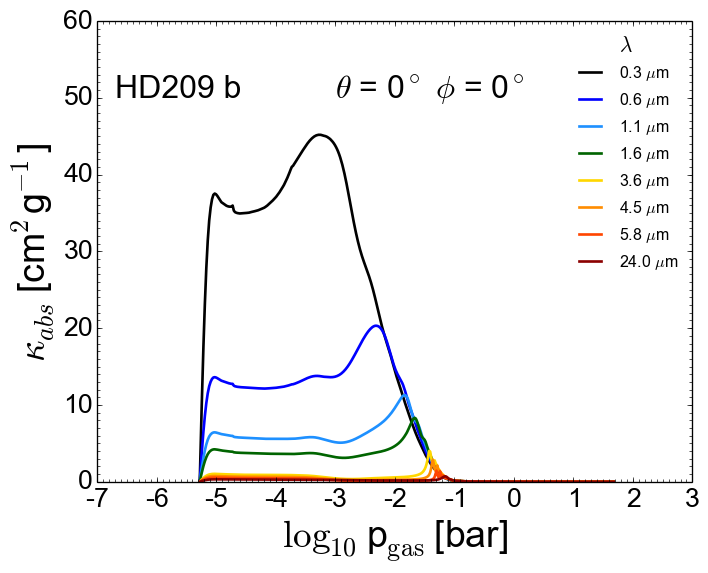}\\
\includegraphics[scale=0.35]{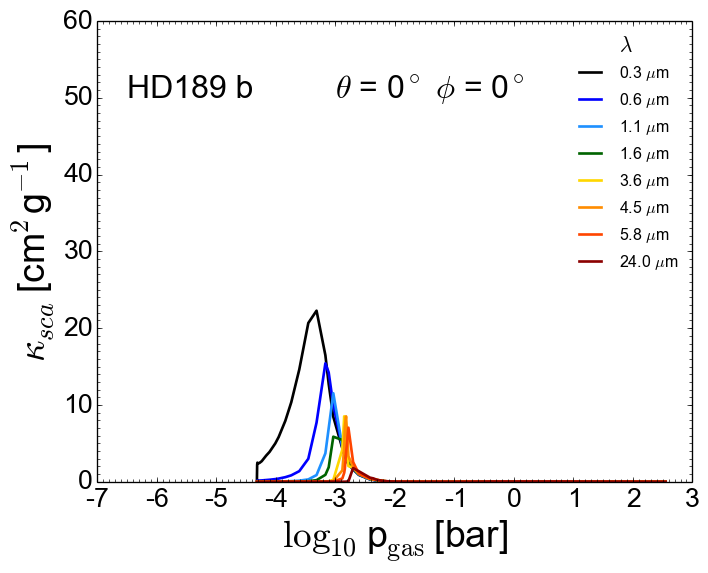} & \includegraphics[scale=0.35]{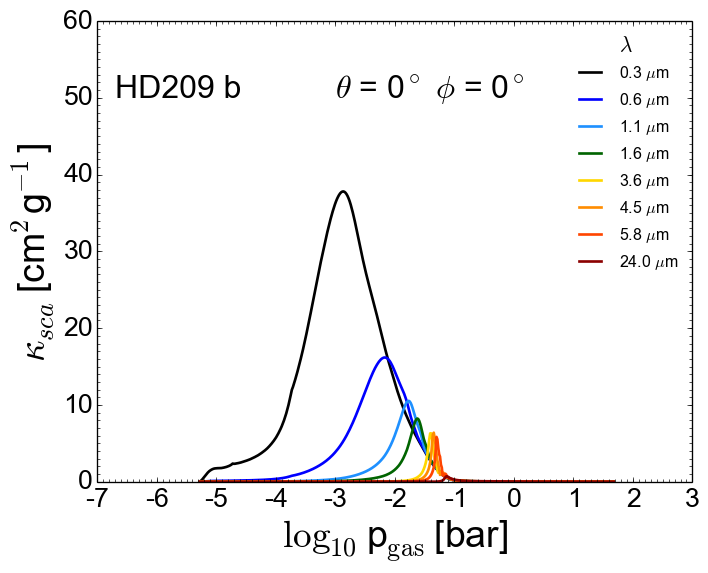}\\
\includegraphics[scale=0.35]{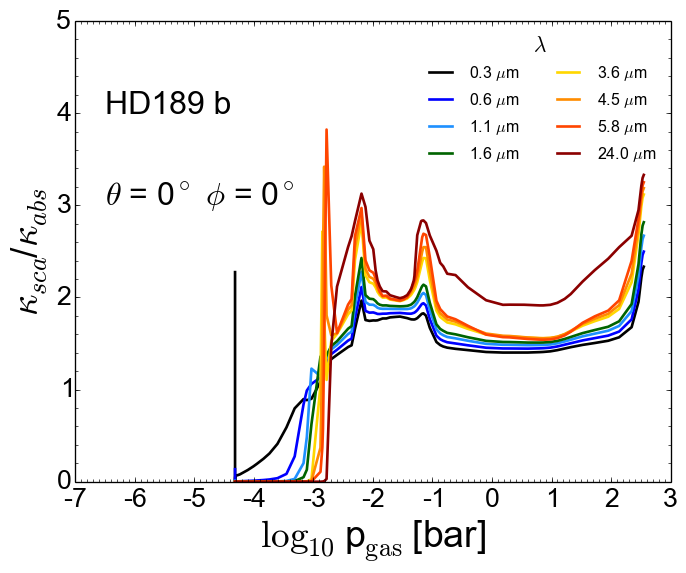} & \includegraphics[scale=0.35]{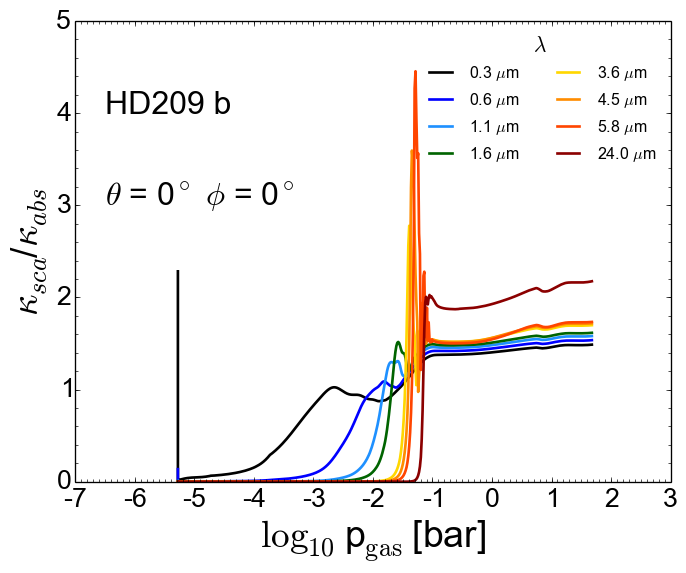}
\end{tabular}
\caption{Cloud extinction (1st row), absorption (2nd row), scattering (3rd row)  and  $\kappa_{\rm sca}/\kappa_{\rm abs}$  (4th row) at the substellar point, Substellar point, $\theta=0^{\circ}$}
\label{fig:opac00detailed}
\end{figure*}

The break-down of the contribution of cloud absorption and cloud scattering to the total cloud extinction at the terminator points for HD189\,733b (Fig.~\ref{fig:opac00detailed}) and for HD209\,458b as complementary material to the results shown in the main text in form of  2D-tables for the total extinction sampled across the globe (Figs.~\ref{fig:opacHD189},~\ref{fig:opacHD209}).

\begin{figure*}
\begin{tabular}{cc}
\includegraphics[scale=0.35]{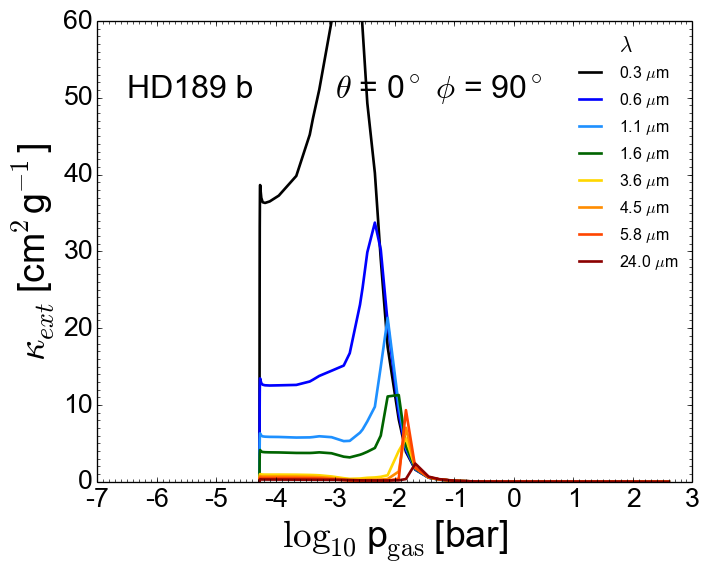} & \includegraphics[scale=0.35]{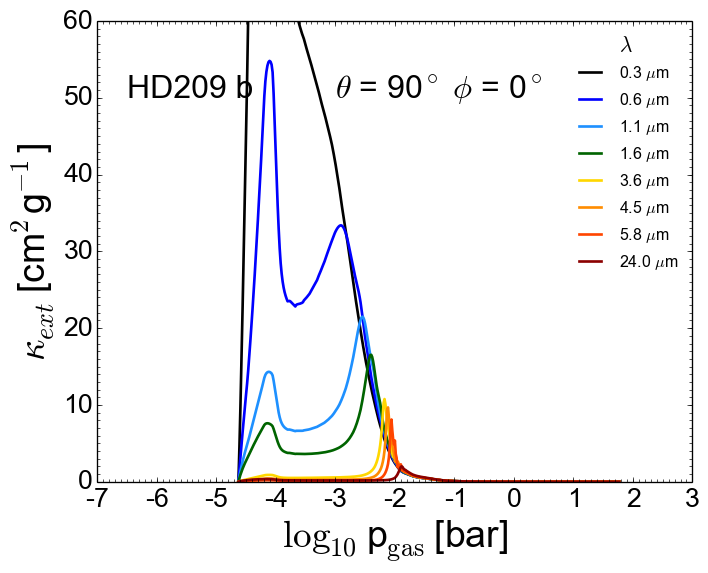}\\
\includegraphics[scale=0.35]{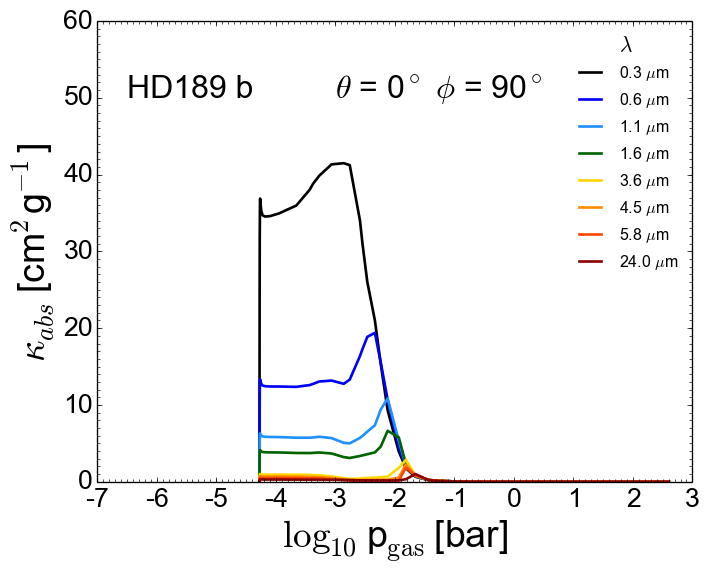} & \includegraphics[scale=0.35]{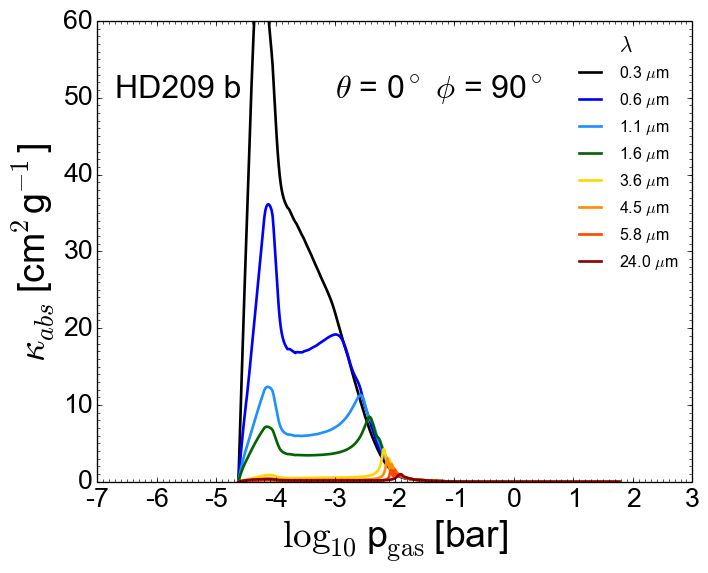}\\
\includegraphics[scale=0.35]{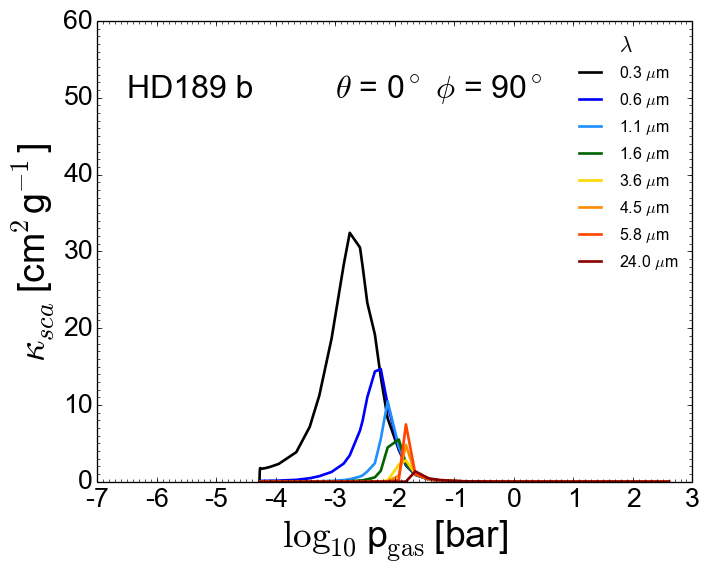} & \includegraphics[scale=0.35]{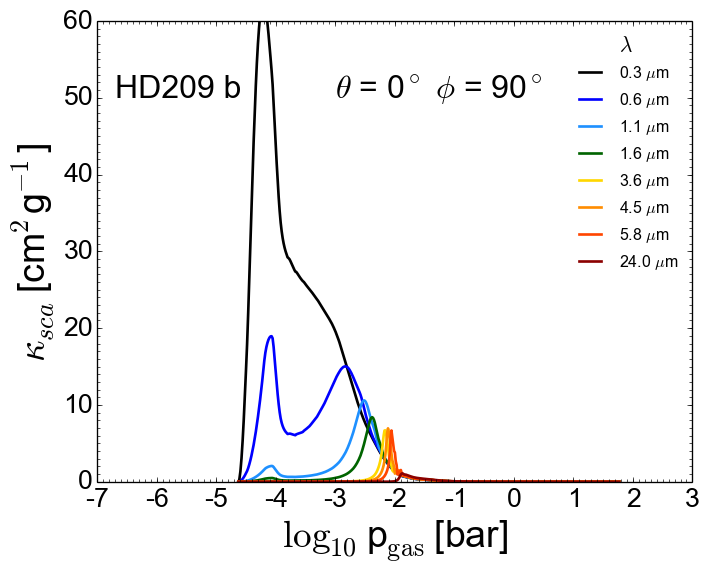}\\
\includegraphics[scale=0.35]{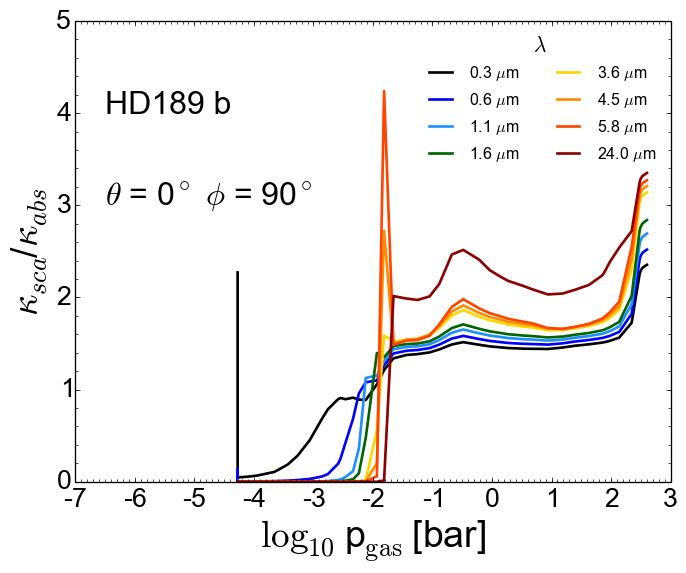} & \includegraphics[scale=0.35]{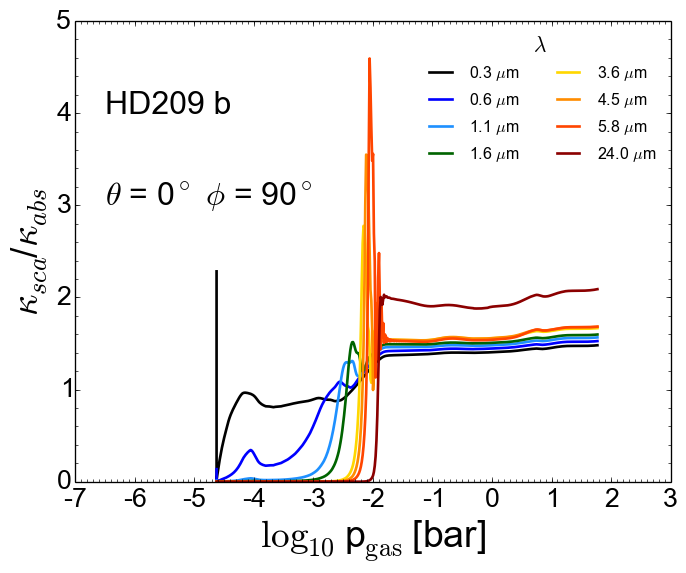}
\end{tabular}
\caption{Cloud extinction (1st row), absorption (2nd row), scattering (3rd row)  and  $\kappa_{\rm sca}/\kappa_{\rm abs}$  (4th row) at the terminator, $\theta=90^{\circ}$, of HD189\,733b.}
\label{fig:opac90detailed}
\end{figure*}

\end{appendix}

\end{document}